\documentstyle[sprocl,epsf,rotate]{article}
\def\lsim{\mathrel{\rlap{\lower4pt\hbox{\hskip1pt$\sim$}}
    \raise1pt\hbox{$<$}}}         
\def\gsim{\mathrel{\rlap{\lower4pt\hbox{\hskip1pt$\sim$}}
    \raise1pt\hbox{$>$}}}         

\input psfig.sty

\def\Journal#1#2#3#4{{#1} {\bf #2}, #3 (#4)}

\def\NPB{{\em Nucl. Phys.} B}
\def\PLB{{\em Phys. Lett.}  B}
\def\PRL{\em Phys. Rev. Lett.}
\def\PRD{{\em Phys. Rev.} D}
\def\PRC{{\em Phys. Rev.} C}

\def\ARAA{\em Ann. Rev. Astron. Astrophys.}
\def\AJ{\em Ap. J.}
\def\AJS{\em Ap. J. Suppl.}
\def\RMP{\em Rev. Mod. Phys.}
\def\N{\em Nature}
\def\SJNP{\em Sov. J. Nucl. Phys.}
\def\PPNP{\em Prog. Part. Nucl. Phys.}
\def\AA{\em Astron. Astrophys.}

\def\be{\begin{equation}}
\def\ee{\end{equation}}
\def\bea{\begin{eqnarray}}
\def\eea{\end{eqnarray}}

\begin{document}

\title{SOLAR, SUPERNOVA, AND ATMOSPHERIC NEUTRINOS}

\author{A.B. BALANTEKIN}

\address{Department of Physics, University of Wisconsin\\
Madison, WI 53706, USA\\
E-mail: baha@nucth.physics.wisc.edu}

\author{W.C. HAXTON}

\address{Institute for Nuclear Theory, Box 351550, and
Department of Physics, Box 351560\\
University of Washington, Seattle, WA 98195, USA\\
E-mail: haxton@phys.washington.edu}


\maketitle\abstracts{ In these Canberra summer school lectures we
  treat a number of topical issues in neutrino astrophysics: the solar
  neutrino problem, including the physics of the standard solar model,
  helioseismology, 
  and the possibility that the solution involves new particle physics;
  atmospheric neutrinos; Dirac and Majorana neutrino masses and their
  consequences for low-energy weak interactions; red giant evolution
  as a test of new particle astrophysics; the supernova mechanism;
  spin-flavor oscillations and oscillations into sterile states,
  including the effects of density fluctuations; and neutrino-induced
  and explosive nucleosynthesis.}

\section{Introduction}

The lectures summarized here were originally delivered by the authors
at a summer school held in Canberra, Australia, in January, 1998.  The
written version is a somewhat updated copy of the original, as we have
taken this opportunity to include some of the exciting neutrino
astrophysics results announced in summer, 1998.

The main theme of these lectures is the interplay between the
properties of the neutrinos --- their mass, mixing, and charge
conjugation properties --- and astrophysical phenomena.  The first
topic is the solar neutrino problem, where we describe the current
status of the measurements, the physics of the standard solar model
including helioseismology tests, the input nuclear microphysics, and
possible solutions.  We discuss the MSW mechanism in some detail as
well as the potential of SuperKamiokande and SNO to demonstrate that
oscillations are occurring.

We then discuss Dirac and Majorana neutrino masses, the seesaw
mechanism, and possibilities such as pseudoDirac neutrinos.  The
associated phenomenology in $\beta$ decay and $\beta \beta$ decay is
briefly described.

We discuss a second astrophysical laboratory for neutrino physics, the
evolution of red giants.  Topics include the nuclear physics of the He
flash and the effects of stellar cooling in red giant and horizontal
branch stellar evolution.  We illustrated how stellar cooling
arguments can place powerful constraints on axions and on anomalous
properties of the neutrino, such as electric and magnetic dipole
moments.

Core-collapse supernovae are the third laboratory.  We describe
currently favored theories of the explosion mechanism and characterize
the spectrum and flavor of the produced neutrinos.  We use supernovae
and the solar neutrino problem to motivate a discussion of more exotic
aspects of the MSW mechanism, including spin-flavor oscillations and
the effects of density fluctuations.  We also discuss the
nucleosynthesis that can arise because of neutrino interactions in the
mantle of the star.

We then turn to the explosive nucleosynthesis associated with the
supernova shock wave and the neutrino-driven wind off the proto-neutron
star surface.  We discuss the s- and r-processes, and constraints the
latter can place on neutrino oscillations.  These constraints prove
relevant to certain terrestrial oscillation experiments, such as
KARMEN and LSND.

The audience for these lectures consisted of advanced graduate
students in nuclear and particle physics: the material is covered at
this level and at a depth appropriate to a survey.

\section{Solar Neutrinos~\protect\cite{haxtonsn}}

More than three decades ago Ray Davis, Jr. and his
collaborators~\cite{davis} constructed a 0.615 kiloton C$_2$Cl$_4$
radiochemical solar neutrino detector in the Homestake Gold Mine, one
mile beneath Lead, South Dakota.  Within a few years it was apparent
that the number of neutrinos detected was considerably below the
predictions of the standard solar model, that is, the standard theory
of main sequence stellar evolution.

Today the results from the $^{37}$Cl detector, which have become quite
accurate due to 30 years of careful measurement, have been augmented
by results from four other experiments, the SAGE~\cite{sage} and
GALLEX~\cite{gallex} gallium experiments and the Kamiokande~\cite{k}
and SuperKamiokande~\cite{sk} water Cerenkov detectors.  It now
appears that the combined results are very difficult to explain ---
some have argued impossible --- by any plausible change in the
standard solar model (SSM).  Thus most believe that the answer to the
solar neutrino problem is new particle physics, most likely some
effect like solar neutrino oscillations associated with massive
neutrinos.  With the recent news that SuperKamiokande sees direct
evidence for $\nu_\mu$ oscillations in the azimuthal dependence of
atmospheric~\cite{atmos} neutrinos, it seems that we may be on the
threshold of a major discovery.

The purpose of this first (and longest) lecture is to summarize the
solar neutrino problem and to present arguments that it represents new
particle physics.

\subsection{The Standard Solar Model~\protect\cite{bbp98}}

Solar models trace the evolution of the Sun over the past 4.6 billion
years of main sequence burning, thereby predicting the present-day
temperature and composition profiles of the solar core that govern
neutrino production.  Standard solar models share four basic
assumptions:
    
\noindent
* The sun evolves in hydrostatic equilibrium, maintaining a local
balance between the gravitational force and the pressure gradient.  To
describe this condition in detail, one must specify the equation of
state as a function of temperature, density, and composition.

\noindent
* Energy is transported by radiation and convection.  While the solar
envelope is convective, radiative transport dominates in the core
region where thermonuclear reactions take place.  The opacity depends
sensitively on the solar composition, particularly the abundances of
heavier elements.

\noindent
* Thermonuclear reaction chains generate solar energy.  The standard
model predicts that over 98\% of this energy is produced from the pp
chain conversion of four protons into $^4$He (see Fig. 1)
\begin{equation}
          4p \rightarrow ^4\mathrm{He} + 2e^+  + 2 \nu_e 
\end{equation}
with proton burning through the CNO cycle contributing the remaining
2\%.  The Sun is a large but slow reactor: the core temperature, $T_c
\sim 1.5 \cdot 10^7$ K, results in typical center-of-mass energies for
reacting particles of $\sim$ 10 keV, much less than the Coulomb
barriers inhibiting charged particle nuclear reactions.  Thus reaction
cross sections are small: in most cases, as laboratory measurements
are only possible at higher energies, cross section data must be
extrapolated to the solar energies of interest.
    
\noindent
* The model is constrained to produce today's solar radius, mass, and
luminosity.  An important assumption of the standard model is that the
Sun was highly convective, and therefore uniform in composition, when
it first entered the main sequence.  It is furthermore assumed that
the surface abundances of metals (nuclei with A $>$ 5) were
undisturbed by the subsequent evolution, and thus provide a record of
the initial solar metallicity.  The remaining parameter is the initial
$^4$He/H ratio, which is adjusted until the model reproduces the
present solar luminosity after 4.6 billion years of evolution.  The
resulting $^4$He/H mass fraction ratio is typically 0.27 $\pm$ 0.01,
which can be compared to the big-bang value of 0.23 $\pm$ 0.01.  Note
that the Sun was formed from previously processed material.

\begin{figure}[htb]
\psfig{bbllx=0.5cm,bblly=4.0cm,bburx=18cm,bbury=18.5cm,figure=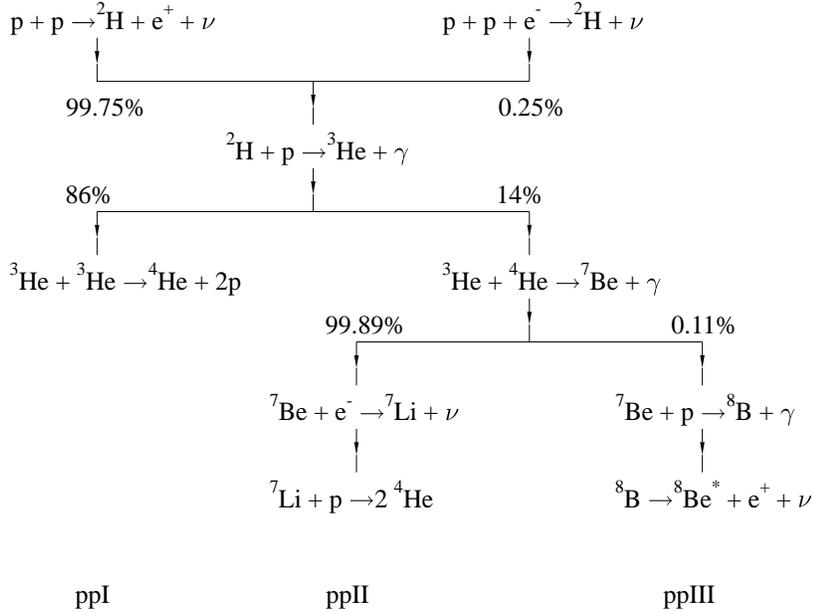,height=3.3in}
\caption{The solar pp chain.}
\end{figure}
  
The model that emerges is an evolving Sun.  As the core's chemical
composition changes, the opacity and core temperature rise, producing
a 44\% luminosity increase since the onset of the main sequence.  The
temperature rise governs the competition between the three cycles of
the pp chain: the ppI cycle dominates below about 1.6 $\cdot 10^7$ K;
the ppII cycle between (1.7-2.3) $\cdot 10^7$K; and the ppIII above
2.4 $\cdot 10^7$K.  The central core temperature of today's SSM is
about 1.55 $\cdot 10^7$K.

The competition between the cycles determines the pattern of neutrino
fluxes.  Thus one consequence of the thermal evolution of our sun is
that the $^8$B neutrino flux, the most temperature-dependent
component, proves to be of relatively recent origin: the predicted
flux increases exponentially with a doubling period of about 0.9
billion years.

A final aspect of SSM evolution is the formation of composition
gradients on nuclear burning timescales.  Clearly there is a gradual
enrichment of the solar core in $^4$He, the ashes of the pp chain.
Another element, $^3$He, is a sort of catalyst for the pp chain, being
produced and then consumed, and thus eventually reaching some
equilibrium abundance.  The timescale for equilibrium to be
established as well as the eventual equilibrium abundance are both
sharply decreasing functions of temperature, and thus increasing
functions of the distance from the center of the core.  Thus a steep
$^3$He density gradient is established over time.

The SSM has had some notable successes.  From helioseismology the
sound speed profile $c(r)$ has been very accurately determined for the
outer 90\% of the Sun, and is in excellent agreement with the SSM.
Such studies verify important predictions of the SSM, such as the
depth of the convective zone.  However the SSM is not a complete model
in that it does not explain all features of solar structure, such as
the depletion of surface Li by two orders of magnitude.  This is
usually attributed to convective processes that operated at some epoch
in our sun's history, dredging Li to a depth where burning takes
place.
  
The principal neutrino-producing reactions of the pp chain and CNO
cycle are summarized in Table 1.  The first six reactions produce
$\beta$ decay neutrino spectra having allowed shapes with endpoints
given by E$_\nu^{\rm max}$.  Deviations from an allowed spectrum occur
for $^8$B neutrinos because the $^8$Be final state is a broad
resonance.  The last two reactions produce line sources of electron
capture neutrinos, with widths $\sim$ 2 keV characteristic of the
temperature of the solar core.  Measurements of the pp, $^7$Be, and
$^8$B neutrino fluxes will determine the relative contributions of the
ppI, ppII, and ppIII cycles to solar energy generation.  As discussed
above, and as later illustrations will show more clearly, this
competition is governed in large classes of solar models by a single
parameter, the central temperature $T_c$.  The flux predictions of the
1998 calculations of Bahcall, Basu, and Pinsonneault~\cite{bbp98}
(BP98) and of Brun, Turck-Chieze and Morel~\cite{tcl} are included in
Table 1.

\begin{table}[t]
\caption{Solar neutrino sources and the flux predictions of the
Bahcall/Pinsonneault (BP98) and Brun/Turck-Chieze/Morel (BTCM98) SSMs in 
cm$^{-2}$s$^{-1}$.}
\vspace{0.2cm}
\begin{center}
\begin{tabular}{|c|c|c|c|}
\hline
 & & & \\
Source & E$_\nu^{max}$ (MeV) & BP98 & BTCM98 \\
& & & \\
\hline
& & & \\
p + p $\rightarrow ^2$H + e$^+ + \nu$ & 0.42 & 5.94E10 & 5.98E10 \\
$^{13}$N $\rightarrow ^{13}$C + e$^+ + \nu$ & 1.20 & 6.05E8 & 4.66E8 \\
$^{15}$O $\rightarrow ^{15}$N + e$^+ + \nu$ & 1.73 & 5.32E8 & 3.97E8 \\
$^{17}$F $\rightarrow ^{17}$O + e$^+ + \nu$ & 1.74 & 6.33E6 & \\
$^8$B $\rightarrow ^8$Be + e$^+ + \nu$ & $\sim$ 15 & 5.15E6 & 4.82E6 \\
$^3$He + p $\rightarrow ^4$He + e$^+ + \nu$ & 18.77 & 2.10E3 & \\
$^7$Be + e$^- \rightarrow ^7$Li + $\nu$ & 0.86 (90\%) & 4.80E9 & 4.70E9 \\
 & 0.38 (10\%) & & \\
p + e$^-$ + p $\rightarrow ^2$H + $\nu$ & 1.44 & 1.39E8 & 1.41E8 \\
 & & & \\
\hline
\end{tabular}
\end{center}
\end{table}
  
\subsection{Solar Neutrino Detection~\protect\cite{haxtontm}}

Let us start with a brief reminder about low energy neutrino-nucleus
interactions in detectors.  Consider the charged current reaction
\begin{equation}
\nu_e + (A,Z) \rightarrow e^- + (A,Z+1)
\end{equation}
Because the momentum transfer to the nucleus is very small for solar
neutrinos, it can be neglected in the weak propagator, leading to an
effective contact current-current interaction.  If we begin with the
simplest case of the free neutron decay $n
\rightarrow p+e^-+\bar{\nu}_e$, the corresponding transition amplitude
is then
\begin{equation}
S_{fi} = {G_F \over \sqrt{2}} \cos \theta_C 
\bar{u}(p) \gamma_\mu (1 - g_A \gamma_5) u(n)
\bar{u}(e) \gamma^\mu (1 - \gamma_5) u(\nu)
\end{equation}
where $G_F$ is the weak coupling constant measured in muon decay and
$\cos \theta_c$ gives the amplitude for the weak interaction to
connect the u quark to its first-generation partner, the d quark.  The
origin of this effective amplitude is the underlying standard model
predictions for the elementary quark and lepton currents.  The weak
interactions at this level are predicted by the standard model to be
exactly left handed.  Experiment shows that the effective coupling of
the W boson to the nucleon is governed by $\gamma_\mu (1 - g_A
\gamma_5)$, as noted above, where $g_A \sim 1.26$.  The axial coupling
is thus shifted from its underlying value by the strong interactions
responsible for the binding of the quarks within the nucleon.

If an isolated nucleon were the target, one could proceed to calculate
the cross section from the effective nucleon current given above.  The
extension to nuclear systems traditionally begins with the observation
that nucleons in the nucleus are rather non-relativistic, $v/c \sim
0.1$.  The amplitude $\bar{u}(p) \gamma^\mu(1-g_A \gamma_5) u(n)$ can
be expanded in powers of $p/M$.  The leading vector and axial
operators are readily found to be
\begin{eqnarray}
 \gamma_0&:&~~~1 \nonumber \\
\vec{\gamma}&:&~~~\vec{p}/M \sim v/c \nonumber \\
\gamma_0 \gamma_5&:&~~~\vec{\sigma} \cdot \vec{p}/M \sim v/c 
\nonumber \\
\vec{\gamma} \gamma_5&:&~~~\vec{\sigma} \nonumber
\end{eqnarray}
Thus it is the time-like part of the vector current and the space-like
part of the axial-vector current that survive in the non-relativistic
limit.\footnote{In a nucleus these currents must be corrected for the
  presence of meson exchange contributions.  The corrections to the
  vector charge and axial three-current, which we just pointed out
  survive in the non-relativistic limit, are of order $(v/c)^2 \sim$
  1\%.  Thus the naive one-body currents are a very good approximation
  to the nuclear currents.  In contrast, exchange current corrections
  to the axial charge and vector three-current operators are of order
  $v/c$, and thus of relative order 1.  This difficulty for the vector
  three-current can be largely circumvented, because current
  conservation as embodied in the generalized Siegert's theorem allows
  one to rewrite important parts of this operator in terms of the
  vector charge operator.  In the long-wavelength limit appropriate to
  solar neutrinos, all terms unconstrained by current conservation do
  not survive.  In effect, one has replaced a current operator with
  large two-body corrections by a charge operator with only small
  corrections.  In contrast, the axial charge operator is
  significantly altered by exchange currents even for long-wavelength
  processes like $\beta$ decay.  Typical axial-charge $\beta$ decay
  rates are enhanced by $\sim$ 2 because of exchange currents.}
  
If such a non-relativistic reduction is done for our single current one
obtains
\begin{eqnarray}
S_{fi}  & \sim &  \cos \theta_c {G_F \over \sqrt{2}} 
( \phi^\dagger (p) \phi(n) \bar{u}(e) \gamma^0(1 -\gamma_5)u(\nu) 
\nonumber \\
  & & - \phi^\dagger (p) g_A \vec{\sigma} \phi(n) \cdot \bar{u}(e)
\vec{\gamma}(1-\gamma_5)u(\nu) ) 
\end{eqnarray}
where the $\phi$'s are now two-component Pauli spinors for the
nucleons.  The above result can be generalized to include
$\bar{\nu}_e$ reactions by introducing the isospin operators
$\tau_\pm$ where $\tau_+$ $\mid$ n$\rangle$ = $\mid$ p$\rangle$ and
$\tau_-$$\mid$ p$\rangle$ = $\mid$ n$\rangle$, with all other matrix
elements being zero.  Thus we can generalize our $n  \rightarrow p$
amplitude to $n  \leftrightarrow p$ by
\[ \phi^\dagger (p) \phi(n) \rightarrow \phi^\dagger (N) \tau_\pm 
\phi(N) \]
\[ \phi^\dagger (p) \vec{\sigma} \phi(n) \rightarrow \phi^\dagger (N)
\vec{\sigma} \tau_\pm \phi(N). \] 
This result easily generalizes to nuclear decay.  Given our comments
about exchange currents, the first step is the replacement
\[ \tau_\pm \rightarrow \sum_{i=1}^A \tau_\pm(i) \]
\[ \sigma \tau_\pm \rightarrow \sum_{i=1}^A \sigma(i)
\tau_\pm(i). \] 
Plugging $S_{fi}$ into the standard cross section formula (which
involves an average over initial and sum over final nuclear spins of
the square of the transition amplitude) then yields the allowed
nuclear matrix element
\begin{equation}
{1 \over 2J_i+1} \left(|\langle f || \sum_{i=1}^A \tau_\pm (i) || i 
\rangle |^2
+ g_A^2 |\langle f || \sum_{i=1}^A \sigma(i) \tau_\pm(i) || i 
\rangle|^2\right).
\end{equation}

Our initial calculation for the nucleon treated that particle as
structureless.  Implicitly we assumed that the momentum transfer is
much smaller than the inverse nucleon size.  If we take 10 MeV as a
typical solar neutrino momentum transfer, these quantities would be in
the ratio 1:20.  For a light nucleus, the corresponding result might
be 1:10.  This long-wavelength approximation in combination with the
non-relativistic approximation yields the allowed result, where only
Fermi and Gamow-Teller operators survive.  These are the
spin-independent and spin-dependent operators appearing above.
  
The Fermi operator is proportional to the isospin raising/lowering
operator: in the limit of good isospin, which typically is good to 5\%
or better in the description of low-lying nuclear states, it can only
connect states in the same isospin multiplet, that is, states with a
common spin-spatial structure.  If the initial state has isospin
$(T_i, M_{Ti})$, this final state has $(T_i, M_{Ti} \pm 1)$ for
$(\nu,e^-)$ and $(\bar{\nu},e^+)$ reactions, respectively, and is
called the isospin analog state (IAS).  In the limit of good isospin
the sum rule for this operator in then particularly simple
\begin{equation}
\sum_f {1 \over 2J_i+1} | \langle f || \sum_{i=1}^A \tau_+(i) || i 
\rangle |^2 =
{1 \over 2J_i+1} | \langle IAS || \sum_{i=1}^A \tau_+(i) || i 
\rangle |^2 = |N-Z|. 
\end{equation}
The excitation energy of the IAS relative to the parent ground state
can be estimated accurately from the Coulomb energy difference
\begin{equation}
E_{IAS} \sim \left({1.728 Z \over 1.12A^{1/3} + 0.78} - 1.293\right) 
\mathrm{MeV}. 
\end{equation}
The angular distribution of the outgoing electron for a pure Fermi
$(N,Z) + \nu \rightarrow (N-1,Z+1) + e^-$ transition is 1 + $\beta
\cos \theta_{\nu e}$, and thus forward peaked.  Here $\beta$ is the
electron velocity.

The Gamow-Teller (GT) response is more complicated, as the operator
can connect the ground state to many states in the final nucleus.  In
general we do not have a precise probe of the nuclear GT response
apart from weak interactions themselves.  However a good approximate
probe is provided by forward-angle (p,n) scattering off nuclei, a
technique that has been developed in particular by experimentalists at
the Indiana University Cyclotron Facility\cite{indiana}.  The (p,n)
reaction transfers isospin and thus is superficially like $(\nu,e^-)$.
At forward angles (p,n) reactions involve negligible three-momentum
transfers to the nucleus.  Thus the nucleus should not be radially
excited.  It thus seems quite plausible that forward-angle (p,n)
reactions probe the isospin and spin of the nucleus, the macroscopic
quantum numbers, and thus the Fermi and GT responses.  For typical
transitions, the correspondence between (p,n) and the weak GT
operators is believed to be accurate to about 10\%.  Of course, in a
specific transition, much larger discrepancies can arise.

The (p,n) studies demonstrate that the GT strength tends to
concentrate in a broad resonance centered at a position $\delta =
E_{GT} - E_{IAS}$ relative to the IAS given by
\begin{equation}
 \delta \sim \left(7.0 -28.9 {N-Z \over A}\right)~\mathrm{MeV}. 
\end{equation}
Thus while the peak of the GT resonance is substantially above the IAS
for $N \sim Z$ nuclei, it drops with increasing neutron excess.  Thus
$\delta \sim 0$ for Pb.  A typical value for the full width at half
maximum $\Gamma$ is $\sim$ 5 MeV.

The approximate Ikeda sum rule constrains the difference
in the $(\nu,e^-)$ and $(\bar{\nu},e^+)$ strengths
\begin{equation}
\sum_f ( |M_{GT}^{fi}(\nu,e^-)|^2 - |M_{GT}^{fi}(\bar{\nu},e^+)|^2 )
= 3(N-Z)
\end{equation}
where
\begin{equation}
|M_{GT}^{fi}(\nu,e^-)|^2 = {1 \over 2J_i+1} 
|\langle f || \sum_{i=1}^A \sigma (i) \tau_+(i) || i \rangle |^2. 
\end{equation}
In many cases of interest in heavy nuclei, the strength in the
$(\bar{\nu},e^+)$ direction is largely blocked.  For example, in a
naive $2s1d$ shell model description of $^{37}$Cl, the p $\rightarrow$
n direction is blocked by the closed neutron shell at N=20.  Thus this
relation can provide an estimate of the total $\beta^-$ strength.
Experiment shows that the $\beta^-$ strength found in and below the GT
resonance does not saturate the Ikeda sum rule, typically accounting
for $\sim (60-70)$ \% of the total.  Measured and shell model
predictions of individual GT transition strengths tend to differ
systematically by about the same factor.  Presumably the missing
strength is spread over a broad interval of energies above the GT
resonance.  This is not unexpected if one keeps in mind that the shell
model is an approximate effective theory designed to describe the long
wavelength modes of nuclei: such a model should require effective
operators, renormalized from their bare values.  Phenomenologically,
the shell model seems to require~\cite{brown} $g_A^{eff} \sim$ 1.0 as
well as a small spin-tensor term $(\sigma \otimes Y_2(\hat{r})
)_{J=1}$ of relative strength $\sim$ 0.1.

The angular distribution of GT $(N,Z) + \nu_e \rightarrow (N-1,Z+1) +
e^-$ reactions is $3 - \beta \cos \theta_{\nu e}$, corresponding to a
gentle peaking in the backward direction.
 
The above discussion of allowed responses can be repeated for neutral
current processes such as $(\nu,\nu')$.  The analog of the Fermi
operator contributes only to elastic processes, where the standard
model nuclear weak charge is approximately the neutron number.  As
this operator does not generate transitions, it is not yet of much
interest for solar or supernova neutrino detection, though there are
efforts to develop low-threshold detectors (e.g., cryogenic
technologies) for recording the modest nuclear recoil energies.  The
analog of the GT response involves
\begin{equation}
|M_{GT}^{fi}(\nu,\nu')|^2 = {1 \over 2J_i+1}
|\langle f || \sum_{i=1}^A \sigma(i) {\tau_3(i) \over 2} || i
\rangle |^2. 
\end{equation}
The operator appearing in this expression is familiar from magnetic
moments and magnetic transitions, where the large isovector magnetic
moment ($\mu_v \sim$ 4.706) often leads to it dominating the orbital
and isoscalar spin operators.

Finally, there is one purely leptonic reaction of great interest,
since it is the reaction exploited by Kamiokande and SuperKamiokande.
Electron neutrinos can scatter off electrons via both charged and
neutral current reactions.  The cross section calculation is
straightforward and will not be repeated here.  Two features of the
result are of importance for our later discussions, however.  Because
of the neutral current contribution, heavy-flavor $(\nu_\mu$ and
$\nu_\tau)$ also scatter off electrons, but with a cross section
reduced by about a factor of seven at low energies.  Second, for
neutrino energies well above the electron rest mass, the scattering is
sharply forward peaked.  Thus this reaction allows one to exploit the
position of the Sun in separating the solar neutrino signal from a
large but isotropic background.
  
As we mentioned earlier, the first experiment performed was one
exploiting the reaction 
\[ ^{37}\mathrm{Cl}(\nu,e^-)^{37}\mathrm{Ar}. \]
As the threshold for this reaction is 0.814 MeV, the important
neutrino sources are the $^7$Be and $^8$B reactions.  The $^7$Be
neutrinos excite just the GT transition to the ground state, the
strength of which is known from the electron capture lifetime of
$^{37}$Ar.  The $^8$B neutrinos can excite all bound states in
$^{37}$Ar, including the dominant transition to the IAS residing at an
excitation of 4.99 MeV.  The strength of excite-state GT transitions
can be determined from the $\beta$ decay $^{37}$Ca$(\beta^+)^{37}$K,
which is the isospin mirror reaction to $^{37}$Cl$(\nu,e^-)^{37}$Ar.
The net result is that, for SSM fluxes, 78\% of the capture rate
should be due to $^8$B neutrinos, and 15\% to $^7$Be neutrinos.  The
measured capture rate~\cite{lande} 2.56 $\pm 0.16 \pm 0.16$ SNU (1 SNU
= 10$^{-36}$ capture/atom/sec) is about 1/3 the standard model value.

Similar radiochemical experiments were done by the SAGE and GALLEX
collaborations using a different target, $^{71}$Ga.  The special
properties of this target include its low threshold and an unusually
strong transition to the ground state of $^{71}$Ge, leading to a large
pp neutrino cross section (see Fig. 2).  The experimental capture
rates are $66 \pm 13 \pm 6$ and $76 \pm 8$ SNU for the SAGE and GALLEX
detectors, respectively.  The SSM prediction is about 130
SNU~\cite{bahcallb}.  Most important, since the pp flux is directly
constrained by the solar luminosity in all steady-state models, there
is a minimum theoretical value for the capture rate of 79 SNU, given
standard model weak interaction physics.  Note there are substantial
uncertainties in the $^{71}$Ga cross section due to $^7$Be neutrino
capture to two excited states of unknown strength.  These
uncertainties were greatly reduced by direct calibrations of both
detectors using $^{51}$Cr neutrino sources.

\begin{figure}[htb]
\psfig{bbllx=0.0cm,bblly=4.0cm,bburx=16cm,bbury=22.5cm,figure=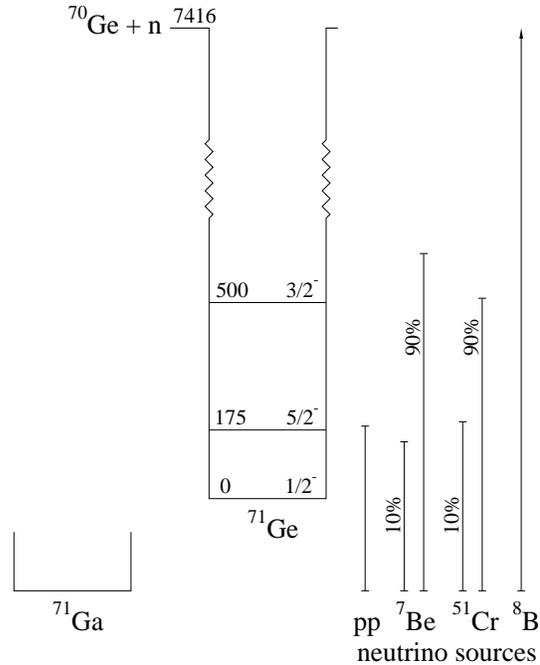,height=3.8in}
\caption{Level scheme for $^{71}$Ge showing the excited states
that contribute to absorption of pp, $^7$Be, $^{51}$Cr, and
$^8$B neutrinos.}
\end{figure}
  
The remaining experiments, Kamiokande II/III and SuperKamiokande,
exploited water Cerenkov detectors to view solar neutrinos on an
event-by-event basis. Solar neutrinos scatter off electrons, with the
recoiling electrons producing the Cerenkov radiation that is then
recorded in surrounding photo-tubes.  Thresholds are determined by
background rates; SuperKamiokande is currently operating with a
trigger at approximately 6 MeV.  The initial experiment, Kamiokande
II/III, found a flux of $^8$B neutrinos of (2.91 $\pm 0.24 \pm 0.35)
\cdot 10^6$/cm$^2$s after about a decade of measurement.  Its much
larger successor SuperKamiokande, with a 22.5 kiloton fiducial volume,
yielded the result $(2.44 \pm 0.05 ^{+0.09}_{-0.07}) \cdot
10^6$/cm$^2$s after the first 504 days of
measurements\cite{superkamnew}.  This is about
half of the SSM flux.  This result continues to improve in accuracy. \\
  
\subsection{Uncertainties in Standard Solar Model Parameters}

The pattern of solar neutrino fluxes that has emerged from these
experiments is
\begin{eqnarray}
\phi (pp) & \sim & 0.9 \, \phi^{\rm {SSM}} (pp)\nonumber \\
\phi (^7{\rm {Be}}) & \sim & 0 \nonumber\\
\phi (^8 {\rm B}) & \sim & 0.43 \, \phi^{\rm {SSM}} (^8{\rm B}).  
\end{eqnarray}
A reduced $^8$B neutrino flux can be produced by lowering the central
temperature of the sun somewhat, as $\phi(^8$B)$\sim T_c^{18}$.
However, such an adjustment, either by varying the parameters of the
SSM or by adopting some nonstandard physics, tends to push the $\phi
(^7$Be)/$\phi(^8$B) ratio to higher values rather than the low one of
eq. (12),
\begin{equation}
{\phi (^7{\rm{Be}}) \over \phi(^8 {\rm B})} \sim T_c^{-10}.
\end{equation}
Thus the observations seem difficult to reconcile with plausible solar
model variations: one observable, $\phi(^8$B), requires a cooler core
while a second, the ratio $\phi(^7$Be)/$\phi(^8$B), requires a hotter
one.

An initial question is whether this problem remains significant when
one takes into account known uncertainties in the parameters of the
SSM.  While a detailed summary of the SSM uncertainties would take us
well beyond the limits of these lectures, a qualitative discussion of
pp chain nuclear uncertainties is appropriate.  This nuclear
microphysics has been the focus of a great deal of experimental work.
The pp chain involves a series of non-resonant charged-particle
reactions occurring at center-of-mass energies that are well below the
height of the inhibiting Coulomb barriers.  As the resulting small
cross sections generally preclude laboratory measurements at the relevant
energies, one must extrapolate higher energy measurements to threshold
to obtain solar cross sections.  This extrapolation is often discussed
in terms of the astrophysical S-factor
\begin{equation}
\sigma (E) = {S(E) \over E} \exp (-2 \pi \eta)
\end{equation}
where $\eta = {Z_1Z_2 \alpha \over \beta}$, with $\alpha$ the fine
structure constant and $\beta = v/c$ the relative velocity of the
colliding particles.  This parameterization removes the gross Coulomb
effects associated with the s-wave interactions of charged, point-like
particles.  The remaining energy dependence of S(E) is gentle and can
be expressed as a low-order polynomial in E.  Usually the variation of
S(E) with E is taken from a direct reaction model and then used to
extrapolate higher energy measurements to threshold.  The model
accounts for finite nuclear size effects, strong interaction effects,
contributions from other partial waves, etc.  As laboratory
measurements are made with atomic nuclei while conditions in the solar
core guarantee the complete ionization of light nuclei, additional
corrections must be made to account for the different electronic
screening environments.

Recently a large working group met at a workshop sponsored by the
Institute for Nuclear Theory, University of Washington, to review past
work on the nuclear reactions of the pp chain and CNO cycle, to
recommend best values and appropriate errors, and to identify specific
issues in experiment and theory where additional work is needed.  The
results are published in Reviews of Modern Physics\cite{rmpint}.  We
will not attempt a summary here, but will give one or two highlights.

The most significant recommend change involves the reaction $^7$Be(p,
$\gamma)\, ^8$B, where the standard S$_{17}$(0)$ \sim$ 22.4 eVb is
that given~\cite{johnson} by Johnson et al.  Measurements of
S$_{17}$(E) are complicated by the need to use radioactive targets and
thus to determine the areal density of the $^7$Be target nuclei.  Two
techniques have been employed, measuring the rate of 478 keV photons
from $^7$Be decay or counting the daughter $^7$Li nuclei via the
reaction $^7$Li (d,p)$^8$Li.  The low-energy data sets for S$_{17}$(E)
disagree by 25\%.  This is a systematic normalization problem as each
data set is consistent with theory in its dependence on E.  The energy
dependence below $\sim$ 500 keV is believed to be quite simple as it
is determined by the asymptotic nuclear wave function.

The Seattle working group on S$_{17}$(E) found that only one
low-energy data set, that of Filippone et al.~\cite{filippone}, was
described in the published literature in sufficient detail to be
evaluated.  The target activity in that experiment had been measured
by both 478 keV gamma rays and by the (d,p) reaction, with consistent
results.  The resulting recommended value was thus based on this
measurement, yielding
\begin{equation}
S_{17} (0) = 19^{+4}_{-2} \mathrm{eV~b},~~1 \sigma . 
\end{equation}
A recent measurement\cite{bogaert} is consistent with this value. 

The $^3$He($\alpha,\gamma) ^7$Be reaction has been measured by two
techniques, by counting the capture $\gamma$ rays and by detecting the
resulting $^7$Be activity.  While the two techniques have been used by
several groups and have yielded separately consistent results, the
capture $\gamma$ ray value S$_{17}$(0) = 0.507 $\pm $ 0.016 keV b is
not in good agreement with the $^7$Be activity value 0.572 $\pm$ 0.026
keV-b.  The Seattle working group concluded that the evidence for a
systematic discrepancy of unknown origin was reasonably strong and
recommended that standard procedures be used in assigning a suitably
expanded error.  The recommended value S$_{34}$ (0) is 0.53 $\pm$
0.05.

These and other recommended values were recently incorporated into the
BP98 and BTCM98 solar model calculations.  While the workshop's
recommended values
involve no qualitative changes, there is some broadening of error
bars.  The downward shift in S$_{17}$(0) leads to a lower $^8$B flux.
The workshop's Reviews of Modern Physics article summarizes a
substantial amount of work on topics not discussed here: screening
effects, weak radiative corrections to and exchange current effects on
p+p, the atomic physics of $^7$Be + e$^-$, etc.  Much of this
discussion was useful in evaluating possible uncertainties in solar
microphysics, and in identifying opportunities for reducing these
uncertainties.

Are uncertainties in the parameters of the SSM a significant source of
uncertainty?  The S-factors discussed above comprise one set of
parameters, but there are others: the solar lifetime, the opacities,
the solar luminosity, etc.  In order to answer this question while
also taking into account correlations among the fluxes when input
parameters are varied, first Bahcall and Ulrich~\cite{bu} and later
Bahcall and Haxton~\cite{bh} constructed 1000 SSMs by randomly varying
five input parameters, the primordial heavy-element-to-hydrogen ratio
Z/X and S(0) for the p-p, $^3$He-$^3$He, $^3$He-$^4$He, and p-$^7$Be
reactions, assuming for each parameter a normal distribution with the
mean and standard deviation.  (These were the parameters assigned the
largest uncertainties.)  Smaller uncertainties from radiative
opacities, the solar luminosity, and the solar age were folded into
the results of the model calculations perturbatively.

The resulting pattern of $^7$Be and $^8$B flux predictions is shown in
Fig. 3.  The elongated error ellipses indicate that the fluxes are
strongly correlated.  Those variations producing $\phi(^8$B) below
0.8$\phi^{\rm{SSM}}(^8$B) tend to produce a reduced $\phi(^7$Be), but
the reduction is always less than 0.8.  Thus a greatly reduced
$\phi(^7$Be) cannot be achieved within the uncertainties assigned to
parameters in the SSM.
    
\begin{figure}[htb]
\psfig{bbllx=0.5cm,bblly=4.0cm,bburx=17cm,bbury=22.5cm,figure=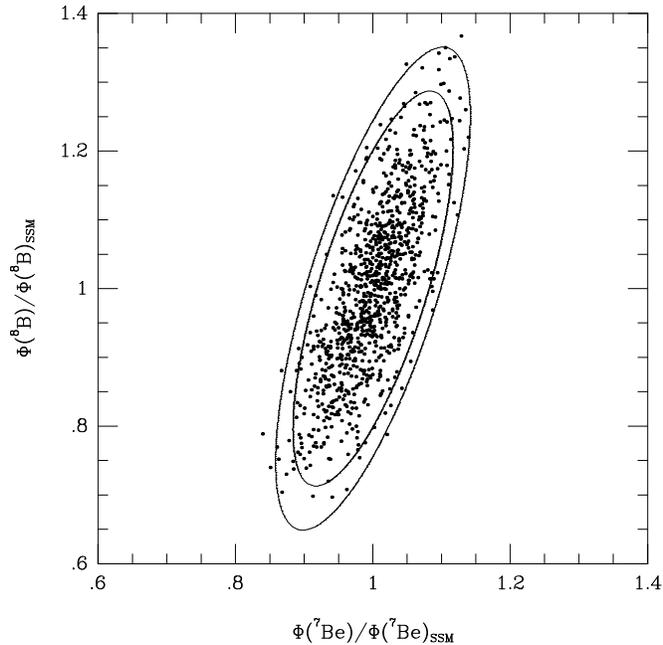,height=3.5in}
\caption{SSM $^7$Be and $^8B$ flux predictions.  The dots represent
the results of SSM calculations where the input parameters were
varied according to their assigned uncertainties, as described
in the text.  The 90\% and 99\% confidence level error ellipses
are shown.}
\end{figure}
  
A similar exploration, but including parameter variations very far
from their preferred values, was carried out by Castellani et
al.~\cite{cast}, who displayed their results as a function of the
resulting core temperature $T_c$.  The pattern that emerges is
striking (see Fig. 4): parameter variations producing the same value
of $T_c$ produce remarkably similar fluxes.  Thus $T_c$ provides an
excellent one-parameter description of standard model perturbations.
Figure 4 also illustrates the difficulty of producing a low ratio of
$\phi(^7$Be)/$\phi(^8$B) when $T_c$ is reduced.

The 1000-solar-model variations were made under the constraint of
reproducing the solar luminosity.  Those variations show a similar
strong correlation with $T_c$
\begin{equation}
\phi(pp) \propto T_c^{-1.2} ~~~~~~~  \phi(^7{\rm {Be}}) \propto T_c^8 ~~~~~~~
 \phi(^8 {\rm B}) \propto T_c^{18}.
\end{equation}
Figures 3 and 4 offer a strong argument that reasonable
variations in the parameters of the SSM, or nonstandard
changes in quantities like the metallicity, opacities, or
solar age, cannot produce the pattern of fluxes deduced
from experiment (eq. (12)).  This would seem to limit 
possible solutions to errors either in the underlying physics
of the SSM or in our understanding of neutrino properties. 

\begin{figure}[htb]
\psfig{bbllx=0.3cm,bblly=4.0cm,bburx=14.5cm,bbury=24.0cm,figure=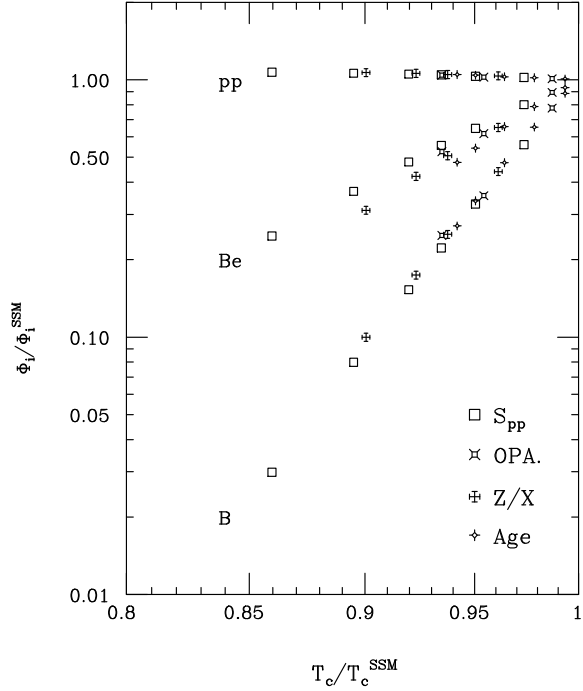,height=3.6in}
\caption{The responses of the pp, $^7$Be, and $^8$B neutrino fluxes
to the indicated variations in solar model input parameters,
displayed as a function of the resulting central temperature
$T_c$.  From Castellani et al.}
\end{figure}
  
\subsection{Nonstandard Solar Models}

Nonstandard solar models include both variations of SSM parameters far
outside the ranges that are generally believed to be reasonable (some
examples of which are given in Figure 4), and changes in the
underlying physics of the model.  The solar neutrino problem has been
a major stimulus to models: in fact, most suggestions were motivated
by the hope of producing a cooler Sun ($T_c \sim 0.95 T_c$) that would
avoid conflict with the results of the $^{37}$Cl experiment.  The
suggestions included models with low heavy element abundances (``low
Z" models), in which one abandons the SSM assumption that the initial
heavy element abundances are those we measure today at the Sun's
surface; periodically mixed solar cores; models where hydrogen is
continually mixed into the core by turbulent diffusion or by
convective mixing; and models where the solar core is partially
supported by a strong central magnetic field or by its rapid rotation,
thereby relaxing the SSM assumption that hydrostatic equilibrium is
achieved only through the gas pressure gradient.  A larger list is
given by Bahcall and Davis~\cite{bd82}.  To illustrate the kinds of
consequences such models have, two of these suggestions are discussed
in more detail below.

In low-Z models one postulates a reduction in the core metallicity
from Z $\sim$ 0.02 to Z $\sim$ 0.002.  This lowers the core opacity
(primarily because metals are very important to free-bound electron
transitions), thus reducing $T_c$ and weakening the ppII and ppIII
cycles.  The attractiveness of low-Z models is due in part to the
existence of additional mechanisms for adding heavier elements to the
Sun's surface.  These include the infall of comets and other debris,
as well as the accumulation of dust as the Sun passes through
interstellar clouds.  However, the increased radiative energy
transport in low-Z models leads to a thin convective envelope, in
contradiction to interpretations of the 5-minute solar surface
oscillations.  A low He mass fraction also results.  As diffusion of
material from a thin convective envelope into the interior would
deplete heavy elements at the surface, investigators have also
questioned whether present abundances could have accumulated in low-Z
models.  Finally, the general consistency of solar heavy element
abundances with those observed in other main sequence stars makes the
model appear contrived.

Models in which the solar core ($\sim$ 0.2 M$_\odot$) is
intermittently mixed break the standard model assumption of a
steady-state Sun: for a period of several million years (the thermal
relaxation time for the core) following mixing, the usual relationship
between the observed surface luminosity and rate of energy (and
neutrino) production is altered as the Sun burns out of equilibrium.
Calculations show that both the luminosity and the $^8$B neutrino flux
are suppressed while the Sun relaxes back to the steady state.  Such
models have been considered seriously because of instabilities
associated with large gradients in the $^3$He abundance, which in
equilibrium varies as $\sim T^{-6}$, where $T$ is the local
temperature.  The resulting steep profile is unstable under finite
amplitude displacements of a volume to smaller r: the energy released
by the increased $^3$He burning at higher T can exceed the energy in
the perturbation.  For a discussion of the plausibility of such a
trigger for core mixing, one can see the original work of Dilke and
Gough~\cite{gough} as well as a more recent critique by
Merryfield~\cite{merry}.  The possibility that continuous mixing on
time scales of $^3$He mixing could produce a flux pattern close to
that observed (e.g., a suppression in both the $^8$B neutrino flux and
the $^7$Be/$^8$B flux ratio) was recently discussed by Cumming and
Haxton~\cite{cumming}.
   
This discussion of two of the more seriously explored nonstandard
model possibilities illustrates how changes motivated by the solar
neutrino problem often produce other, unwanted consequences.  In
particular, many experts feel that the good SSM agreement with
helioseismology is likely to be destroyed by changes such as those
discussed above.

Figure 5 is an illustration by Hata et al.~\cite{hata} of the flux
predictions of several nonstandard models, including a low-Z model
consistent with the $^{37}$Cl results.  As in the Castellani et al.
exploration, the results cluster along a track that defines the naive
$T_c$ dependence of the $\phi (^7$Be)/$\phi(^8$B) ratio, well
separated from the experimental contours.
   
\begin{figure}[htb]
\psfig{bbllx=2.3cm,bblly=7.5cm,bburx=17cm,bbury=19.8cm,figure=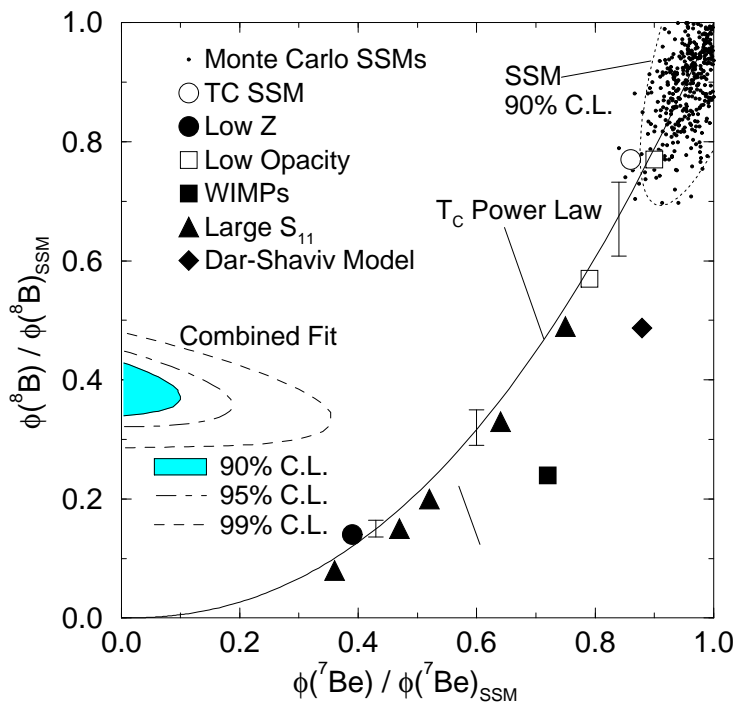,height=3.6in}
\caption{The fluxes allowed by the combined results of the various
solar neutrino experiments compared to the results of SSM 
variations and various nonstandard solar models.  The solid line
in the naive $T_c$ power law discussed in the text.  From
Hata et al.}
\end{figure}
  
There is now a popular argument that no such nonstandard model can
solve the solar neutrino problem: if one assumes undistorted neutrino
spectra, no combination of pp, $^7$Be, and $^8$B neutrino fluxes fits
the experimental results well~\cite{karsten}.  In fact, in an
unconstrained fit, the required $^7$Be flux is unphysical, negative by
almost 3$\sigma$.  Thus, barring some unfortunate experimental error,
it appears we are forced to look elsewhere for a solution.

If experimental error, SSM parameter uncertainties, and nonstandard
solar physics are ruled out as potential solutions, new particle
physics is left as the leading possibility.  Suggested particle
physics solutions of the solar neutrino problem include neutrino
oscillations, neutrino decay, neutrino magnetic moments, and weakly
interacting massive particles.  Among these, the
Mikheyev-Smirnov-Wolfenstein effect --- neutrino oscillations enhanced
by matter interactions --- is widely regarded as the most plausible.

\subsection{Helioseismology}

Earlier it was mentioned that measurements of the sound velocity
within the Sun, deduced from observations of surface oscillations,
provide a powerful check on the SSM.  In this section the
basic physics of helioseismology is reviewed. 

A static, stable star at spherically-symmetric equilibrium can be
characterized with pressure $p(r)$, mass density $\rho(r)$, the
gravitational potential $\phi(r)$, the rate of nuclear energy
generation $\epsilon(r)$, temperature $T(r)$, the energy flux {\bf F}
and the entropy $s$. Introducing the adiabatic indices
\begin{equation}
  \label{eq:he1}
  \Gamma_1 = \left( {\partial \log p \over \partial \log \rho} 
\right)_s,
\end{equation}
\begin{equation}
  \label{eq:he2}
  \Gamma_3 -1= \left( {\partial \log T \over \partial \log \rho}
  \right)_s,
\end{equation}
and the total derivative
\begin{equation}
  \label{eq:he3}
  {D \over Dt}={\partial \over \partial t} + {\bf v} \cdot \nabla
\end{equation}
one can write down the equation of motion 
\begin{equation}
  \label{eq:he4}
  \rho {D {\bf v} \over Dt} = - \nabla p - \rho \nabla \phi, 
\end{equation}
the equation of continuity
\begin{equation}
  \label{eq:he5}
  {D \rho \over Dt} + \rho \nabla \cdot {\bf v} = 0,
\end{equation}
Poisson's equation for gravitational attraction
\begin{equation}
  \label{eq:he6}
  \nabla^2 \phi = 4 \pi G \rho,
\end{equation}
and an equation describing energy conservation
\begin{equation}
  \label{eq:he7}
  {1 \over p}{D p \over Dt} - \Gamma_1 {1\over\rho} {D \rho \over Dt} 
  = {{\Gamma_3 -1}\over p} (\rho \epsilon - \nabla \cdot {\bf F}). 
\end{equation}
These equations describe a static star. To do stellar seismology one
introduces Eulerian (i.e. at a given point) perturbations on the
physical quantities, e.g.
\begin{equation}
  \label{eq:he8}
 \rho({\bf r}, t) = \rho_0 ({\bf r}) + \rho'({\bf r}, t) 
\end{equation}
where $\rho_0$ denotes the equilibrium value and the displacement is
calculated from the velocity amplitude
\begin{equation}
  \label{eq:he9}
  {\bf v} = {\partial \over \partial t} (\delta{\bf r}).
\end{equation}
Inserting expressions like Eq. (\ref{eq:he8}) in the above equations
and subtracting the equilibrium equations one obtains equations that
describe the perturbations. From the conservation of momentum one gets 
\begin{equation}
  \label{eq:he10}
  \rho {\partial^2 \delta{\bf r} \over \partial t^2 } = - \nabla p' 
+{\rho' \over \rho} \nabla p - \rho \nabla \phi'.
\end{equation}
The equation of continuity gives
\begin{equation}
  \label{eq:11}
  \rho'+ \nabla \cdot ( \rho \delta{\bf r}) = 0.
\end{equation}
Poisson's equation becomes
\begin{equation}
  \label{eq:12}
  \nabla^2 \phi' = 4 \pi G \rho',
\end{equation}
and the energy equation yields
\begin{equation}
  \label{eq:13}
  {\rho'\over\rho} +{1\over \rho} \delta{\bf r} \cdot \nabla \rho = 
{1 \over \Gamma_1} \left( {p' \over p} + {1\over p} \delta{\bf r} 
\cdot \nabla p \right).
\end{equation}
In these equations for convenience we dropped the subscript zero in
writing down the equilibrium values. To obtain the normal modes of a
star we assume a time dependence of $\exp (-i \omega t)$ for the
perturbations: 
\begin{equation}
  \label{eq:14}
  \rho'({\bf r},t) \sim  \rho'(r) Y_{\ell m}(\theta,\phi) 
\exp (-i \omega t). 
\end{equation}
Using Eq. (\ref{eq:14}) and introducing the auxiliary quantity
\begin{equation}
  \label{eq:15}
  \Psi(r) = c^2 \rho^{1/2} \nabla \cdot \delta{\bf r},
\end{equation}
where $c$ is the adiabatic sound speed
\begin{equation}
  \label{eq:16}
  c^2 = {\Gamma_1 p \over \rho},
\end{equation}
Eqs. (26) through (29) can be written in a compact form
\begin{equation}
  \label{eq:17}
  {d^2 \Psi \over dr^2} + {1\over c^2} \left[ \omega^2 -  
\omega^2_{\rm co} - {\ell(\ell+1) c^2 \over r^2} \left( 1 - 
{N^2 \over \omega^2} \right) \right] \Psi \simeq 0.
\end{equation}
In Eq. (\ref{eq:17}) we used the buoyancy frequency:
\begin{equation}
  \label{eq:18}
  N^2 = {G m_0(r) \over r} \left( {1\over\Gamma_1} 
{d \log p \over dr} - {d \log \rho \over dr} \right), 
\end{equation}
and the acoustical cut-off frequency
\begin{equation}
  \label{eq:19}
  \omega^2_{\rm co} = {c^2 \over 4H^2} \left( 1 - 2 {dH \over dr}
  \right), 
\end{equation}
where the density scale height is 
\begin{equation}
  \label{eq:20}
  H = -(d\log\rho/dr)^{-1}. 
\end{equation}
One observes that oscillation frequencies are determined by the sound
speed profile. The sound speed profile calculated using the
Bahcall-Pinsonneault 1998 SSM is shown in Figure 6. 

\begin{figure}[t]
\vspace{8pt} \centerline{\hbox{\epsfxsize=3 in \epsfbox[35 83 526
691]{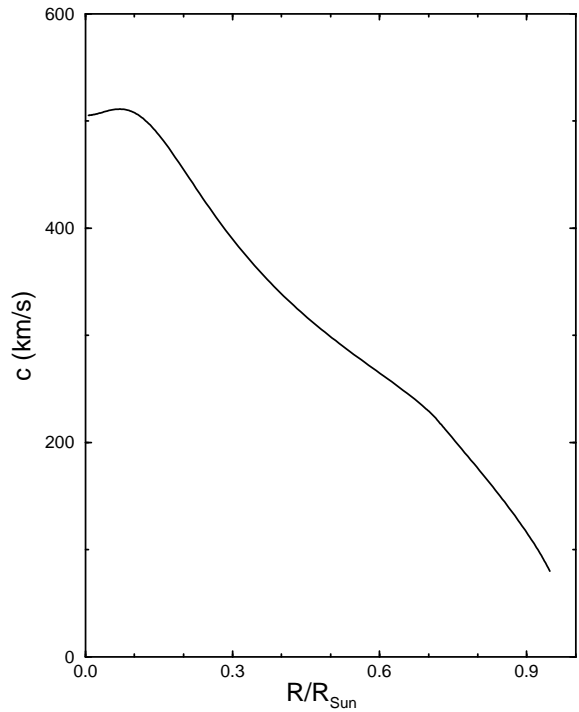}}}
\caption{Sound speed profile in the Sun.  From Bahcall and
  Pinsonneault.}
\vspace{8pt}
\end{figure}

Eq. (\ref{eq:17}) resembles the Schroedinger equation in quantum
mechanics. For the present Sun the buoyancy frequency is approximately
constant in the radiative zone except very near the core, but vanishes
in the convective zone.  The acoustical cut-off frequency is a
monotonically decreasing function of the distance from the center of
the Sun. Inserting this ``potential'' to the ``Schroedinger
equation'', Eq. (\ref{eq:17}), one observes that for $N^2/\omega^2 \ll
1 $ the amplitude of the oscillations die out in the radiative zone
(``classically forbidden'') of the Sun. The resulting oscillations are
confined to the convective (outer) zone of the Sun and called p-modes
(for pressure). For ${\ell(\ell+1) c^2 \over r^2 \omega^2} \gg 1$ the
situation is reversed. The ``classically allowed'' region is the core
of the Sun and the amplitudes die out in the convective zone. These
oscillations are called g-modes (for gravity). Note that g-mode
oscillation amplitudes vanish at the solar surface, hence it is very
difficult to directly observe g-modes. On the other hand p-mode
oscillations are readily measurable by observing the solar surface.
Eq. (\ref{eq:17}) indicates that p-mode oscillations with different
$\ell$ values penetrate to different depths, the observed frequencies
$\omega_{\ell}$ are determined by conditions at different parts of the
Sun.

It is possible to gain insight to the properties of solar oscillations
by regarding the equations outlined above as an eigenvalue problem in
a linear Hilbert space. Hence it is possible to directly relate
perturbations in the sound speed to the perturbations in
$\omega_{\ell}$. One starts with the sound speed profile and
oscillation frequencies calculated in a reference solar model. These
quantities are taken as the unperturbed quantities. Then using
standard Raleigh-Schroedinger perturbation theory one relates the
difference between observed and calculated frequencies to the deviation
of the sound speed from the model prediction. A very readable
introduction the the theoretical aspects of helioseismology is
available on the world wide web \cite{cdhelio}. There is very good
agreement between calculated and observed sound speeds in the
Sun. Figure 7 shows the fractional difference between the predicted
and observed sound speed profiles \cite{bbp98}. Sound speed profiles
deduced from helioseismology provides an important constraint on solar
models. 

\begin{figure}[t]
\vspace{8pt} \centerline{\rotate[r]{\epsfxsize=3in
\epsfbox{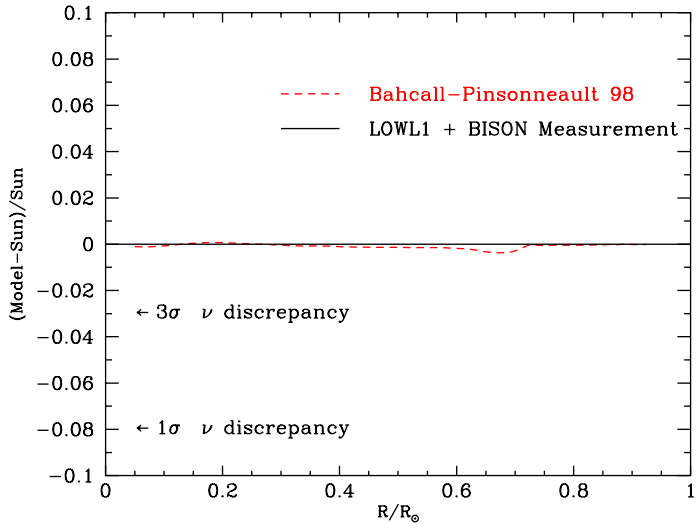}}} 
\caption{Fractional difference between the predicted
and observed sound speed profiles. From Bahcall, Basu, and
Pinsonneault.}
\vspace{8pt}
\end{figure}

\subsection{Neutrino Oscillations}

One odd feature of particle physics is that neutrinos, which are not
required by any symmetry to be massless, nevertheless must be much
lighter than any of the other known fermions.  For instance, the
current limit on the $\overline{\nu}_e$ mass is $\lsim$ 5 eV.  The
standard model requires neutrinos to be massless, but the reasons are
not fundamental.  Dirac mass terms $m_D$, analogous to the mass terms
for other fermions, cannot be constructed because the model contains
no right-handed neutrino fields.  Neutrinos can also have Majorana
mass terms
\begin{equation}
\overline{\nu^c_L} m_L \nu_L ~~~ \mathrm{and} 
~~~ \overline{\nu^c_R} m_R \nu_R 
\end{equation}
where the subscripts $L$ and $R$ denote left- and right-handed
projections of the neutrino field $\nu$, and the superscript $c$
denotes charge conjugation.  The first term above is constructed from
left-handed fields, but can only arise as a nonrenormalizable
effective interaction when one is constrained to generate $m_L$ with
the doublet scalar field of the standard model.  The second term is
absent from the standard model because there are no right-handed
neutrino fields.

None of these standard model arguments carries over to the more
general, unified theories that theorists believe will supplant the
standard model.  In the enlarged multiplets of extended models it is
natural to characterize the fermions of a single family, e.g.,
$\nu_e$, e, u, d, by the same mass scale $m_D$.  Small neutrino masses
are then frequently explained as a result of the Majorana neutrino
masses.  In the seesaw mechanism\cite{seesaw},
\begin{equation}
M_\nu \sim \left(\begin{array}{cc}
0 & m_D \\
m^T_D & m_R \end{array}\right) .  
\end{equation}
Diagonalization of the mass matrix produces one light neutrino,
$m_{\mathrm{light}}\sim {m_D^2 \over m_R}$, and one unobservably
heavy, $m_{\mathrm{heavy}} \sim m_R$.  The factor ($m_D$/$m_R$) is the
needed small parameter that accounts for the distinct scale of
neutrino masses.  The masses for the $\nu_e, \nu_\mu$, and $\nu_\tau$
are then related to the squares of the corresponding quark masses
$m_u$, $m_c$, and $m_t$.  Taking $m_R \sim 10^{16}$ GeV, a typical
grand unification scale for models built on groups like SO(10), the
seesaw mechanism gives the crude relation
\begin{equation}
m_{\nu_e}: m_{\nu_\mu}: m_{\nu_\tau} \leftrightarrow 2 \cdot 
10^{-12}: 2 
\cdot 10^{-7}: 3 \cdot 10^{-3} \mathrm{eV}. 
\end{equation}
The fact that solar neutrino experiments can probe small neutrino
masses, and thus provide insight into possible new mass scales $m_R$
that are far beyond the reach of direct accelerator measurements, has
been an important theme of the field.
    
One of the most interesting possibilities for solving the solar
neutrino problem has to do with neutrino masses.  For simplicity we
will discuss just two neutrinos.  If a neutrino has a mass $m$, we
mean that as it propagates through free space, its energy and momentum
are related in the usual way for this mass.  Thus if we have two
neutrinos, we can label those neutrinos according to the eigenstates
of the free Hamiltonian, that is, as mass eigenstates.

But neutrinos are produced by the weak interaction.  In this case, we
have another set of eigenstates, the flavor eigenstates.  We can
define a $\nu_e$ as the neutrino that accompanies the positron in
$\beta$ decay.  Likewise we label by $\nu_\mu$ the neutrino produced
in muon decay.

Now the question: are the eigenstates of the free Hamiltonian and of
the weak interaction Hamiltonian identical?  Most likely the answer is
no: we know this is the case with the quarks, since the different
families (the analog of the mass eigenstates) do interact through the
weak interaction.  That is, the up quark decays not only to the down
quark, but also occasionally to the strange quark.  (This is why we
had a $\cos \theta_c$ in our weak interaction amplitude: the amplitude
for $u \rightarrow s$ is proportional to $\sin \theta_c$.)  Thus we
suspect that the weak interaction and mass eigenstates, while spanning
the same two-neutrino space, are not coincident: the mass eigenstates
$|\nu_1 \rangle$ and $|\nu_2 \rangle$ (with masses $m_1$ and $m_2$)
are related to the weak interaction eigenstates by
\begin{eqnarray}
|\nu_e\rangle  &=& \cos \theta_v |\nu_1\rangle  
+ \sin \theta_v|\nu_2 \rangle  \nonumber \\
|\nu_\mu\rangle &=& - \sin \theta_v |\nu_1 \rangle 
+ \cos \theta_v |\nu_2 
\rangle 
\end{eqnarray}
where $\theta_v$ is the (vacuum) mixing angle. 
  
An immediate consequence is that a state produced as a $|\nu_e\rangle$
or a $|\nu_\mu\rangle$ at some time $t$ --- for example, a neutrino
produced in $\beta$ decay --- does not remain a pure flavor eigenstate
as it propagates away from the source.  This is because the different
mass eigenstates comprising the neutrino will accumulate different
phases as they propagate downstream, a phenomenon known as vacuum
oscillations (vacuum because the experiment is done in free space).
To see the effect, suppose we produce a neutrino in some $\beta$ decay
where we measure the momentum of the initial nucleus, final nucleus,
and positron.  Thus the outgoing neutrino is a momentum
eigenstate~\cite{nauenberg}.  At time $t$=0
\begin{equation}
|\nu(t=0)\rangle  = |\nu_e \rangle = \cos \theta_v |\nu_1\rangle  
+ \sin \theta_v|\nu_2 \rangle . 
\end{equation}
Each eigenstate subsequently propagates with a phase
\begin{equation}
e^{i(\vec{k} \cdot \vec{x} - \omega t)} =
e^{i(\vec{k} \cdot \vec{x} - \sqrt{m_i^2 + k^2}t)} . 
\end{equation}
But if the neutrino mass is small compared to the neutrino
momentum/energy, one can write
\begin{equation}
\sqrt{m_i^2+k^2} \sim k(1 + {m_i^2 \over 2k^2}) . 
\end{equation}
Thus we conclude
\begin{eqnarray}
|\nu(t) \rangle &=& e^{i(\vec{k} \cdot \vec{x} - kt
-(m_1^2+m_2^2)t/4k)} \nonumber \\
& & \times [\cos \theta_v |\nu_1 \rangle e^{i \delta m^2 t/4k}
+ \sin \theta_v |\nu_2 \rangle e^{-i \delta m^2 t/4k} ] . 
\label{eq:24}
\end{eqnarray}
We see there is a common average phase (which has no physical
consequence) as well as a beat phase that depends on
\begin{equation}
\delta m^2 = m_2^2 - m_1^2 .
\end{equation}
Now it is a simple matter to calculate the probability that 
our neutrino state remains a $|\nu_e\rangle$ at time t
\begin{eqnarray}
P_{\nu_e} (t) &=& | \langle \nu_e | \nu(t) \rangle |^2 \nonumber \\ 
 &=& 1 - \sin^2 2 \theta_v \sin^2 \left({\delta m^2 t \over 4 
k}\right) \rightarrow 1 - {1 \over 2} \sin^2 2 \theta_v 
\end{eqnarray}
where the limit on the right is appropriate for large $t$.  Now $E
\sim k$, where E is the neutrino energy, by our assumption that the
neutrino masses are small compared to $k$.  Thus we can reinsert the
units above to write the probability in terms of the distance $x$ of
the neutrino from its source,
\begin{equation}
P_{\nu} (x) =1 - \sin^2 2 \theta_v \sin^2 \left({\delta m^2c^4 
x\over 4 \hbar c E} \right) . 
\end{equation}
(When one properly describes the neutrino state as a wave packet, the
large-distance behavior follows from the eventual separation of the
mass eigenstates.)  If the the oscillation length
\begin{equation}
L_o = {4 \pi \hbar c E \over \delta m^2 c^4} 
\end{equation}
is comparable to or shorter than one astronomical unit, a reduction in
the solar $\nu_e$ flux would be expected in terrestrial neutrino
oscillations.
  
The suggestion that the solar neutrino problem could be explained by
neutrino oscillations was first made by Pontecorvo in 1958, who
pointed out the analogy with $K_0 \leftrightarrow \bar K_0$
oscillations.  From the point of view of particle physics, the sun is
a marvelous neutrino source.  The neutrinos travel a long distance and
have low energies ($\sim$ 1 MeV), implying a sensitivity down to
\begin{equation}
\delta m^2 \gsim 10^{-12} eV^2.
\end{equation}
In the seesaw mechanism, $\delta m^2 \sim m^2_2$, so neutrino masses
as low as $m_2 \sim 10^{-6}$ eV could be probed.  In contrast,
terrestrial oscillation experiments with accelerator or reactor
neutrinos are typically limited to $\delta m^2 \gsim 0.1 $ eV$^2$.

From the expressions above one expects vacuum oscillations to affect
all neutrino species equally, if the oscillation length is small
compared to an astronomical unit.  This is somewhat in conflict with
the solar neutrino data, as we have argued that the $^7$Be neutrino
flux is quite suppressed.  Furthermore, there is a weak theoretical
prejudice that $\theta_v$ should be small, like the Cabibbo angle.
The first objection, however, can be circumvented in the case of
``just so" oscillations where the oscillation length is comparable to
one astronomical unit.  In this case the oscillation probability
becomes sharply energy dependent, and one can choose $\delta m^2$ to
preferentially suppress one component (e.g., the monochromatic $^7$Be
neutrinos).  This scenario has been explored by several groups and
remains an interesting possibility.  However, the requirement of large
mixing angles remains.
  
Below we will see that stars allow us to ``get around" this problem
with small mixing angles.  In preparation for this, we first present
the results above in a slightly more general way.  The analog of Eq.
(\ref{eq:24}) for an initial muon neutrino ($|\nu(t=0)\rangle =
|\nu_\mu\rangle$) is
\begin{eqnarray}
|\nu(t) \rangle &=& e^{i(\vec{k} \cdot \vec{x} - kt
-(m_1^2+m_2^2)t/4k)} \nonumber \\
&& \times [-\sin \theta_v |\nu_1 \rangle e^{i \delta m^2 t/4k}
+ \cos \theta_v |\nu_2 \rangle e^{-i \delta m^2 t/4k} ]
\label{eq:30}
\end{eqnarray}
Now if we compare Eqs. (\ref{eq:24}) and (\ref{eq:30}) we see that
they are special cases of a more general problem.  Suppose we write
our initial neutrino wave function as
\begin{equation}
 |\nu(t=0)\rangle = a_e(t=0) |\nu_e \rangle + a_\mu(t=0) 
|\nu_\mu \rangle . 
\label{eq:31}
\end{equation}
Then Eqs. (\ref{eq:24}) and (\ref{eq:30}) tell us that the subsequent
  propagation is described by changes in $a_e(x)$ and $a_\mu(x)$
  according to (this takes a bit of algebra)
\begin{equation}
i {d \over dx} \left( \matrix { a_{\textstyle e} \cr
a_{\textstyle \mu} \cr} \right) = {1 \over 4E} \left ( \matrix{
- \delta m^2 \cos 2 \theta_{\textstyle v}
~~~~~~~~~~~\delta m^2\sin
2\theta_{\textstyle v} \cr 
\delta m^2\sin 2 \theta_{\textstyle v} ~~~~~~~~~~~ 
\delta m^2
\cos 2\theta_{\textstyle v} \cr} \right) \left( \matrix {
a_{\textstyle e} \cr
a_{\textstyle \mu} \cr} \right) . 
\end{equation}
Note that the common phase has been ignored: it can be absorbed into
the overall phase of the coefficients $a_e$ and $a_\mu$, and thus has
no consequence.  Also, we have equated $x = t,$ that is, set $c$ = 1.

\subsection{The Mikheyev-Smirnov-Wolfenstein Mechanism}

The view of neutrino oscillations changed when Mikheyev and
Smirnov~\cite{ms} showed in 1985 that the density dependence of the
neutrino effective mass, a phenomenon first discussed by Wolfenstein
in 1978, could greatly enhance oscillation probabilities: a $\nu_e$ is
adiabatically transformed into a $\nu_\mu$ as it traverses a critical
density within the sun.  It became clear that the sun was not only an
excellent neutrino source, but also a natural regenerator for cleverly
enhancing the effects of flavor mixing.
   
While the original work of Mikheyev and Smirnov was numerical, their
phenomenon was soon understood analytically as a level-crossing
problem.  If one writes the neutrino wave function in matter as in Eq.
(\ref{eq:31}), the evolution of $a_e(x)$ and $a_\mu(x)$ is governed by
\begin{equation}
i {d \over dx} \left( \matrix { a_{\textstyle e} \cr
a_{\textstyle \mu} \cr} \right) = {1 \over 4E} \left ( \matrix{
2E \sqrt2 G_F \rho(x) - \delta m^2 \cos 2 \theta_{\textstyle v}
~~~~~~\delta m^2\sin
2\theta_{\textstyle v} \cr 
\delta m^2\sin 2 \theta_{\textstyle v} ~~~ -2E \sqrt2 G_F \rho(x) +
\delta m^2
\cos 2\theta_{\textstyle v} \cr} \right) \left( \matrix {
a_{\textstyle e} \cr
a_{\textstyle \mu} \cr} \right) 
\end{equation}
where G$_F$ is the weak coupling constant and $\rho (x)$ the solar
electron density.  If $\rho (x)$ = 0, this is exactly our previous
result and can be trivially integrated to give the vacuum oscillation
solutions of Sec. 2.5.  The new contribution to the diagonal elements,
$2 E \sqrt2 G_F \rho(x)$, represents the effective contribution to
$M^2_\nu$ that arises from neutrino-electron scattering.  The indices
of refraction of electron and muon neutrinos differ because the former
scatter by charged and neutral currents, while the latter have only
neutral current interactions.  The difference in the forward
scattering amplitudes determines the density-dependent splitting of
the diagonal elements of the new matter equation.

It is helpful to rewrite this equation in a basis consisting of the
light and heavy local mass eigenstates (i.e., the states that
diagonalize the right-hand side of the equation),
\begin{eqnarray}
|\nu_L (x)\rangle &=& \cos \theta (x)|\nu_e\rangle - \sin \theta 
(x)|\nu_\mu\rangle \nonumber \\
|\nu_H(x)\rangle &=& \sin \theta (x)|\nu_e\rangle + \cos \theta 
(x)|\nu_\mu \rangle . 
\end{eqnarray}
The local mixing angle is defined by
\begin{eqnarray}
\sin 2 \theta (x)  &=& {\sin 2 \theta_{\textstyle v} \over 
\sqrt{X^2 (x) + \sin^2
2\theta_{\textstyle v}}} \nonumber \\
\cos 2\theta (x)  &=& {-X (x) \over \sqrt{X^2 (x) + 
\sin^2 2\theta_{\textstyle v}}} 
\end{eqnarray}
where $X(x) = 2 \sqrt2G_F \rho(x) E/\delta m^2 - \cos
2\theta_{\textstyle v}$.  Thus $\theta(x)$ ranges from
$\theta_{\textstyle v}$ to $\pi/2$ as the density $\rho(x)$ goes from
0 to $\infty$.

If we define
\begin{equation}
|\nu (x) \rangle = a_H(x)|\nu_H(x)\rangle + a_L(x)|\nu_L(x)\rangle,
\end{equation}
the neutrino propagation can be rewritten in terms of the local
mass eigenstates
\begin{equation}
i {d \over dx} \pmatrix{
a_H \cr
a_L \cr} = \pmatrix {
\lambda(x) & i \alpha (x) \cr
-i \alpha (x) & - \lambda (x) \cr }
\pmatrix
{a_H \cr
a_L }
\end{equation}
with the splitting of the local mass eigenstates determined by
\begin{equation}
2 \lambda (x) = {\delta m^2 \over 2E} \sqrt{X^2 (x) + \sin^2 2 
\theta_{\textstyle v}} 
\end{equation}
and with mixing of these eigenstates governed by the density gradient
\begin{equation}
\alpha (x) = \left({E \over \delta m^2}\right)
 \, {\sqrt2 \, G_F {d \over dx}
\rho(x)
\sin 2 \theta_{\textstyle v} \over X^2 (x) + \sin^2 2 
\theta_{\textstyle v}}.
\end{equation}
The results above are quite interesting: the local mass eigenstates
diagonalize the matrix if the density is constant.  In such a limit,
the problem is no more complicated than our original vacuum
oscillation case, although our mixing angle is changed because of the
matter effects.  But if the density is not constant, the mass
eigenstates in fact evolve as the density changes.  This is the crux
of the MSW effect.  Note that the splitting achieves its minimum
value, ${\delta m^2 \over 2E} \sin 2 \theta_v$, at a critical density
$\rho_c = \rho (x_c)$
\begin{equation}
2 \sqrt2 E G_F \rho_c = \delta m^2 \cos 2 \theta_v 
\end{equation}
that defines the point where the diagonal elements of the original
flavor matrix cross.

Our local-mass-eigenstate form of the propagation equation can be
trivially integrated if the splitting of the diagonal elements is
large compared to the off-diagonal elements (see Eq. (57)),
\begin{equation}
\gamma (x) = \left|{\lambda (x) \over \alpha (x)}\right| = {\sin^2
2\theta_{\textstyle v} \over \cos
2\theta_{\textstyle v}} \, {\delta m^2 \over 2 E} \, {1 \over 
|{1 \over \rho_c}
{d \rho (x) \over
dx}|} {[X (x)^2 + \sin^2 2\theta_v]^{3/2} \over \sin^3 2\theta_v} 
\gg 1, 
\end{equation}
a condition that becomes particularly stringent near the crossing
point,
\begin{equation}
\gamma_c = \gamma (x_c) = {\sin^2 2\theta_v \over \cos 2\theta_v} 
\, {\delta
m^2 \over 2 E} \, {1 \over \left|{1 \over \rho_c} {d \rho (x) 
\over dx}|_{x = x_c}\right|} \gg 1. 
\end{equation}
The resulting adiabatic electron neutrino survival
probability~\cite{bethe}, valid when $\gamma_c \gg 1$, is
\begin{equation}
P^{\rm adiab}_{\nu_e} = {1 \over 2} + {1 \over 2} \cos 2 \theta_v \cos 2
\theta_i 
\end{equation}
where $\theta_i = \theta (x_i)$ is the local mixing angle at the
density where the neutrino was produced.

\begin{figure}[htb]
\psfig{bbllx=1.2cm,bblly=2.0cm,bburx=18cm,bbury=14.5cm,figure=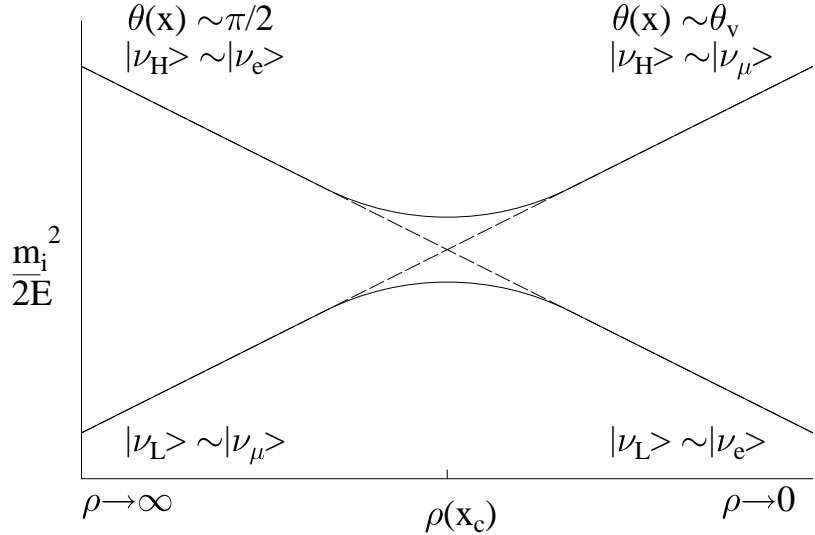,height=3.0in}
\caption{Schematic illustration of the MSW crossing.  The dashed 
lines correspond to the electron-electron and muon-muon diagonal
elements of the $M^2$ matrix in the flavor basis.  Their 
intersection defines the level-crossing density $\rho_c$.
The solid lines are the trajectories of the light and heavy
local mass eigenstates.  If the electron neutrino is produced 
at high density and propagates adiabatically, it will follow
the heavy-mass trajectory, emerging from the sun as a $\nu_\mu$.}
\end{figure}
  
The physical picture behind this derivation is illustrated in Figure
8.  One makes the usual assumption that, in vacuum, the $\nu_e$ is
almost identical to the light mass eigenstate, $\nu_L(0)$, i.e., $m_1
< m_2$ and $\cos \theta_v \sim$ 1.  But as the density increases, the
matter effects make the $\nu_e$ heavier than the $\nu_\mu$, with
$\nu_e \to \nu_H (x)$ as $\rho(x)$ becomes large.  The special
property of the Sun is that it produces $\nu_e$s at high density that
then propagate to the vacuum where they are measured.  The adiabatic
approximation tells us that if initially $\nu_e \sim \nu_H (x)$, the
neutrino will remain on the heavy mass trajectory provided the density
changes slowly.  That is, if the solar density gradient is
sufficiently gentle, the neutrino will emerge from the sun as the
heavy vacuum eigenstate, $ \sim \nu_\mu$.  This guarantees nearly
complete conversion of $\nu_e$s into $\nu_\mu$s, producing a flux that
cannot be detected by the Homestake or SAGE/GALLEX detectors.
   
But this does not explain the curious pattern of partial flux
suppressions coming from the various solar neutrino experiments.  The
key to this is the behavior when $\gamma_c \lsim$ 1.  Our expression
for $\gamma(x)$ shows that the critical region for non-adiabatic
behavior occurs in a narrow region (for small $\theta_v$) surrounding
the crossing point, and that this behavior is controlled by the
derivative of the density.  This suggests an analytic strategy for
handling non-adiabatic crossings: one can replace the true solar
density by a simpler (integrable!) two-parameter form that is
constrained to reproduce the true density and its derivative at the
crossing point $x_c$. Two convenient choices are the linear $(\rho (x)
= a + bx)$ and exponential $(\rho (x) = ae^{-bx})$ profiles.  As the
density derivative at $x_c$ governs the non-adiabatic behavior, this
procedure should provide an accurate description of the hopping
probability between the local mass eigenstates when the neutrino
traverses the crossing point.  The initial and ending points $x_i$ and
$x_f$ for the artificial profile are then chosen so that $\rho(x_i)$
is the density where the neutrino was produced in the solar core and
$\rho(x_f) = 0$ (the solar surface), as illustrated in in Figure 9.
Since the adiabatic result ($P_{\nu_e}^{\mathrm{adiab}}$) depends only
on the local mixing angles at these points, this choice builds in that
limit.  But our original flavor-basis equation can then be integrated
exactly for linear and exponential profiles, with the results given in
terms of parabolic cylinder and Whittaker functions, respectively.

\begin{figure}[htb]
\psfig{bbllx=0.0cm,bblly=2.8cm,bburx=16cm,bbury=21.3cm,figure=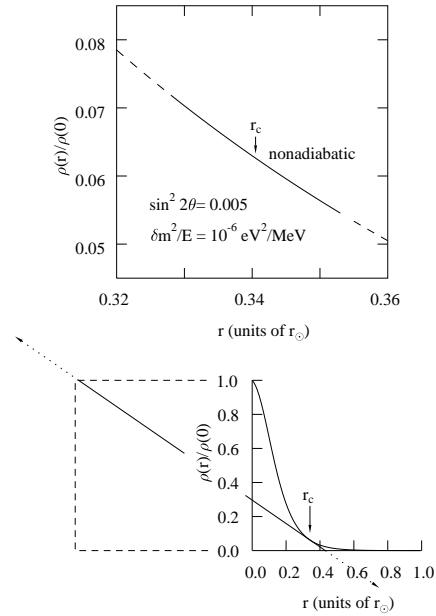,height=3.3in}
\caption{The top figure illustrates, for one choice of sin$^2 2\theta$
  and $\delta m^2$, that the region of non-adiabatic propagation (solid
  line) is usually confined to a narrow region around the crossing
  point $r_c$.  In the lower figure, the solid lines represent the
  solar density and a linear approximation to that density that has
  the correct initial and final values, as well as the correct density
  and density derivative at $r_c$.  Thus the linear profile is a very
  good approximation to the sun in the vicinity of the crossing point.
  The MSW equations can be solved analytically for this wedge.  By
  extending the wedge to $\pm \infty$ (dotted lines) and assuming
  adiabatic propagation in these regions of unphysical density, one
  obtains the simple Landau-Zener result discussed in the text.}
\end{figure}
  
That result can be simplified further by observing that the
non-adiabatic region is generally confined to a narrow region around
$x_c$, away from the endpoints $x_i$ and $x_f$.  We can then extend
the artificial profile to $x = \pm \infty$, as illustrated by the
dashed lines in Figure 9.  As the neutrino propagates adiabatically in
the unphysical region $x < x_i$, the exact solution in the physical
region can be recovered by choosing the initial boundary conditions
\begin{eqnarray}
a_L (- \infty) &=& - a_\mu (- \infty) = \cos \theta_i e^{- i 
\int^{x_i}_{- 
\infty} \lambda (x) dx} \nonumber\\
a_H (- \infty) &=& a_e (- \infty) = \sin \theta_i 
e^{i \int^{x_i}_{- \infty} 
\lambda (x) dx} . 
\end{eqnarray}
That is, $|\nu (-\infty)\rangle$ will then adiabatically evolve to
$|\nu (x_i)\rangle = |\nu_e\rangle$ as $x$ goes from $- \infty$ to
$x_i$.  The unphysical region $x > x_f$ can be handled similarly.

With some algebra a simple generalization of the adiabatic
result emerges that is valid for all $\delta m^2/E$ and $\theta_v$
\begin{equation}
P_{\nu_e} = {1 \over 2} + {1 \over 2} \cos 2 \theta_v \cos 2 
\theta_i ( 1 - 2P_{\rm {hop}}) 
\end{equation}
where P$_{\rm {hop}}$ is the Landau-Zener probability of hopping from
the heavy mass trajectory to the light trajectory on traversing the
crossing point.  For the linear approximation to the
density~\cite{hlz,plz},
\begin{equation}
P^{\rm {lin}}_{\rm {hop}} = e^{- \pi \gamma_c/2} . 
\end{equation}
As it must by our construction, $P_{\nu_e}$ reduces to P$^{\rm
  {adiab}}_{\nu_e}$ for $\gamma_c \gg$ 1.  When the crossing becomes
non-adiabatic (e.g., $\gamma_c \ll 1$ ), the hopping probability goes
to 1, allowing the neutrino to exit the sun on the light mass
trajectory as a $\nu_e$, i.e., no conversion occurs.

Thus there are two conditions for strong conversion of solar
neutrinos: there must be a level crossing (that is, the solar core
density must be sufficient to render $\nu_e \sim \nu_H (x_i)$ when it
is first produced) and the crossing must be adiabatic.  The first
condition requires that $\delta m^2/E$ not be too large, and the
second $\gamma_c \gsim$ 1.  The combination of these two constraints,
illustrated in Fig. 10, defines a triangle of interesting parameters in
the ${\delta m^2 \over E} - \sin^2 2\theta_v$ plane, as Mikheyev and
Smirnov found by numerically integration.  A remarkable feature of
this triangle is that strong $\nu_e \to \nu_\mu$ conversion can occur
for very small mixing angles $(\sin^2 2 \theta \sim10^{-3}$), unlike
the vacuum case.

\begin{figure}[htb]
\psfig{bbllx=-1.5cm,bblly=0.0cm,bburx=15cm,bbury=22.0cm,figure=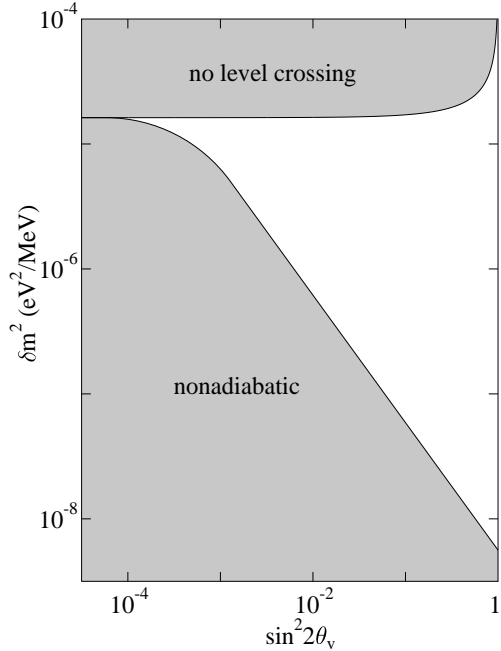,height=3.6in}
\caption{MSW conversion for a neutrino produced at the sun's
center.  The upper shaded region indicates those $\delta m^2/E$
where the vacuum mass splitting is too great to be overcome
by the solar density.  Thus no level crossing occurs.  The
lower shaded region defines the region where the level crossing
is non-adiabatic ($\gamma_c$ less than unity).  The unshaded
region corresponds to adiabatic level crossings where strong
$\nu_e \rightarrow \nu_\mu$ will occur.}
\end{figure}
  
One can envision superimposing on Fig. 10 the spectrum of solar
neutrinos, plotted as a function of ${\delta m^2 \over E}$ for some
choice of $\delta m^2$.  Since Davis sees {\it some} solar neutrinos,
the solutions must correspond to the boundaries of the triangle in
Fig. 10.  The horizontal boundary indicates the maximum ${\delta m^2
  \over E}$ for which the sun's central density is sufficient to cause
a level crossing.  If a spectrum properly straddles this boundary, we
obtain a result consistent with the Homestake experiment in which low
energy neutrinos (large 1/E) lie above the level-crossing boundary
(and thus remain $\nu_e$'s), but the high-energy neutrinos (small 1/E)
fall within the unshaded region where strong conversion takes place.
Thus such a solution would mimic nonstandard solar models in that only
the $^8$B neutrino flux would be strongly suppressed.  The diagonal
boundary separates the adiabatic and non-adiabatic regions.  If the
spectrum straddles this boundary, we obtain a second solution in which
low energy neutrinos lie within the conversion region, but the
high-energy neutrinos (small 1/E) lie below the conversion region and
are characterized by $\gamma \ll 1$ at the crossing density.  (Of
course, the boundary is not a sharp one, but is characterized by the
Landau-Zener exponential).  Such a non-adiabatic solution is quite
distinctive since the flux of pp neutrinos, which is strongly
constrained in the standard solar model and in any steady-state
nonstandard model by the solar luminosity, would now be sharply
reduced.  Finally, one can imagine ``hybrid" solutions where the
spectrum straddles both the level-crossing (horizontal) boundary and
the adiabaticity (diagonal) boundary for small $\theta$, thereby
reducing the $^7$Be neutrino flux more than either the pp or $^8$B
fluxes.

What are the results of a careful search for MSW solutions satisfying
the Homestake, Kamiokande/SuperKamiokande, and SAGE/GALLEX
constraints?  This has been explored in detail by several groups.  One
solution, corresponding to a region surrounding $\delta m^2 \sim 6
\cdot 10^{-6} $eV$^2$ and $\sin^2 2\theta_v \sim 6 \cdot 10^{-3}$, is
the hybrid case described above.  It is commonly called the
small-angle solution.  A second, large-angle solution exists,
corresponding to $\delta m^2 \sim 10^{-5} $eV$^2$ and $\sin^2 2
\theta_v \sim$ 0.6.  These solutions can be distinguished by their
characteristic distortions of the solar neutrino spectrum.  The
survival probabilities $P_{\nu_e}^{\rm MSW}$(E) for the small- and
large-angle parameters given above are shown as a function of E in
Fig. 11.

\begin{figure}[htb]
\psfig{bbllx=0.5cm,bblly=1.3cm,bburx=18cm,bbury=13.7cm,figure=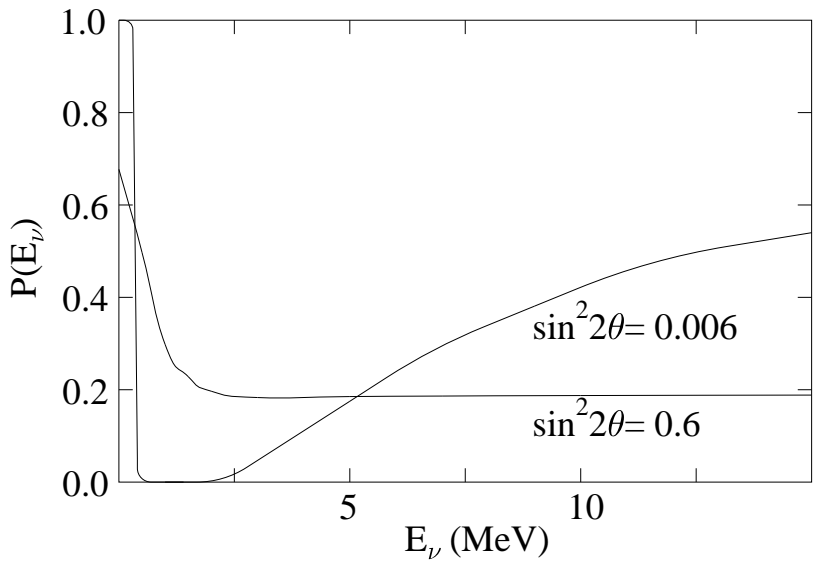,height=3.0in}
\caption{MSW survival probabilities P(E$_\nu$) for typical small
angle and large angle solutions.}
\end{figure}
  
The MSW mechanism provides a natural explanation for the pattern of
observed solar neutrino fluxes.  While it requires profound new
physics, both massive neutrinos and neutrino mixing are expected in
extended models.  The small-angle solution corresponds to $\delta m^2
\sim 10^{-5}$ eV$^2$, and thus is consistent with $m_2 \sim$ few
$\cdot 10^{-3}$ eV.  This is a typical $\nu_\tau$ mass in models where
$m_R \sim m_{\rm {GUT}}$.  This mass is also reasonably close to
atmospheric neutrino values.  On the other hand, if it is the
$\nu_\mu$ participating in the oscillation, this gives $m_R \sim
10^{12}$ GeV and predicts a heavy $\nu_\tau \sim$ 10 eV.  Such a mass
is of great interest cosmologically as it would have consequences for
supernova physics, the dark matter problem, and the formation of
large-scale structure.

\subsection{SuperKamiokande, SNO, and the MSW Mechanism}

SuperKamiokande and Sudbury Neutrino Observatory (SNO) detectors are
real-time counting detectors in contrast to the radiochemical
detectors such as Homestake, GALLEX, and SAGE, which can only
determine a time- and energy-integral of the flux (cf. Section 2.2).
Both the SuperKamiokande and SNO can detect neutrinos through the
reaction 
\begin{equation}
  \label{eq:sno1}
   \nu_x + e^- \rightarrow \nu_x + e^-.
\end{equation}
The electrons coming from this reaction are confined to a forward
cone. Hence detecting the Cerenkov radiation from the final electron
one can determine neutrino's direction. In this reaction it is
very difficult to determine the energy of the neutrino from the
measured energy of the final electron because of the kinematical
broadening. However the measured energy spectra of the recoil
electrons can nevertheless yield valuable information about the
neutrino energy spectrum. Indeed recent SuperKamiokande measurement of
the recoil electron energy spectrum is consistently below the SSM
predictions at electron energies up to 14 MeV (which itself is
consistent with other neutrino experiments). However for the higher
energy bins the data are anomalously high as compared to the
predictions. Although it can be fitted by assuming a very large hep
neutrino flux \cite{plam1} the origin of this anomalous behavior is not
understood.

In addition to the reaction Eq. (\ref{eq:sno1}) SNO can detect
neutrinos through the charged-current reaction
\begin{equation}
  \label{eq:sno2}
  \nu_e + d \rightarrow p + p + e^-, 
\end{equation}
and the neutral current reaction
\begin{equation}
  \label{eq:sno3}
  \nu_x (\overline{\nu}_x)+ d \rightarrow \nu_x (\overline{\nu}_x)+ p
  + n. 
\end{equation}
The neutrons produced in Eq. (\ref{eq:sno3}) can be detected either
using $(n, \gamma )$ reactions on salt dissolved in the heavy water or
by using $^3$He proportional counters. The electrons coming from the
reaction (\ref{eq:sno2}) are essentially monochromatic with energies
$\sim E_{\nu} - 1.44$ MeV and they have a very different angular
distribution ($1- \cos \theta_e /3$) with respect to the neutrino
direction than that of the electrons coming from the reaction
(\ref{eq:sno1}) (which are constrained to the forward cone).

It should be emphasized that since matter-enhanced neutrino conversion
is energy-dependent, MSW mechanism will distort the neutrino energy
spectra. However since only the energy of the final state electron can
be measured, if a broad range of electron energies correspond to a
given neutrino energy this distortion will be smeared. The charged
current break-up reaction, Eq. (\ref{eq:sno2}), is more suitable for
this purpose since the final electron energy is distributed around a
very narrow peak centered at the initial neutrino energy. Spectrum
distortion at SNO for the small-angle MSW solution ($\delta m^2 \sim 5
\times 10^{-6}$ eV$^2$ and $\sin 2 \theta \sim 0.01$) is shown in
Figure 12.
\begin{figure}[t]
  \vspace{8pt} \centerline{\hbox{\epsfxsize=3 in \epsfbox[19 65 526
      685]{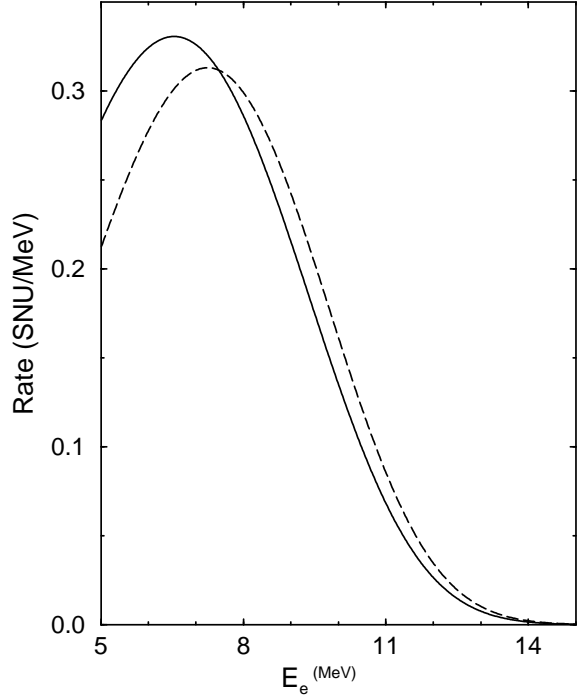}}}
\caption{The dashed line shows the spectrum distortion at SNO for 
  the small-angle MSW solution ($\delta m^2 \sim 5 \times 10^{-6}$
  eV$^2$ and $\sin 2 \theta \sim 0.01$). The solid line is the
  spectrum without MSW oscillations, normalized to the same total rate
  as with MSW oscillations.} 
\vspace{8pt}
\end{figure}

The reaction (\ref{eq:sno3}) measures the total flux of all the
neutrinos that participate in the low-energy weak interactions. Hence
if there are no sterile neutrinos (more about this later) it can be
used to test the neutrino oscillation hypothesis. For example if the
electron neutrinos are being converted to muon neutrinos in the Sun,
flux measured using the reaction (\ref{eq:sno3}) should agree with the
SSM predictions. Finally it is worth mentioning that the electron
antineutrinos coming from a supernova or a possible spin-flavor
precession in the Sun (cf. Section 2.9) can also be detected through
the reaction \be \overline{\nu}_e + d \rightarrow n + n + e^+. \ee

\subsection{Other Particle Physics Solutions}

If the MSW mechanism proves not to be the solution of the solar
neutrino problem, it still will have greatly enhanced the importance
of solar neutrino physics: the existing experiments have ruled out
large regions in the $\delta m^2 - \sin^2 2\theta_v$ plane
(corresponding to nearly complete $\nu_e \to \nu_\mu$ conversion) that
remain hopelessly beyond the reach of accelerator neutrino oscillation
experiments.
 
A number of other particle physics solutions have been considered,
such as neutrino decay, matter-catalyzed neutrino decay, and solar
energy transport by weakly interacting massive particles.  But perhaps
the most interesting possibility, apart from the MSW mechanism, was
stimulated by suggestions that the $^{37}$Cl signal might be varying
with a period comparable to the 11-year solar cycle.  While the
evidence for this has weakened, the original claims generated renewed
interest in neutrino magnetic moment interactions with the solar
magnetic field.

The original suggestions by Cisneros and by
Okun, Voloshyn, and Vysotsky envisioned the rotation
\begin{equation}
\nu_{e_L} \to \nu_{e_R} 
\end{equation}
producing a right-handed neutrino with sterile interactions in the
standard model.  With the discovery of the MSW mechanism, it was
realized that matter effects would break the vacuum degeneracy of the
$\nu_{e_L}$ and $\nu_{e_R}$, suppressing the spin precession.  Lim and
Marciano~\cite{lim} and Akhmedov~\cite{ak} pointed out that this
difficulty was naturally circumvented for
\begin{equation}
\nu_{e_L} \to \nu_{\mu_R} 
\end{equation}
as the different matter interactions of the $\nu_e$ and $\nu_\mu$ can
compensate for the vacuum $\nu_e - \nu_\mu$ mass difference, producing
a crossing similar to the usual MSW mechanism.  Such spin-flavor
precession can then occur at full strength due to an off-diagonal (in
flavor) magnetic moment.

Quite relevant to this suggestion is the very strong limit on both
diagonal and off-diagonal magnetic moments is imposed by studies of
the red giant cooling process of plasmon decay into neutrinos
\begin{equation}
\gamma^* \to \nu_i \bar \nu_j. 
\end{equation}
The result is $|\mu_{ij}| \lsim 3 \cdot 10^{-12} \mu_B$, where $\mu_B$
is an electron Bohr magneton~\cite{raff}.  (This can be compared to a
simple one-loop estimate~\cite{fuj} of the neutrino magnetic moment of
$\sim 10^{-18} \mu_B$, taking a typical dark matter value for the
neutrino mass of a few eV.)  If a magnetic moment at the red giant
limit is assumed, it follows that solar magnetic field strengths of
$B_\odot \gsim 10^6 G$ are needed to produce interesting effects.
Since the location of the spin-flavor level crossing depends on the
neutrino energy, such fields have to be extensive to affect an
appreciable fraction of the neutrino spectrum.  It is unclear whether
these conditions can occur in the sun.
    
We will return to a discussion of spin-flavor oscillations and
associated effects later in these lectures.  However, this brief
introduction leads naturally to the next two topics, neutrino mass
possibilities and the evolution of red giants.

\section{Dirac and Majorana Neutrinos and Stellar Cooling}
\subsection{The Neutrino Mass Matrix}

Consider a general 4n $\times$ 4n neutrino mass matrix where 
n is the number of flavors
\begin{equation}
\begin{array}{c} (\bar{\Psi}^c_L,\bar{\Psi}^R,
\bar{\Psi}_L,\bar{\Psi}^c_R) \\ \\ \\ \end{array}
\left( \begin{array}{cccc} 0 & 0 &  M_L & M^T_D \\
0 & 0 & M_D &  M_R^\dag \\  M_L^\dag & M_D^\dag & 0 & 0 \\ M_D^* &  M_R & 0 & 0
\end{array} \right) 
\left( \begin{array}{c} \Psi^c_L \\ \Psi_R \\ \Psi_L \\ \Psi^c_R
\end{array} \right)
\end{equation}
where each entry in this matrix is understood to be a n $\times$ n
matrix operating in flavor space.  The entries $M_D$ are the Dirac
mass terms, while the $M_L$ and $M_R$ are the Majorana terms.  The
latter break the invariance of the Dirac equation under the
transformation $\psi(x) \rightarrow e^{i\alpha} \psi(x)$ associated
with a conserved lepton number.  Thus it is these terms that govern
lepton-number-violating processes like double beta decay.

One can proceed to diagonalize this matrix
\begin{equation}
\Psi_{\nu(e)}^L = \sum_{i=1}^{2n} U_{ei}^L \tilde{\nu}_i(x) 
~~\mathrm{with~masses}~m_i .
\end{equation}
The eigenstates are two-component Majorana neutrinos~\cite{haxtonbb},
yielding the proper $2 \times 2n = 4n$ degrees of freedom,
where $n$ is the number of flavors.
We can recover the Majorana and Dirac limits:\\
$\bullet$ If $M_R$ = $M_L$ = 0, the eigenstates of this matrix
become pairwise degenerate, allowing the $2n$ two-component 
eigenstates to be paired to form $n$ four-component Dirac
eigenstates.\\
$\bullet$ If $M_D$ = 0, the left- and right-handed components
decouple, yielding $n$ left-handed Majorana eigenstates with
standard model interactions. 
  
There are interesting physical effects associated with these limits.
Dirac neutrinos can have magnetic dipole, electric dipole (CP and T
violating), and anapole (P violating) moments, as well as nonzero
charge radii.  Majorana neutrinos can have anapole moments but only
transition magnetic and electric dipole moments. We mentioned in the
previous lecture that transition magnetic moments were quite
interesting in the context of MSW effects, as well as the use of $M_R$
in the seesaw mechanism.
  
In most extended models both Dirac and Majorana mass terms arise, with
possibilities for interesting physics.  We previously described the
seesaw mechanism, which explains the lightness of the neutrinos as a
effect of Dirac masses mixing with a heavy right-handed Majorana mass
associated with scales far beyond the standard model.  Another
scenario often discussed in the pseudoDirac limit, where larger Dirac
masses are accompanied by small Majorana masses.  In the limit that
the Majorana masses vanish, one recovers the degeneracy that allows
one to patch two Majorana neutrinos with opposite CP into a
four-component Dirac state, as described above.  But if the Majorana
terms remain non-vanishing, two Majorana states remain split by an
amount $M_L$.  If $M_L$ is much less than $M_D$, one obtains a nearly
degenerate pair of states at mass $M_D$.  

There are interesting low-energy consequences of these scenarios.
Studies of the shape of the $\beta$ decay spectrum provides one test
of neutrino masses: such tests are clearly limited to masses less than
the Q-value of the decay.  Since the different mass eigenstates
correspond to distinct final states, the mass eigenstates do not
interfere.  Each neutrino that couples to the electron contributes to
the decay according to its mixing probability $|U_{ei}|^2$ and
according to the Fermi function $F(\omega_e,m_i)$, where $\omega_e$ is
the electron energy and $m_i$ the mass of the $ith$ neutrino mass
eigenstates.  Thus massive neutrinos can appear as increments to the
$\beta$ spectrum, turning on once $\omega_e$ drops below Q - $m_i$.

Another interesting process is neutrinoless double $\beta$ decay,
which can be pictured as $\beta$ decay in which the produced neutrino
is reabsorbed on a second nucleon, leading to a final state with no
neutrinos and two electrons (and thus a nuclear charge change of two
units).  The neutrino mass eigenstates appear virtually, and thus
clearly interfere.  The process is also manifestly lepton-number
violating, and thus must require Majorana masses.  It can be shown
that the amplitude, for light neutrino masses, depends on $|U_{ei}|^2
m_i \eta^{CP}_i$, where $\eta^{CP}_i$ is the relative CP eigenvalue of
the $ith$ mass eigenstate.  (We are assuming CP conservation.)  Thus
in the Dirac limit, where the mass eigenstates of opposite CP become
pairwise degenerate, this $\beta \beta$ decay mass vanishes.  This
then guarantees that the $\beta \beta$ amplitude vanishes in the
absence of Majorana mass terms.

\subsection{Red Giants and Helium Burning}

We now consider the evolution off the main sequence of a solar-like
star, with a mass above half a solar mass.  As the hydrogen burning in
the core progresses to the point that no more hydrogen is available,
the stellar core consists of the ashes from this burning, $^4$He.  The
star then goes through an
interesting evolution:\\
$\bullet$ With no further means of producing energy, the core slowly
contracts, thereby increasing in temperature as gravity
does work on the core.\\
$\bullet$ Matter outside the core is still hydrogen rich, and can
generate energy through hydrogen burning.  Thus burning in this shell
of material supports the outside layers of the star.  Note as the core
contracts, this matter outside the core also is pulled deeper into the
gravitational potential.  Furthermore, the shell H burning continually
adds more mass to the core.  This means the burning in the shell must
intensify to generate the additional gas pressure to fight gravity.
The shell also thickens as this happens, since more hydrogen is above
the burning temperature.\\
$\bullet$ The resulting increasing gas pressure causes the outer
envelope of the star to expand by a larger factor, up to a factor of
50.  The increase in radius more than compensates for the increased
internal energy generation, so that a cooler surface results.  The
star reddens.  Thus this class of
star is named a red supergiant.\\
$\bullet$ This evolution is relatively rapid, perhaps a few hundred
million years: the dense core requires large energy production.  The
helium core is supported by its degeneracy pressure, and is
characterized by densities $\sim 10^6$ g/cm$^3$.  This stage ends when
the core reaches densities and temperatures that allow helium burning
through the reaction
\begin{equation}
\alpha + \alpha + \alpha \rightarrow ^{12}C + \gamma .
\end{equation}
As this reaction is very temperature dependent (see below), the
conditions for ignition are very sharply defined.  This has the
consequence that the core mass at the helium flash point
is well determined. \\
$\bullet$ The onset of helium burning produces a new source of support
for the core.  The energy release elevates the temperature and the
core expands: He burning, not electron degeneracy, now supports the
core.  The burning shell and envelope have moved outward, higher in
the gravitational potential.  Thus shell hydrogen burning slows (the
shell cools) because less gas pressure is needed to satisfy
hydrostatic equilibrium.  All of this means the evolution of the star
has now slowed: the red giant moves along the ``horizontal branch", as
interior temperatures slowly elevate much as in the main sequence.

The 3$\alpha$ process depends on some rather interesting nuclear
physics.  The first interesting ``accident" involves the near
degeneracy of the $^8$Be ground state and two separated $\alpha$s: The
$^8$Be $0^+$ ground state is just 92 keV above the 2$\alpha$
threshold.  The measured width of the $^8$Be ground state is 2.5 eV,
which corresponds to a lifetime of
\begin{equation}
\tau_m \sim 2.6 \cdot 10^{-16} \mathrm{s} . 
\end{equation}
One can compare this number to the typical time for one $\alpha$
to pass another.  The red giant core temperature is $T_7 \sim 10
\rightarrow E \sim 8.6$ keV.  Thus v/c $\sim$ 0.002.  So the
transit time is
\begin{equation}
\tau \sim {d \over v} \sim {5f \over 0.002} {1 \over
3 \cdot 10^{10} \mathrm{cm/sec}} {10^{-13} \mathrm{cm} \over
f} \sim 8 \cdot 10^{-21} \mathrm{s} . 
\end{equation}
This is more than five orders of magnitude shorter than $\tau_m$
above.  Thus when a $^8$Be nucleus is produced, it lives for a
substantial time compared to this naive estimate.

To quantify this, we calculate the flux-averaged cross section
assuming resonant capture
\begin{equation}
\langle \sigma v \rangle = ( {2 \pi \over \mu k T} )^{3/2}
{\Gamma \Gamma \over \Gamma} e^{-E_r/kT} 
\end{equation}
where $\Gamma$ is the 2$\alpha$ width of the $^8$Be ground state.
This is the cross section for the $\alpha+\alpha$ reaction to form the
compound nucleus then decay by $\alpha + \alpha$.  But since there is
only one channel, this is clearly also the result for producing the
compound nucleus $^8$Be.

By multiplying the rate/volume for producing $^8$Be by the lifetime of
$^8$Be, one gets the number of $^8$Be nuclei per unit volume
\begin{eqnarray}
N(Be) &=& {N_\alpha N_\alpha \over 1 + \delta_{\alpha \alpha}}
\langle \sigma v \rangle \tau_m \nonumber \\
&=& {N_\alpha N_\alpha \over 1 + \delta_{\alpha \alpha}}
\langle \sigma v \rangle {1 \over \Gamma} \nonumber \\
 &=&{N_\alpha^2 \over 2} ({2 \pi \over \mu k T})^{3/2} 
e^{-E_r/kT} . 
\end{eqnarray}
Notice that the concentration is {\it independent} of $\Gamma$.  So a
small $\Gamma$ is not the reason we obtain a substantial buildup of
$^8$Be.  This is easily seen: if the width is small, then the
production rate of $^8$Be goes down, but the lifetime of the nucleus
once it is produced is longer.  The two effects cancel to give the
same $^8$Be concentration.  One sees that the significant $^8$Be
concentration results from two effects: 1) $\alpha+\alpha$ is the only
open channel and 2) the resonance energy is low enough that some small
fraction of the $\alpha+\alpha$ reactions have the requisite energy.
As $E_r = 92$ keV, $E_r/kT$ = 10.67/$T_8$ (where $T_8$ is the
temperature in $10^8$K) so that
\begin{equation}
N(Be) = N_\alpha^2 T_8^{-3/2} e^{-10.67/T_8} (0.94 \cdot
10^{-33} \mathrm{cm^3} ). 
\end{equation}
So plugging in typical values of $N_\alpha \sim 1.5 \cdot 
10^{28}$/cm$^3$ (corresponding to $\rho_\alpha \sim 10^5$
g/cm$^3$) and $T_8 \sim$ 1 yields
\begin{equation}
 {N(^8Be) \over N(\alpha)} \sim 3.2 \times 10^{-10} . 
\end{equation}

Salpeter suggested that this concentration would then allow $\alpha +
^8$Be$ \rightarrow ^{12}$C to take place.  Hoyle then argued that this
reaction would not be fast enough to produce significant burning
unless it was also resonant.  Now the mass of $^8$Be + $\alpha$,
relative to $^{12}$C,is 7.366 MeV, and each nucleus has $J^\pi = 0^+$.
Thus s-wave capture would require a $0^+$ resonance in $^{12}$C at
$\sim$ 7.4 MeV.  No such state was then known, but a search by Cook,
Fowler, Lauritsen, and Lauritsen revealed a $0^+$ level at 7.644 MeV,
with decay channels $^8$Be$ + \alpha$ and $\gamma$ decay to the $2^+$
4.433 level in $^{12}$C.  The parameters are
\begin{equation}
 \Gamma_\alpha \sim 8.9 \mathrm{eV} ~~~~~~~~
\Gamma_\gamma \sim 3.6 \cdot 10^{-3} \mathrm{eV} . 
\end{equation}
The resonant cross section formula gives
\begin{equation}
 r_{48} = N_8 N_\alpha ({2 \pi \over \mu kT})^{3/2}
{\Gamma_\alpha \Gamma_\gamma \over \Gamma} e^{-E_r/kT}. 
\end{equation}
Plugging in our previous expression for $N(^8$Be) yields
\begin{equation}
r_{48} = N_\alpha^3 T_8^{-3} e^{-42.9/T_8}
(6.3 \cdot 10^{-54} \mathrm{cm^6/s}).
\end{equation}
If we denote by $\omega_{3 \alpha}$ the decay rate of an $\alpha$ 
in our plasma, then
\begin{eqnarray}
 \omega_{3 \alpha} &=& 3 N_\alpha^2 T_8^{-3} 
e^{-42.9/T_8} (6.3 \cdot 10^{-54} \mathrm{cm^6/sec}) \nonumber \\
 &=& ({N_\alpha \over 1.5 \cdot 10^{28}/\mathrm{cm}^3})^2
(4.3 \cdot 10^3/\mathrm{sec}) T_8^{-3} e^{-42.9/T_8} . 
\end{eqnarray}

Now the energy release per reaction is 7.27 MeV.
Thus we can calculate the energy produced per gram, $\epsilon$:
\begin{eqnarray}
 \epsilon &=& \omega_{3 \alpha} {7.27 \mathrm{MeV} \over 3}
{1.5 \cdot 10^{23} \over \mathrm{g}} \nonumber \\ 
 &=& (2.5 \cdot 10^{21} \mathrm{erg/g~sec}) ({N_\alpha \over
1.5 \cdot 10^{28}/\mathrm{cm}^3})^2 T_8^{-3} 
e^{-42.9/T_8} . 
\end{eqnarray}
We can evaluate this at a temperature of $T_8 \sim$ 1 to find
\begin{equation}
 \epsilon \sim (584 \mathrm{ergs/g~sec}) ({N_\alpha \over
1.5 \cdot 10^{28}/\mathrm{cm}^3})^2 . 
\end{equation}
Typical values found in stellar calculations are in good agreement
with this: typical red giant energy production is $\sim$ 100
ergs per gram per second.

To get a feel for the temperature sensitivity of this process,
we can do a Taylor series expansion, finding
\begin{equation}
 \epsilon(T) \sim ({T \over T_o})^{40} N^2_\alpha . 
\end{equation}
This steep temperature dependence is the reason the He flash
is delicately dependent on conditions in the core.\\

\subsection{Neutrino Magnetic Moments and He Ignition}

Prior to the helium flash, the degenerate He core radiates energy
largely by neutrino pair emission.  The process is the decay of a
plasmon --- which one can think of as a photon ``dressed" by
electron-hole excitations --- thereby acquiring an effective mass of
about 10 keV.  The photon couples to a neutrino pair through a
electron particle-hole pair that then decays into a $Z_o \rightarrow
\nu \bar{\nu}$.

If this cooling is somehow enhanced, the degenerate helium core would
be kept cooler, and would not ignite at the normal time.  Instead it
would continue to grow until it overcame the enhanced cooling to
reach, once again, the ignition temperature.

One possible mechanism for enhanced cooling is a neutrino magnetic
moment.  Then the plasmon could directly couple to a neutrino pair.
The strength of this coupling would depend on the size of the magnetic
moment.

A delay in the time of He ignition has several observable
consequences, including changing the ratio of red giant to horizontal
branch stars.  Thus, using the standard theory of red giant evolution,
investigators have attempted to determine what size of magnetic moment
would produce unacceptable changes in the astronomy.  The result is a
limit~\cite{raff} on the neutrino magnetic moment of
\begin{equation}
 \mu_{ij} \lsim 3 \cdot 10^{-12} \mathrm{electron~Bohr~magnetons} 
\end{equation}
as was mentioned earlier.  This limit is more than two orders of
magnitude more stringent than that from direct laboratory tests.

This example is just one of a number of such constraints that can be
extracted from similar stellar cooling arguments.  The arguments
above, for example, can be repeated for neutrino electric dipole
moments.  More interesting, it can be repeated for axion emission from
red giants.  Axions, the pseudoGolstone bosons associate with the
solution of the strong CP problem suggested by Peccei and Quinn, are
very light and can be produced radiatively within the red giant by the
Compton process, by the Primakoff process off nuclei, or by emission
from low-lying nuclear levels, such at from the 14 keV transition in
$^{57}$Fe.  The net result is that axions of mass above a few eV are
excluded; if axions have a direct coupling to electrons, so that the
Compton process off electrons operates, the constraint is considerably
tighter.

A similar argument can be formulated for supernova cooling.  During
SN1987A the neutrino burst detected by IMB and by Kamiokande was
consistent with cooling on a timescale of about 4 seconds.  Thus any
process cooling the star more efficiently than neutrino emission would
have shortened this time, while also reducing the flux in neutrinos.
Large Dirac neutrino masses allow trapped neutrinos to scatter into
sterile right-handed states.  Right-handed neutrinos, lacking standard
model interactions, would then escape the star (provided they do not
scatter back into interacting left-handed states).  Unfortunately the
upper bounds imposed on the neutrino mass are quite model dependent,
ranging over (1-25) keV.

The supernova cooling argument can also be repeated for axions.  The
window of sensitive runs from 1 eV (above this mass they are more
strongly coupled than neutrinos, and thus cannot compete with neutrino
cooling) to about 0.01 eV (below this mass they are too weakly coupled
to be produced on the timescale of supernova cooling).  It is
interesting that the supernova and red giant cooling limit on axions
nearly meet: a small window may still exist around a few eV if the
axion has no coupling to electrons.

\section{Atmospheric Neutrinos}

Atmospheric neutrinos arise from the decay of secondary pions, kaons,
and muons produced by the collisions of primary cosmic rays with the
oxygen and nitrogen nuclei in the upper atmosphere. For energies less
than 1 GeV all the secondaries decay :
\begin{eqnarray}
\pi^{\pm} (K^{\pm}) &\rightarrow &\mu^{\pm} + \nu_{\mu}
(\overline{\nu}_{\mu}), \nonumber\\ \mu^{\pm} &\rightarrow & e^{\pm} +
\nu_e (\overline{\nu}_e) +  \overline{\nu}_{\mu} (\nu_{\mu}).
\end{eqnarray}
Consequently one expects the ratio
\begin{equation}
r = (\nu_e + \overline{\nu}_e) / (\nu_{\mu} + \overline{\nu}_{\mu})
\end{equation}
to be approximately 0.5 in this energy range. Detailed Monte Carlo
calculations \cite{gaisser}, including the effects of muon
polarization, give $  r \sim 0.45$. Since one is evaluating a ratio of
similarly calculated processes, systematic errors are significantly
reduced. Different groups estimating  this ratio, even though they
start with neutrino fluxes which can differ in magnitude by up to
25\%, all agree within a few percent \cite{bludman}. As the shower
energy increases more muons survive due to time dilation. Hence one
expects the ratio $r$ to decrease as the energy increases. The ratio
(observed to predicted) of ratios
\begin{equation}
R = {(\nu_{\mu} / \nu_e)_{\rm data} \over (\nu_{\mu} / \nu_e)_{\rm
Monte  Carlo} }
\end{equation}
was determined in several experiments. There seems to  be a persistent
discrepancy between theory and experiment. The most recent measurement
of this ratio at SuperKamiokande \cite{skatm} gives 
\begin{equation}
  \label{atm1}
  R=0.63 \pm 0.03 ({\rm stat}) \pm 0.05 ({\rm syst}) \nonumber
\end{equation}
for sub-GeV events which were fully contained in the detector and
\begin{equation}
  \label{atm2}
  R=0.65 \pm 0.05 ({\rm stat}) \pm 0.08 ({\rm syst}) \nonumber
\end{equation}
for fully- and partially-contained multi-GeV events. As the agreement
between data and theoretical expectation would give $R=1$, neutrino
oscillations are invoked to explain the discrepancy. Another evidence
for this explanation comes from measuring $R$ as a function of the
zenith angle, $\Theta$, between the vertical and neutrino direction. A
down-going neutrino ($\Theta \sim 0^o$) travels through the atmosphere
above the detector (a distance of about 20 km), whereas an up-going
neutrino ($\Theta \sim 180^o$) has traveled through the entire Earth
(a distance of about 13000 km). Hence a measurement of number of
neutrinos as a function of the zenith angle yields information about
their numbers as a function of the distance traveled. 

\begin{figure}[t]
  \vspace{8pt} \centerline{\hbox{\epsfxsize=3 in 
\epsfbox[8 -3 502 485]{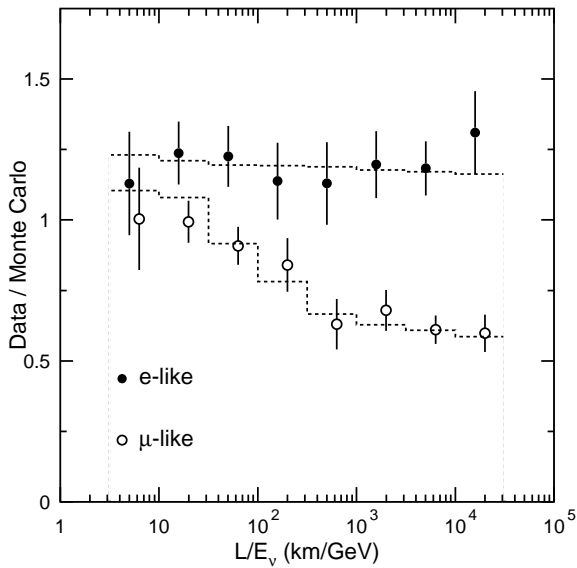}}}
\caption{The ratio of fully contained events measured
  at SuperKamiokande versus reconstructed $L/E_{\nu}$. the dashed
  lines show the expected shape for $\nu_{\mu} \leftrightarrow
  \nu_{\tau}$ oscillations with $\delta m^2 = 2.2 \cdot 10^{-3}$
  eV$^2$ and $\sin^2 2 \theta = 1$. } 
\vspace{8pt}
\end{figure}

The SuperKamiokande collaboration measured the zenith angle dependence
of not only $R$, but of the electron and muon neutrino fluxes
separately \cite{skatm}. This information is shown in Figure 13, where
the data is plotted as a function of the reconstructed $L/E_{\nu}$
instead of the zenith angle. The data exhibit a zenith-angle
(distance) dependent deficit of muon neutrinos, but not of electron
neutrinos. This behavior is consistent with $\nu_{\mu} \leftrightarrow
  \nu_{\tau}$ oscillations. This interpretation is consistent with the
  recent measurements of up-going muons at Kamiokande \cite{katm} and
  MACRO \cite{matm} detectors. These muons are produced by the very
  high energy up-going
  muon neutrinos in the rocks surrounding the detectors. Both of these
  experiments also observe a deficit. 

\section{Supernovae, Supernova Neutrinos, and Nucleosynthesis}

Consider a massive star, in excess of 10 solar masses, burning the
hydrogen in its core under the conditions of hydrostatic equilibrium.
When the hydrogen is exhausted, the core contracts until the density
and temperature are reached where 3$\alpha \rightarrow ^{12}$C can
take place.  The He is then burned to exhaustion.  This pattern (fuel
exhaustion, contraction, and ignition of the ashes of the previous
burning cycle) repeats several times, leading finally to the explosive
burning of $^{28}$Si to Fe.  For a heavy star, the evolution is rapid:
the star has to work harder to maintain itself against its own
gravity, and therefore consumes its fuel faster.  A 25 solar mass star
would go through all of these cycles in about 7 My, with the final
explosive Si burning stage taking a few days.  The result is an
``onion skin" structure of the pre-collapse star in which the star's
history can be read by looking at the surface inward: there are
concentric shells of H, $^4$He, $^{12}$C, $^{16}$O and $^{20}$Ne,
$^{28}$Si, and $^{56}$Fe at the center.

\subsection{The Explosion Mechanism~\protect\cite{mezz}}
The source of energy for this evolution is nuclear binding energy.  A
plot of the nuclear binding energy $\delta$ as a function of nuclear
mass shows that the minimum is achieved at Fe.  In a scale where the
$^{12}$C mass is picked as zero:
\begin{center}
$^{12}$C~~~~~$\delta$/nucleon = 0.000 MeV \\
$^{16}$O~~~~~$\delta$/nucleon = -0.296 MeV \\
$^{28}$Si~~~~$\delta$/nucleon = -0.768 MeV \\
$^{40}$Ca~~~~$\delta$/nucleon = -0.871 MeV \\
$^{56}$Fe~~~~$\delta$/nucleon = -1.082 MeV \\
$^{72}$Ge~~~~$\delta$/nucleon = -1.008 MeV \\
$^{98}$Mo~~~~$\delta$/nucleon = -0.899 Mev
\end{center}
Thus once the Si burns to produce Fe, there is no further source of
nuclear energy adequate to support the star.  So as the last remnants
of nuclear burning take place, the core is largely supported by
degeneracy pressure, with the energy generation rate in the core being
less than the stellar luminosity.  The core density is about 2 $\times
10^9$ g/cc and the temperature is kT $\sim$ 0.5 MeV.

Thus the collapse that begins with the end of Si burning is not halted
by a new burning stage, but continues.  As gravity does work on the
matter, the collapse leads to a rapid heating and compression of the
matter.  As the nucleons in Fe are bound by about 8 MeV, sufficient
heating can release $\alpha$s and a few nucleons.  At the same time,
the electron chemical potential is increasing.  This makes electron
capture on nuclei and any free protons favorable,
\begin{equation}
 e^- + p \rightarrow \nu_e + n. 
\end{equation}
Note that the chemical equilibrium condition is
\begin{equation}
 \mu_e + \mu_p = \mu_n + \langle E_\nu \rangle. 
\end{equation}
Thus the fact that neutrinos are not trapped plus the rise in the
electron Fermi surface as the density increases, lead to increased
neutronization of the matter.  The escaping neutrinos carry off energy
and lepton number.  Both the electron capture and the nuclear
excitation and disassociation take energy out of the electron gas,
which is the star's only source of support.  This means that the
collapse is very rapid.  Numerical simulations find that the iron core
of the star ($\sim$ 1.2-1.5 solar masses) collapses at about 0.6 of the
free fall velocity.

In the early stages of the infall the $\nu_e$s readily escape.  But
neutrinos are trapped when a density of $\sim$ 10$^{12}$g/cm$^3$ is
reached.  At this point the neutrinos begin to scatter off the matter
through both charged current and coherent neutral current processes.
The neutral current neutrino scattering off nuclei is particularly
important, as the scattering cross section is off the total nuclear
weak charge, which is approximately the neutron number.  This process
transfers very little energy because the mass energy of the nucleus is
so much greater than the typical energy of the neutrinos.  But
momentum is exchanged.  Thus the neutrino ``random walks" out of the
star.  When the neutrino mean free path becomes sufficiently short,
the ``trapping time" of the neutrino begins to exceed the time scale
for the collapse to be completed.  This occurs at a density of about
10$^{12}$ g/cm$^3$, or somewhat less than 1\% of nuclear density.
After this point, the energy released by further gravitational
collapse and the star's remaining lepton number are trapped within the
star.

If we take a neutron star of 1.4 solar masses and a radius of
10 km, an estimate of its binding energy is
\begin{equation}
 {G M^2 \over 2R} \sim 2.5 \times 10^{53} \mathrm{ergs}. 
\end{equation}
Thus this is roughly the trapped energy that will later be radiated in
neutrinos.

The trapped lepton fraction $Y_L$ is a crucial parameter in the
explosion physics: a higher trapped $Y_L$ leads to a larger homologous
core, a stronger shock wave, and easier passage of the shock wave
through the outer core, as will be discussed below.  Most of the
lepton number loss of an infalling mass element occurs as it passes
through a narrow range of densities just before trapping.  The reasons
for this are relatively simple: on dimensional grounds weak rates in a
plasma go as $T^5$, where T is the temperature.  Thus the electron
capture rapidly turns on as matter falls toward the trapping radius,
and lepton number loss is maximal just prior to trapping.  Inelastic
neutrino reactions have an important effect on these losses, as the
coherent trapping cross section goes as $E_\nu^2$ and is thus least
effective for the lowest energy neutrinos.  As these neutrinos escape,
inelastic reactions repopulate the low energy states, allowing the
neutrino emission to continue.

The velocity of sound in matter rises with increasing density.  The
inner homologous core, with a mass $M_{HC} \sim 0.6-0.9 $ solar
masses, is that part of the iron core where the sound velocity exceeds
the infall velocity.  This allows any pressure variations that may
develop in the homologous core during infall to even out before the
collapse is completed.  As a result, the homologous core collapses as
a unit, retaining its density profile.  That is, if nothing were to
happen to prevent it, the homologous core would collapse to a point.

The collapse of the homologous core continues until nuclear densities
are reached.  As nuclear matter is rather incompressible ($\sim$ 200
MeV/f$^3$), the nuclear equation of state is effective in halting the
collapse: maximum densities of 3-4 times nuclear are reached, e.g.,
perhaps $6 \cdot 10^{14}$ g/cm$^3$.  The innermost shell of matter
reaches this supernuclear density first, rebounds, sending a pressure
wave out through the homologous core.  This wave travels faster than
the infalling matter, as the homologous core is characterized by a
sound speed in excess of the infall speed.  Subsequent shells follow.
The resulting series of pressure waves collect near the sonic point
(the edge of the homologous core).  As this point reaches nuclear
density and comes to rest, a shock wave breaks out and begins its
traversal of the outer core.

Initially the shock wave may carry an order of magnitude more energy
than is needed to eject the mantle of the star (less than 10$^{51}$
ergs).  But as the shock wave travels through the outer iron core, it
heats and melts the iron that crosses the shock front, at a loss of
$\sim$ 8 MeV/nucleon.  The enhanced electron capture that occurs off
the free protons left in the wake of the shock, coupled with the
sudden reduction of the neutrino opacity of the matter (recall
$\sigma_{coherent} \sim N^2$), greatly accelerates neutrino emission.
This is another energy loss.\footnote{Many numerical models predict a
  strong ``breakout" burst of $\nu_e$s in the few milliseconds
  required for the shock wave to travel from the edge of the
  homologous core to the neutrinosphere at $\rho \sim 10^{12}$
  g/cm$^3$ and $r \sim 50$ km.  The neutrinosphere is the term from
  the neutrino trapping radius, or surface of last scattering.}  The
summed losses from shock wave heating and neutrino emission are
comparable to the initial energy carried by the shock wave.  Thus most
numerical models fail to produce a successful ``prompt" hydrodynamic
explosion.

Most of the attention in the past decade focused on two explosion
scenarios.  In the prompt mechanism described above, the shock wave is
sufficiently strong to survive the passage of the outer iron core with
enough energy to blow off the mantle of the star.  The most favorable
results were achieved with smaller stars (less than 15 solar masses)
where there is less overlying iron, and with soft equations of state,
which produce a more compact neutron star and thus lead to more energy
release.  In part because of the lepton number loss problems discussed
earlier, now it is widely believed that this mechanism fails for all
but unrealistically soft nuclear equations of state.

The delayed mechanism begins with a failed hydrodynamic explosion;
after about 0.01 seconds the shock wave stalls at a radius of 200-300
km.  It exists in a sort of equilibrium, gaining energy from matter
falling across the shock front, but loosing energy to the heating of
that material.  However, after perhaps 0.5 seconds, the shock wave is
revived due to neutrino heating of the nucleon ``soup" left in the
wake of the shock.  This heating comes primarily from charged current
reactions off the nucleons in that nucleon gas; quasi-elastic
scattering also contributes.  This high entropy radiation-dominated
gas may reach two MeV in temperature.  The pressure exerted by this
gas helps to push the shock outward. It is important to note that
there are limits to how effective this neutrino energy transfer can
be: if matter is too far from the core, the coupling to neutrinos is
too weak to deposit significant energy.  If too close, the matter may
be at a temperature (or soon reach a temperature) where neutrino
emission cools the matter as fast or faster than neutrino absorption
heats it.  The term ``gain radius" is used to describe the region
where useful heating is done.

This subject is still controversial and unclear.  The problem is
numerically challenging, forcing modelers to handle the difficult
hydrodynamics of a shock wave; the complications of the nuclear
equation of state at densities not yet accessible to experiment;
modeling in two or three dimensions; handling the slow diffusion of
neutrinos; etc.  Not all of these aspects can be handled reasonably at
the same time, even with existing supercomputers.  Thus there is
considerable disagreement about whether we have any supernova model
that succeeds in ejecting the mantle.

However the explosion proceeds, there is agreement that 99\% of the 3
$\cdot 10^{53}$ ergs released in the collapse is radiated in neutrinos
of all flavors.  The time scale over which the trapped neutrinos leak
out of the protoneutron star is about three seconds.  Through most of
their migration out of the protoneutron star, the neutrinos are in
flavor equilibrium
\begin{equation}
 \mathrm{e.g.},~~ \nu_e + \bar{\nu}_e \leftrightarrow \nu_\mu + \bar{\nu}_\mu. 
\end{equation}
As a result, there is an approximate equipartition of energy among the
neutrino flavors.  After weak decoupling, the $\nu_e$s and
$\bar{\nu_e}$s remain in equilibrium with the matter for a longer
period than their heavy-flavor counterparts, due to the larger cross
sections for scattering off electrons and because of the
charge-current reactions
\begin{eqnarray}
 \nu_e + n &&\leftrightarrow p + e^- \nonumber \\ 
 \bar{\nu_e} + p &&\leftrightarrow n + e^+. 
\end{eqnarray}
Thus the heavy flavor neutrinos decouple from deeper within the star,
where temperatures are higher.  Typical calculations yield
\begin{equation}
 T_{\nu_\mu} \sim T_{\nu_\tau} \sim 8 \mathrm{MeV} ~~~~
 T_{\nu_e} \sim 3.5 \mathrm{MeV}~~~~T_{\bar{\nu_e}} \sim 4.5 \mathrm{MeV}. 
\end{equation}
The difference between the $\nu_e$ and $\bar{\nu_e}$ temperatures is a
result of the neutron richness of the matter, which enhances the rate
for charge-current reactions of the $\nu_e$s, thereby keeping them
coupled to the matter somewhat longer.

This temperature hierarchy is crucially important to nucleosynthesis
and also to possible neutrino oscillation scenarios.  The three-flavor
MSW level-crossing diagram is shown in Fig. 14.  One very popular
scenario attributes the solar neutrino problem to $\nu_\mu
\leftrightarrow \nu_e$ transmutation; this means that a second
crossing with a $\nu_\tau$ could occur at higher density.  It turns
out plausible seesaw mass patterns suggest a $\nu_\tau$ mass on the
order of a few eV, which would be interesting cosmologically.  The
second crossing would then occur outside the neutrino sphere, that is,
after the neutrinos have decoupled and have fixed spectra with the
temperatures given above.  Thus a $\nu_e \leftrightarrow \nu_\tau$
oscillation would produce a distinctive $T \sim 8$ MeV spectrum of
$\nu_e$s.  This has dramatic consequences for terrestrial detection
and for nucleosynthesis in the supernova.

\begin{figure}[htb]
\psfig{bbllx=1.0cm,bblly=4.0cm,bburx=18cm,bbury=18.5cm,figure=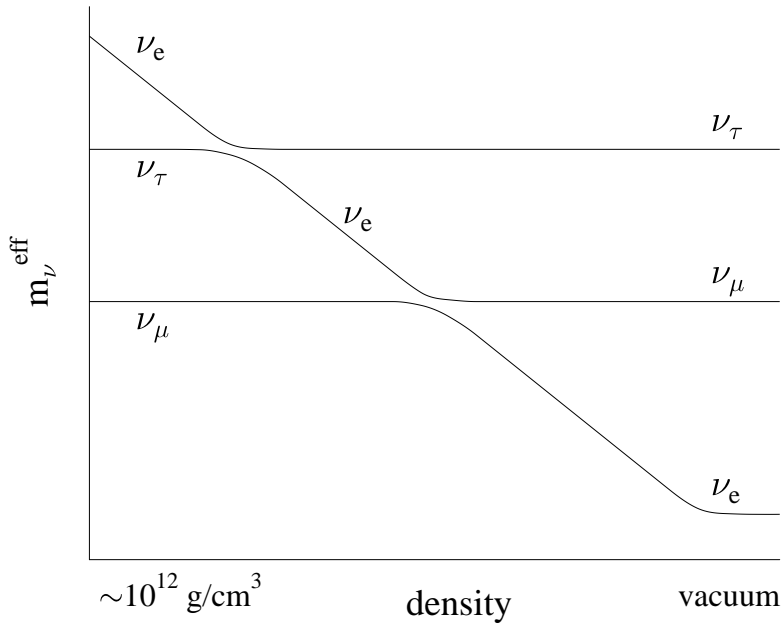,height=3.5in}
\caption{Three-flavor neutrino level-crossing diagram.  One 
popular scenario associates the solar neutrino problem with
$\nu_e \leftrightarrow \nu_\mu$ oscillations and predicts 
a cosmologically interested massive $\nu_\tau$ with 
$\nu_e \leftrightarrow \nu_\tau$ oscillations near the
supernova neutrinosphere.}
\end{figure}
  
\subsection{The Neutrino Process~\protect\cite{nupro}}

Core-collapse supernovae are one of the major engines driving galactic
chemical evolution, producing and ejecting the metals that enrich our
galaxy.  The discussion of the previous section described the
hydrostatic evolution of a presupernova star in which large quantities
of the most abundant metals (C, O, Ne,...) are synthesized and later
ejected during the explosion.  During the passage of the shock wave
through the star's mantle, temperature of $\sim (1-3) \cdot 10^9$K and
are reached in the silicon, oxygen, and neon shells.  This shock wave
heating induces $(\gamma,\alpha) \leftrightarrow (\alpha,\gamma)$ and
related reactions that generate a mass flow toward highly bound
nuclei, resulting in the synthesis of iron peak elements as well as
less abundant odd-A species.  Rapid neutron-induced reactions are
thought to take place in the high-entropy atmosphere just above the
mass cut, producing about half of the heavy elements above A $\sim$
80.  This is the subject of the Sec. 4.3.  Finally, the $\nu$-process
described below is responsible for the synthesis of rare species such
as $^{11}$B and $^{19}$F.  This process involves the response of
nuclei at momentum transfers where the allowed approximation is no
longer valid.  Thus we will use the $\nu$-process in this section to
illustrate some of the relevant nuclear physics.

One of the problems -- still controversial -- that may be connected
with the neutrino process is the origin of the light elements Be, B
and Li, elements which are not produced in sufficient amounts in the
big bang or in any of the stellar mechanisms we have discussed.  The
traditional explanation has been cosmic ray spallation interactions
with C, O, and N in the interstellar medium.  In this picture, cosmic
ray protons collide with C at relatively high energy, knocking the
nucleus apart.  So in the debris one can find nuclei like $^{10}$B,
$^{11}$B, and $^7$Li.

But there are some problems with this picture.  First of all, this is
an example of a secondary mechanism: the interstellar medium must be
enriched in the C, O, and N to provide the targets for these
reactions.  Thus cosmic ray spallation must become more effective as
the galaxy ages.  The abundance of boron, for example, would tend to
grow quadratically with metallicity, since the rate of production goes
linearly with metallicity.  But observations, especially recent
measurements with the HST, find a linear growth~\cite{timmes} in the
boron abundance.

A second problem is that the spectrum of cosmic ray protons peaks near
1 GeV, leading to roughly comparable production of the two isotopes
$^{10}$B and $^{11}$B.  That is, while it takes more energy to knock
two nucleons out of carbon than one, this difference is not
significant compared to typical cosmic ray energies.  More careful
studies lead to the expectation that the abundance ratio of $^{11}$B
to $^{10}$B might be $\sim$ 2.  In nature, it is greater than 4.

Fans of cosmic ray spallation have offered solutions to these
problems, e.g., similar reactions occurring in the atmospheres of
nebulae involving lower energy cosmic rays.  As this suggestion was
originally stimulated by the observation of nuclear $\gamma$ rays from
Orion, now retracted, some of the motivation for this scenario has
evaporated.  Here we focus on an alternative explanation, synthesis
via neutrino spallation.

Previously we described the allowed Gamow-Teller (spin-flip) and Fermi
weak interaction operators.  These are the appropriate operators when
one probes the nucleus at a wavelength -- that is, at a size scale --
where the nucleus responds like an elementary particle.  We can then
characterize its response by its macroscopic quantum numbers, the spin
and charge.  On the other hand, the nucleus is a composite object and,
therefore, if it is probed at shorter length scales, all kinds of
interesting radial excitations will result, analogous to the
vibrations of a drumhead.  For a reaction like neutrino scattering off
a nucleus, the full operator involves the additional factor
\begin{equation}
e^{i \vec{k} \cdot \vec{r}} \sim 1 + i \vec{k} \cdot \vec{r} 
\end{equation}
where the expression on the right is valid if the magnitude of
$\vec{k}$ is not too large.  Thus the full charge operator 
includes a ``first forbidden" term
\begin{equation}
 \sum_{i=1}^A \vec{r}_i \tau_3(i) 
\end{equation}
and similarly for the spin operator
\begin{equation}
 \sum_{i=1}^A [\vec{r}_i \otimes \vec{\sigma}(i)]_{J=0,1,2} \tau_3(i). 
\end{equation}
These operators generate collective radial excitations, leading to the
so-called ``giant resonance" excitations in nuclei.  The giant
resonances are typically at an excitation energy of 20-25 MeV in light
nuclei.  One important property is that these operators satisfy a sum
rule (Thomas-Reiche-Kuhn) of the form
\begin{equation}
 \sum_f | \langle f | \sum_{i=1}^A r(i) \tau_3(i) | i \rangle |^2
\sim {N Z \over A} \sim {A \over 4} 
\end{equation}
where the sum extends over a complete set of final nuclear states.
These first-forbidden operators tend to dominate the cross sections
for scattering the high energy supernova neutrinos ($\nu_{\mu}$s and
$\nu_\tau$s), with $E_\nu \sim$ 25 MeV, off light nuclei. From the sum
rule above, it follows that nuclear cross sections per target {\it
  nucleon} are roughly constant.

The E1 giant dipole mode described above is depicted qualitatively in
Fig. 15a.  This description, which corresponds to an early model of
the giant resonance response by Goldhaber and Teller, involves the
harmonic oscillation of the proton and neutron fluids against one
another.  The restoring force for small displacements would be linear
in the displacement and dependent on the nuclear symmetry energy.
There is a natural extension of this model to weak interactions, where
axial excitations occur.  For example, one can envision a mode similar
to that of Fig. 15a where the spin-up neutrons and spin-down protons
oscillate against spin-down neutrons and spin-up protons, the
spin-isospin mode of Fig. 15b.  This mode is one that arises in a
simple SU(4) extension of the Goldhaber-Teller model, derived by
assuming that the nuclear force is spin and isospin independent, at
the same excitation energy as the E1 mode.  In full, the
Goldhaber-Teller model predicts a degenerate 15-dimensional
supermultiplet of giant resonances, each obeying sum rules analogous
to the TRK sum rule.  While more sophisticated descriptions of the
giant resonance region are available, of course, this crude picture is
qualitatively accurate.
  
\begin{figure}[htb]
\psfig{bbllx=0.3cm,bblly=2.8cm,bburx=13cm,bbury=11.5cm,figure=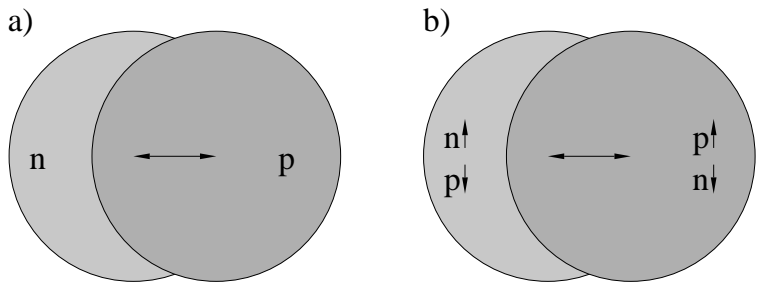,height=1.9in}
\caption{Schematic illustration of a) the E1 giant dipole mode
familiar from electromagnetic interactions and b) a spin-isospin
giant dipole mode associated with the first-forbidden weak
axial response.}
\end{figure}
  
This nuclear physics is important to the $\nu$-process.  The simplest
example of $\nu$-process nucleosynthesis involves the Ne shell in a
supernova.  Because of the first-forbidden contributions, the cross
section for inelastic neutrino scattering to the giant resonances in
Ne is $\sim 3 \cdot 10^{-41}$ cm$^2$/flavor for the more energetic
heavy-flavor neutrinos.  This reaction
\begin{equation}
 \nu + A \rightarrow \nu' + A^* 
\end{equation}
transfers an energy typical of giant resonances, $\sim$ 20 MeV.  A
supernova releases about 3 $\times 10^{53}$ ergs in neutrinos, which
converts to about $4 \times 10^{57}$ heavy flavor neutrinos.  The Ne
shell in a 20 M$_\odot$ star has at a radius $\sim$ 20,000 km.  Thus
the neutrino fluence through the Ne shell is
\begin{equation}
 \phi \sim { 4 \cdot 10^{57} \over 4 \pi (20,000 \mathrm{km})^2 }
\sim 10^{38}/\mathrm{cm}^2. 
\end{equation}
Thus folding the fluence and cross section, one concludes that
approximately 1/300th of the Ne nuclei interact.

This is quite interesting since the astrophysical origin of $^{19}$F
had not been understood.  The only stable isotope of fluorine,
$^{19}$F has an abundance
\begin{equation}
 {^{19}\mathrm{F} \over ^{20}\mathrm{Ne}} \sim {1 \over 3100}. 
\end{equation}
This leads to the conclusion that the fluorine found in a tube of
toothpaste was created by neutral current neutrino reactions deep
inside some ancient supernova.

The calculation of the final $^{19}$F/$^{20}$Ne ratio is
more complicated than the simple 1/300 ratio given above: \\
$\bullet$ When Ne is excited by $\sim$ 20 MeV through inelastic
neutrino scattering, it breaks up in two ways
\begin{eqnarray}
 ^{20}\mathrm{Ne}(\nu,\nu')^{20}\mathrm{Ne}^* 
&&\rightarrow ^{19}\mathrm{Ne} + n 
\rightarrow ^{19}\mathrm{F} + e^+ + \nu_e + n \nonumber \\
 ^{20}\mathrm{Ne}(\nu,\nu')^{20}\mathrm{Ne}^* 
&&\rightarrow ^{19}\mathrm{F}
+ p 
\end{eqnarray}
with the first reaction occurring half as frequently as the second.
As both channels lead to $^{19}$F, we have correctly estimated the
instantaneous abundance ratio in the Ne shell of
\begin{equation}
 {^{19}\mathrm{F} \over ^{20}\mathrm{Ne}} \sim {1 \over 300}. 
\end{equation}
$\bullet$ We must also address the issue of whether the produced
$^{19}$F survives.  In the first 10$^{-8}$ sec the co-produced neutrons
in the first reaction react via
\begin{equation}
^{15}\mathrm{O}(n,p)^{15}\mathrm{N}~~
^{19}\mathrm{Ne}(n,\alpha)^{16}\mathrm{O}~~
^{20}\mathrm{Ne}(n,\gamma)^{21}\mathrm{Ne}~~
^{19}\mathrm{Ne}(n,p)^{19}\mathrm{F} 
\end{equation}
with the result that about 70\% of the $^{19}$F produced via
spallation of neutrons is then immediate destroyed, primarily by the
$(n,\alpha)$ reaction above.  In the next $10^{-6}$ sec the co-produced
protons are also processed
\begin{equation}
 ^{15}\mathrm{N}(p,\alpha)^{12}\mathrm{C}~~
^{19}\mathrm{F}(p,\alpha)^{16}\mathrm{O}~~
^{23}\mathrm{Na}(p,\alpha)^{20}\mathrm{Ne} 
\end{equation}
with the latter two reactions competing as the primary proton poisons.
This makes an important prediction: stars with high Na abundances
should make more F, as the $^{23}$Na acts as a proton
poison to preserve the produced F.\\
$\bullet$ Finally, there is one other destruction mechanism, the
heating associated with the passage of the shock wave.  It turns out
the the F produced prior to shock wave passage can survive if it is in
the outside half of the Ne shell.  The reaction
\begin{equation}
 ^{19}\mathrm{F}(\gamma,\alpha)^{15}\mathrm{N} 
\end{equation}
destroys F for peak explosion temperatures exceeding $1.7 \cdot
10^9$K.  Such a temperature is produced at the inner edge of the Ne
shell by the shock wave heating, but not at the outer edge.

If all of this physics in handled is a careful network code that
includes the shock wave heating and F production both before and
after shock wave passage, the following are the results:
 \[ \begin{array}{cc} \underline{[^{19}\mathrm{F}/^{20}\mathrm{Ne}]/
[^{19}\mathrm{F}/^{20}\mathrm{Ne}]_\odot} & 
\underline{T_{\mathrm{heavy}~\nu} \mathrm{(MeV)}} \\
0.14 & 4 \\ 0.6 & 6 \\ 1.2 & 8 \\ 1.1 & 10 \\ 1.1 & 12 \end{array} \]
where the abundance ratio in the first column has been normalized to
the solar value. One sees that the attribution of F to the neutrino
process argues that the heavy flavor $\nu$ temperature must be greater
than 6 MeV, a result theory favors.  One also sees that F cannot be
overproduced by this mechanism: although the instantaneous production
of F continues to grow rapidly with the neutrino temperature, too much
F results in its destruction through the $(p,\alpha)$ reaction, given
a solar abundance of the competing proton poison $^{23}$Na.  Indeed,
this illustrates an odd quirk: although in most cases the neutrino
process is a primary mechanism, one needs $^{23}$Na present to produce
significant F. Thus in this case the neutrino process is a secondary
mechanism.

While there are other significant neutrino process products ($^7$Li,
$^{138}$La, $^{180}$Ta, $^{15}$N ...), the most important product is
$^{11}$B, produced by spallation off carbon.  A calculation by Timmes
et al.\cite{timmes} found that the combination of the neutrino
process, cosmic ray spallation and big-bang nucleosynthesis together
can explain the evolution of the light elements.  The neutrino
process, which produces a great deal of $^{11}$B but relatively little
$^{10}$B, combines with the cosmic ray spallation mechanism to yield
the observed isotope ratio.  Again, one prediction of this picture is
that early stars should be $^{11}$B rich, as the neutrino process is
primary and operates early in our galaxy's history; the cosmic ray
production of $^{10}$B is more recent.  There is hope that HST studies
will soon be able to discriminate between $^{10}$B and $^{11}$B: as
yet this has not been done.

\subsection{The r-process}

Beyond the iron peak nuclear Coulomb barriers become so high that
charged particle reactions become ineffective, leaving neutron capture
as the mechanism responsible for producing the heaviest nuclei.  If
the neutron abundance is modest, this capture occurs in such a way
that each newly synthesized nucleus has the opportunity to $\beta$
decay, if it is energetically favorable to do so.  Thus weak
equilibrium is maintained within the nucleus, so that synthesis is
along the path of stable nuclei.  This is called the s- or
slow-process.  However a plot of the s-process in the (N,Z) plane
reveals that this path misses many stable, neutron-rich nuclei that
are known to exist in nature.  This suggests that another mechanism is
at work, too.  Furthermore, the abundance peaks found in nature near
masses A $\sim$ 130 and A $\sim$ 190, which mark the closed neutron
shells where neutron capture rates and $\beta$ decay rates are slower,
each split into two sub-peaks.  One set of sub-peaks corresponds to the
closed-neutron-shell numbers N $\sim$ 82 and N $\sim$ 126, and is
clearly associated with the s-process.  The other set is shifted to
smaller N, $\sim$ 76 and $\sim$ 116, respectively, and is suggestive
of a much more explosive neutron capture environment where neutron
capture can be rapid.
  
This second process is the r- or rapid-process, characterized by: \\
$\bullet$ The neutron capture is fast compared to $\beta$ decay
rates. \\
$\bullet$ The equilibrium maintained within a nucleus is established
by $(n,\gamma) \leftrightarrow (\gamma,n)$: neutron capture fills up
the available bound levels in the nucleus until this equilibrium sets
in.  The new Fermi level
depends on the temperature and the relative $n/\gamma$ abundance.\\
$\bullet$ The nucleosynthesis rate is thus controlled by the $\beta$
decay rate: each $\beta^-$ capture converting n $\rightarrow$ p opens
up a hole in the neutron Fermi sea, allowing another neutron
to be captured. \\
$\bullet$ The nucleosynthesis path is along exotic, neutron-rich
nuclei that would be highly unstable under normal laboratory
conditions. \\
$\bullet$ As the nucleosynthesis rate is controlled by the $\beta$
decay, mass will build up at nuclei where the $\beta$ decay rates are
slow.  It follows, if the neutron flux is reasonable steady over time
so that equilibrated mass flow is reached, that the resulting
abundances should be inversely proportional to these $\beta$ decay
rates.
  
Let's first explore the $(n,\gamma) \leftrightarrow (\gamma,n)$
equilibrium condition, which requires that the rate for $(n,\gamma)$
balances that for $(\gamma,n)$ for an average nucleus.  So consider
the formation cross section
\begin{equation}
 A + n \rightarrow (A+1) + \gamma . 
\end{equation}
This is an exothermic reaction, as the neutron drops into the nuclear
well.  Our averaged cross section, assuming a resonant reaction (the
level density is high in heavy nuclei) is
\begin{equation}
\langle \sigma v \rangle_{(n,\gamma)} = 
\left( {2 \pi \over \mu kT} \right)^{3/2} {\Gamma_n \Gamma_\gamma
\over \Gamma} e^{-E/KT} 
\end{equation}
where E $\sim$ 0 is the resonance energy,
and the $\Gamma$s are the indicated partial and total widths.
Thus the rate per unit volume is
\begin{equation}
r_{(n,\gamma)} \sim N_n N_A \left( {2 \pi \over \mu kT} \right)^{3/2}
{\Gamma_n \Gamma_\gamma \over \Gamma}
\end{equation}
where $N_n$ and $N_A$ are the neutron and nuclear number densities
and $\mu$ the reduced mass.
This has to be compared to the $(\gamma,n)$ rate. 

The $(\gamma,n)$ reaction requires the photon number density in
the gas.  This is given by the Bose-Einstein distribution
\begin{equation}
N(\epsilon) = {8 \pi \over c^3 h^3} {\epsilon^2 d \epsilon
\over e^{\epsilon/kT} -1} .
\end{equation}
The high-energy tail of the normalized distribution can thus
be written
\begin{equation}
 \sim {1 \over N_\gamma \pi^2} \epsilon^2 e^{-\epsilon/kT} d \epsilon 
\end{equation}
where in the last expression we have set $\hbar = c = 1$. 

Now we need the resonant cross section in the $(\gamma,n)$ direction.
For photons the wave number is proportional to the energy, so
\begin{equation}
\sigma_{(\gamma,n)} = {\pi \over \epsilon^2}
{\Gamma_\gamma \Gamma_n \over (\epsilon-E_r)^2 + (\Gamma/2)^2 } .
\end{equation}
As the velocity is c =1,
\begin{equation}
\langle \sigma v \rangle = {1 \over \pi^2 N_\gamma}
\int_0^\infty \epsilon^2 e^{-\epsilon/kT} d \epsilon
{\pi \over \epsilon^2} {\Gamma_\gamma \Gamma_n \over 
(\epsilon-E_r)^2 +(\Gamma/2)^2} . 
\end{equation}
We evaluate this in the usual way for a sharp resonance, remembering
that the energy integral over just the denominator above (the sharply
varying part) is $2 \pi/ \Gamma$
\begin{equation}
 \sim {\Gamma_\gamma \Gamma_n \over N_\gamma} e^{-E_r/kT}
{2 \over \Gamma} . 
\end{equation}
So that the rate becomes
\begin{equation}
r_{(\gamma,n)} \sim 2 N_{A+1} {\Gamma_\gamma \Gamma_n
\over \Gamma} e^{-E_r/kT} .  
\end{equation}
Equating the $(n,\gamma)$ and $(\gamma,n)$ rates and taking $N_A \sim
N_{A-1}$ then yields
\begin{equation}
N_n \sim {2 \over (\hbar c)^3} \left( {\mu c^2 kT \over
2 \pi} \right)^{3/2} e^{-E_r/kT} 
\end{equation}
where the $\hbar$s and $c$s have been properly inserted to give the
right dimensions.  Now $E_r$ is essentially the binding energy.  So
plugging in the conditions $N_n \sim 3 \times 10^{23}$/cm$^3$ and $T_9
\sim 1$, we find that the binding energy is $\sim$ 2.4 MeV.  Thus
neutrons are bound by about 30 times $kT$, a value that is still small
compared to a typical binding of 8 MeV for a normal nucleus.  (In this
calculation the neutron reduced mass is calculated by assuming a
nuclear target with A=150.)

The above calculation fails to count spin states for the photons and
nuclei and is thus not quite correct.  But it makes the essential
point: the r-process involves very exotic species largely unstudied in
any terrestrial laboratory.  It is good to bear this in mind, as in
the following section we will discuss the responses of such nuclei to
neutrinos.  Such responses thus depend on the ability of theory to
extrapolate responses from known nuclei to those quite unfamiliar.

The path of the r-process is along neutron-rich nuclei, where the
neutron Fermi sea is just $\sim$ (2-3) MeV away from the neutron drip
line (where no more bound neutron levels exist).  After the r-process
finishes (the neutron exposure ends) the nuclei decay back to the
valley of stability by $\beta$ decay.  This can involve some neutron
spallation ($\beta$-delayed neutrons) that shift the mass number A to
a lower value.  But it certainly involves conversion of neutrons into
protons, and that shifts the r-process peaks at N $\sim$ 82 and 126 to
a lower N, off course.  This effect is clearly seen in the abundance
distribution: the r-process peaks are shifted to lower N relative to
the s-process peaks.  This is the origin of the second set of
``sub-peaks" mentioned at the start of the section.

It is believed that the r-process can proceed to very heavy nuclei (A
$\sim$ 270) where it is finally ended by $\beta$-delayed and n-induced
fission, which feeds matter back into the process at an A $\sim$
A$_{max}$/2.  Thus there may be important cycling effects in the upper
half of the r-process distribution.
  
What is the site(s) of the r-process?  This has been debated 
many years and still remains a controversial subject:\\
$\bullet$ The r-process requires exceptionally explosive conditions 
\begin{center}
$\rho$(n) $\sim 10^{20}$ cm$^{-3}$~~~T $\sim 10^9$K~~~t $\sim$ 1s.
\end{center}
$\bullet$ Both primary and secondary sites proposed.  Primary sites
are those not requiring preexisting metals.  Secondary sites are those
where the neutron capture occurs
on preexisting s-process seeds.\\
$\bullet$ Suggested primary sites include the the neutronized
atmosphere above the proto-neutron star in a Type II supernova,
neutron-rich jets produced in supernova explosions or in neutron star
mergers, inhomogeneous big
bangs, etc. \\
$\bullet$ Secondary sites, where $\rho$(n) can be lower for successful
synthesis, include the He and C zones in Type II supernovae, the red
giant He flash, etc.

The balance of evidence favors a primary site, so one requiring
no pre-enrichment of heavy s-process metals.  Among the evidence: \\
  
\noindent
1) HST studies of very-metal-poor halo stars: The most important
evidence are the recent HST measurements of Cowan, Sneden et
al.~\cite{sneden} of very metal-poor stars ([Fe/H] $\sim$ -1.7 to
-3.12) where an r-process distribution very much like that of our sun
has been seen for Z $\gsim$ 56.  Furthermore, in these stars the iron
content is variable.  This suggests that the ``time resolution"
inherent in these old stars is short compared to galactic mixing times
(otherwise Fe would be more constant).  The conclusion is that the
r-process material in these stars is most likely from one or a few
local supernovae.  The fact that the distributions match the solar
r-process (at least above charge 56) strongly suggests that there is
some kind of unique site for the r-process: the solar r-process
distribution did not come from averaging over many different kinds of
r-process events.  Clearly the fact that these old stars are enriched
in r-process metals also strongly argues for a primary process: the
r-process works quite well in an
environment where there are few initial s-process metals.\\

\noindent
2) There are also fairly good theoretical arguments that a primary
r-process occurring in a core-collapse supernova might be
viable~\cite{hotbub}.  First, galactic chemical evolution studies
indicate that the growth of r-process elements in the galaxy is
consistent with low-mass Type II supernovae in rate and distribution.
More convincing is the fact that modelers have shown that the
conditions needed for an r-process (very high neutron densities,
temperatures of 1-3 billion degrees) might be realized in a supernova.
The site is the last material expelled from the supernova, the matter
just above the mass cut.  When this material is blown off the star
initially, it is a very hot neutron-rich, radiation-dominated gas
containing neutrons and protons, but an excess of the neutrons.  As it
expands off the star and cools, the material first goes through a
freeze-out to $\alpha$ particles, a step that essentially locks up all
the protons in this way.  Then the $\alpha$s interact through
reactions like
\begin{eqnarray}
 \alpha + \alpha +\alpha &&\rightarrow ^{12}C  \nonumber \\
 \alpha + \alpha + n &&\rightarrow ^9Be \nonumber
\end{eqnarray}
to start forming heavier nuclei.  Note, unlike the big bang,
that the density is high enough to allow such three-body 
interactions to bridge the mass gaps at A = 5,8.  The
$\alpha$ capture continues up to heavy nuclei,
to A $\sim$ 80, in the network calculations.  
The result is a small number of ``seed" nuclei,
a large number of $\alpha$s, and excess neutrons.  These 
neutrons preferentially capture on the heavy seeds to
produce an r-process.  Of course, what is necessary is to
have $\sim$ 100 excess neutrons per seed in order to 
successfully synthesize heavy mass nuclei.  Some of the
modelers find conditions where this almost happens. 
  
There are some very nice aspects of this site: the amount of matter
ejected is about 10$^{-5} - 10^{-6}$ solar masses, which is just about
what is needed over the lifetime of the galaxy to give the integrated
r-process metals we see, taking a reasonable supernova rate.  But
there are also a few problems, especially the fact that with
calculated entropies in the nucleon soup above the proto-neutron star,
neutron fractions appear to be too low to produce a successful A
$\sim$ 190 peak.  There is some interesting recent work invoking
neutrino oscillations to cure this problem: charge current reactions
on free protons and neutrons determine the n/p ratio in the gas.
Then, for example, an oscillation of the type $\nu_e \rightarrow
\nu_{\mathrm{sterile}}$ can alter this ratio, as it would turn off the
$\nu_e$s that destroy neutrons by charged-current
reactions\cite{gail1}.  Unfortunately, a full discussion of such
possibilities would take us too far afield today.

The nuclear physics of the r-process tells us that the synthesis
occurs when the nucleon soup is in the temperature range of (3-1)
$\cdot 10^9$K, which, in the hot bubble r-process described above,
corresponds to a freeze-out radius of (600-100) km and a time $\sim$ 10
seconds after core collapse.  The neutrino fluence after freeze-out
(when the temperature has dropped below 10$^9$K and the r-process
stops) is then $\sim$ (0.045-0.015) $\cdot 10^{51}$ ergs/(100km).
Thus, after completion of the r-process, the newly synthesized
material experiences an intense flux of neutrinos.  This brings up the
question of whether the neutrino flux could have any effect on the
r-process.

\subsection{Neutrinos and the r-process~\protect\cite{qian}}

Rather than describe the exotic effects of neutrino oscillations on
the r-process, mentioned briefly above, we will examine standard-model
effects that are nevertheless quite interesting.  The nuclear physics
of this section -- neutrino-induced neutron spallation reactions -- is
also relevant to recently proposed supernova neutrino observatories
such as OMNIS and LAND.  In contrast to our first discussion of the
$\nu$-process in Sec. 4.2, it is apparent that neutrino effects could
be much larger in the hot bubble r-process: the synthesis occurs {\it
  much} closer to the star than our Ne radius of 20,000 km: estimates
are 600-1000 km.  The r-process is completed in about 10 seconds (when
the temperature drops to about one billion degrees), but the neutrino
flux is still significant as the r-process freezes out.  The net
result is that the ``post-processing" neutrino fluence - the fluence
that can alter the nuclear distribution after the r-process is
completed - is about 100 times larger than that responsible for
fluorine production in the Ne zone.  Recalling that 1/300 of the
nuclei in the Ne zone interacted with neutrinos, and remembering that
the relevant neutrino-nucleus cross sections scale as A, one quickly
sees that the probability of a r-process nucleus interacting with the
neutrino flux is approximately unity.

Because the hydrodynamic conditions of the r-process are highly
uncertain, one way to attack this problem is to work backward in time.
We know the final r-process distribution (what nature gives us) and we
can calculate neutrino-nucleus interactions relatively well.  Thus
from the observed r-process distribution (including neutrino
post-processing) we can work backward to find out what the r-process
distribution looked like at the point of freeze-out.  In Figs. 16 and
17, the ``real" r-process distribution - that produced at freeze-out -
is given by the dashed lines, while the solid lines show the effects
of the neutrino post-processing for a particular choice of fluence.
The nuclear physics input into these calculations is precisely that
previously described: GT and first-forbidden cross sections, with the
responses centered at excitation energies consistent with those found
in ordinary, stable nuclei, taking into account the observed
dependence on $|N-Z|$.

\begin{figure}[htb]
\psfig{bbllx=-2.0cm,bblly=4.5cm,bburx=18cm,bbury=23.0cm,figure=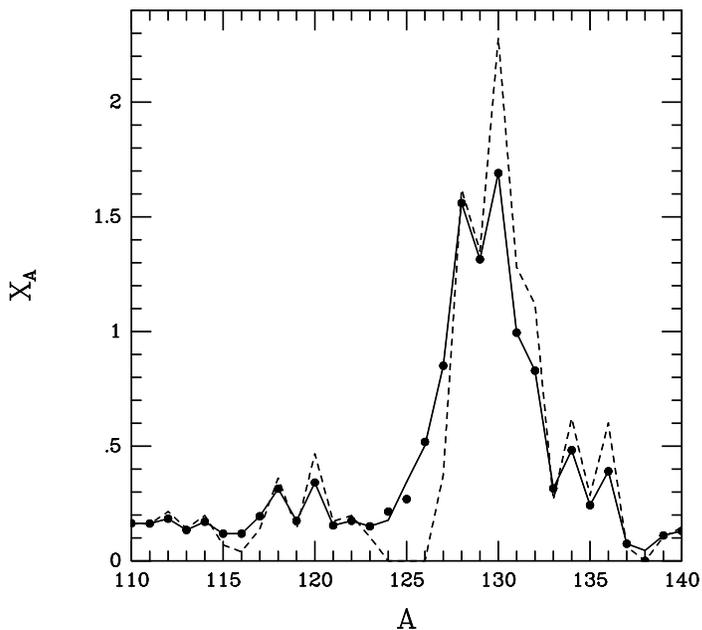,height=3.5in}
\caption{Comparison of the r-process distribution that would 
  result from the freeze-out abundances near the A $\sim$ 130 mass peak
  (dashed line) to that where the effects of neutrino post-processing
  have been include (solid line).  The fluence has been fixed by
  assuming that the A = 124-126 abundances are entirely due to the
  $\nu$-process.}
\end{figure}
  
One important aspect of the figures is that the mass shift is
significant.  This has to do with the fact that a 20 MeV excitation of
a neutron-rich nucleus allows multiple neutrons ( $\sim$ 5) to be
emitted.  (Remember we found that the binding energy of the last
neutron in an r-process neutron-rich nuclei was about 2-3 MeV under
typical r-process conditions.)  The second thing to notice is that the
relative contribution of the neutrino process is particularly
important in the ``valleys" beneath the mass peaks: the reason is that
the parents on the mass peak are abundant, and the valley daughters
rare.  In fact, it follows from this that the neutrino process effects
can be dominant for precisely seven isotopes (Te, Re, etc.) lying in
these valleys.  Furthermore if an appropriate neutrino fluence is
picked, these isotope abundances are produced perfectly (given the
abundance errors).  The fluences are
\begin{eqnarray}
     \mathrm{N} &=& 82~ \mathrm{peak}~~~~~0.031 \cdot 10^{51} 
\mathrm{ergs/(100km)^2/flavor} \nonumber \\
     \mathrm{N} &=& 126~ \mathrm{peak}~~~~0.015 \cdot 10^{51} 
\mathrm{ergs/(100km)^2/flavor}, \nonumber
\end{eqnarray}
values in fine agreement with those that would be found in a hot
bubble r-process.  So this is circumstantial but significant evidence
that the material near the mass cut of a Type II supernova is the site
of the r-process: there is a neutrino fingerprint.

\begin{figure}[htb]
\psfig{bbllx=-2.0cm,bblly=4.5cm,bburx=18cm,bbury=23.0cm,figure=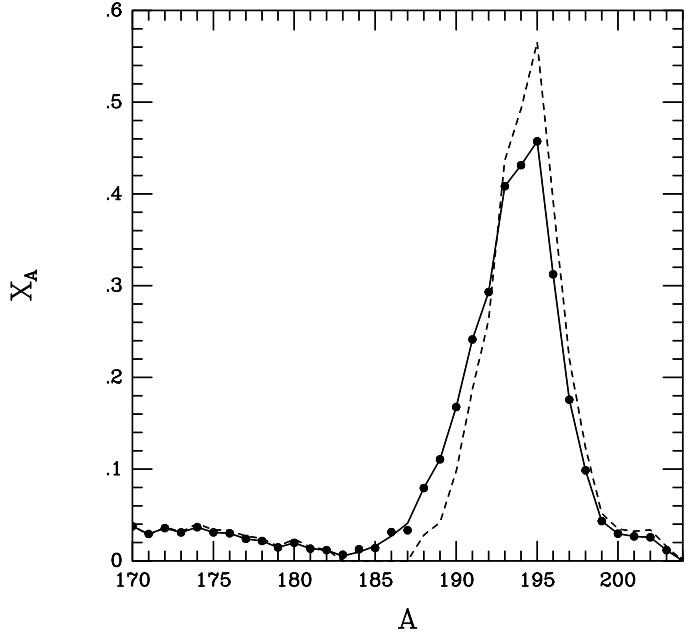,height=3.5in}
\caption{As in Fig. 16, but for the A $\sim$ 195 mass peak.
The A = 183-187 abundances are entirely attributed to the 
$\nu$-process.}
\end{figure}

In the following sections we will briefly return to the effects of
supernova density and electron fraction fluctuations and sterile
neutrino mixings on the supernova r-process. 

\section{MSW Exotica}

\subsection{Spin-Flavor Oscillations}

In section 2.9 we briefly touched upon the effects of neutrino
magnetic moment on neutrino propagation through matter. The chiral
components of neutrino propagating in matter satisfy the
equation \cite{cisneros}
\begin{equation}
  \label{sp1}
i\frac{\partial}{\partial t} \left[\begin{array}{cc} \nu_L \\ \\ \nu_R
  \end{array}\right] =
\left[\begin{array}{cc} \frac{G_F}{\sqrt{2}} (2N_e-N_n) & \mu B \\ \\
\mu B & 0
\end{array}\right]
\left[\begin{array}{cc} \nu_L \\ \\ \nu_R \end{array}\right]\,,
\end{equation}
where $N_e$ and $N_n$ are the electron and neutron densities in the
media, respectively. As Eq. (\ref{sp1}) demonstrates the chiral
rotation is actually suppressed in matter. It was soon
realized \cite{lam}, however, that combining the chiral precession and
the MSW flavor transformation by including non-diagonal magnetic
moments circumvents this problem. One then
should consider the evolution of left- and right-handed components of
the electron and muon neutrinos together:
\begin{equation}
  \label{sp2}
  i \frac{\partial}{\partial t} \left[\begin{array}{c} \nu_e^L \\ 
\nu_{\mu}^L \\ \nu_e^R \\ \nu{_\mu}^R \end{array}\right] = 
\left[\begin{array}{cc}  H_L & BM^{\dagger} \\ \\ BM & H_R
  \end{array}\right] 
\left[\begin{array}{c} \nu_e^L \\ 
\nu_{\mu}^L \\ \nu_e^R \\ \nu{_\mu}^R \end{array}\right].
\end{equation}
For Dirac neutrinos the block matrices appearing in the evolution 
Hamiltonian in Eq. (\ref{sp2}) are given by
\begin{equation}
  \label{sp3}
  H_L = \left[\begin{array}{cc}  \frac{\delta m^2}{2E}  \sin^2{\theta_v}
    + \frac{G_F}{\sqrt{2}} (2N_e-N_n) & \frac{\delta m^2}{4E}
    \sin{2\theta_v} \\ \\ \frac{\delta m^2}{4E}
    \sin{2\theta_v} & \frac{\delta m^2}{2E}  \sin^2{\theta_v}
    - \frac{G_F}{\sqrt{2}} N_n \end{array}\right], 
\end{equation}
\begin{equation}
  \label{sp4}
  H_R = \left[\begin{array}{cc}  \frac{\delta m^2}{2E}  \sin^2{\theta_v}
    & \frac{\delta m^2}{4E}
    \sin{2\theta_v} \\ \\ \frac{\delta m^2}{4E}
    \sin{2\theta_v} & \frac{\delta m^2}{2E}  \sin^2{\theta_v}
     \end{array}\right] ,
\end{equation}
and
\begin{equation}
  \label{sp5}
  M =  \left[\begin{array}{cc} \mu_{ee} & \mu_{e\mu} \\ \\
 \mu_{\mu e} & \mu_{\mu \mu} \end{array}\right]. 
\end{equation}
In these equations $B$ is the component of the magnetic field
transverse to the direction of neutrino
propagation. The last equation contains both diagonal and transition
magnetic moments for neutrinos. For Majorana neutrinos diagonal
moments vanish and, since neutrinos are their own antiparticles, both 
left-handed and right-handed neutrinos must mix  with the same mixing
matrix.  Hence for Majorana neutrinos the evolution matrix takes the
form
\begin{equation}
  \label{sp6}
  \left[\begin{array}{cccc} \frac{G_F}{\sqrt{2}} (2N_e-N_n) & 
\frac{\delta m^2}{4E}  \sin{2\theta_v} & 0 & \mu^* B \\ \\ \\ \\ 
\frac{\delta m^2}{4E}  \sin{2\theta_v} & \frac{\delta m^2}{2E}  
\cos{2\theta_v}  - \frac{G_F}{\sqrt{2}} N_n & - \mu^* B & 0 \\ \\ \\
\\ 0 & - \mu B & - \frac{G_F}{\sqrt{2}} (2N_e-N_n) & \frac{\delta
  m^2}{4E}  \sin{2\theta_v} \\ \\ \\ \\ 
\mu B & 0 & \frac{\delta m^2}{4E}  \sin{2\theta_v} & 
\frac{\delta m^2}{2E}  
\cos{2\theta_v}  + \frac{G_F}{\sqrt{2}} N_n \end{array}\right] .
\end{equation}
These evolution equations were numerically investigated in detail 
both for the
Sun \cite{baha1} and supernovae \cite{qiannunu}. This was motivated by
the Homestake data which suggested an anti-correlation between sunspot 
number and the capture rate of solar neutrinos. It is 
rather difficult to understand such a behavior in a theoretical
framework since the most straightforward explanation for short-term 
time-variation of the neutrino flux is to assume the existence of a 
rather large neutrino magnetic moment, inconsistent with the bounds
discussed in Section 3.3. Also Kamiokande and SuperKamiokande solar
neutrino data do not show such time-variations. 

\subsection{Neutrino Propagation in Stochastic Media}

In implementing the MSW solution to the solar neutrino problem one
typically assumes that the electron density of the Sun is a
monotonically decreasing function of the distance from the core and
ignores potentially de-cohering effects \cite{sawyer}. To understand
such effects one possibility is to study  
parametric changes in the density or the role of matter currents 
\cite{othernoise}. In this regard, Loreti and Balantekin 
\cite{orignoise} considered neutrino propagation in stochastic
media. They studied the situation where the electron density in the
medium has two components, one average component given by the Standard
Solar Model or Supernova Model, etc. and one fluctuating
component. Then the Hamiltonian in Eq. (53) takes the form 
\begin{equation}
\hat H =
\left({{-\delta m^2}\over 4E} \cos 2\theta + {1\over \sqrt{2}}
G_F(N_e(r) + N^r_e(r))\right){\sigma_z} + \left({{\delta m^2}\over 4E}
\sin 2\theta \right) {\sigma_x}.  
\end{equation}
where one imposes for
consistency 
\begin{equation}
\langle N^r_e(r)\rangle = 0, 
\end{equation}
 and a two-body
correlation function 
\begin{equation}
\langle N^r_e(r)N^r_e(r^{\prime}) \rangle =
{\beta}^2 \ N_e(r) \ N_e(r^{\prime}) \ \exp(-|r-r^{\prime}|/\tau_c).
\end{equation}
\begin{figure}[t]
 \centerline{\rotate[r]{\epsfxsize=3in
\epsfbox{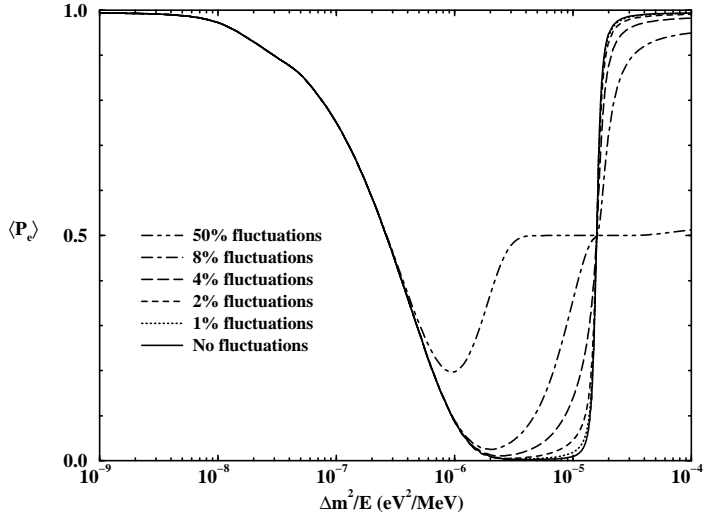}}} 
\caption{
Mean electron neutrino survival probability in the sun with
fluctuations. The average electron density is given by the Standard
Solar Model of Bahcall and Pinsonneault and $\sin^2 2
\theta=0.01$.} 
\label{Fig3}
\end{figure}
In the calculations of the Wisconsin group the fluctuations are
typically taken to be subject to colored noise, i.e. higher order
correlations 
\begin{equation}
f_{12 \cdots }=\langle N^r_e(r_1)N^r_e(r_2) \cdots
\rangle
\end{equation}
are taken to be 
\begin{equation}
 f_{1234}= f_{12}f_{34} + f_{13}f_{24} +
f_{14}f_{23},
\end{equation}
 and so on.

\begin{figure}[htb]
\vspace{8pt} \centerline{\rotate[r]{\epsfxsize=3in
\epsfbox{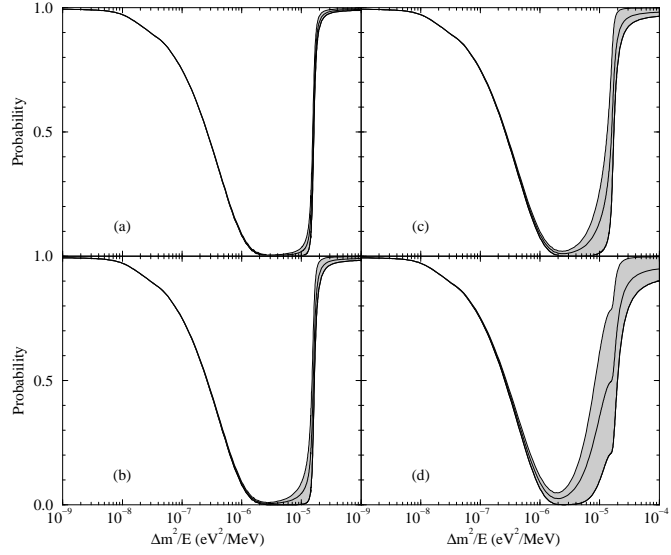}}} 
\caption{ Mean electron neutrino survival probability plus minus
$\sigma$ in the sun with fluctuations. The average electron density is
given by the Standard Solar Model of Bahcall and Pinsonneault and  
$\sin^2 2 \theta=0.01$. Panels (a), (b), (c), and (d) correspond to an
average fluctuation of 1\%, 2\%, 4\%, and 8\% respectively. 
}
\label{Fig4}
\vspace{18pt}
\end{figure}

The mean survival probability \cite{newnoise}  for the electron
neutrino in the Sun is shown in Figure \ref{Fig3} where fluctuations
are imposed on the average solar electron density given by the
Bahcall-Pinsonneault model.  One notes that for very large
fluctuations complete flavor de-polarization should be achieved,
i.e. the neutrino survival probability is 0.5, the same as the vacuum
oscillation probability for long distances. To illustrate this
behavior the results from the physically unrealistic case of 50\%
fluctuations are shown. Also the effect of the fluctuations is largest
when the neutrino propagation in their absence is adiabatic. This
scenario was applied to the neutrino convection in a core-collapse
supernova where the adiabaticity condition is satisfied
\cite{supernoise}.  Similar results were also obtained by other
authors \cite{nunokawa}. 
It may be possible to test solar
matter density fluctuations at the BOREXINO detector currently under
construction \cite{borex}. Propagation of a
neutrino with a magnetic moment in a random magnetic moment has also
been investigated \cite{orignoise,ranmagnetic}. Also if
the magnetic field in a polarized medium has a domain structure with
different strength and direction in different domains, the
modification of the potential felt by the neutrinos due polarized
electrons will have a random character \cite{polarized}.  Using the
formalism sketched above, it is possible to calculate not only the
mean survival probability, but also the variance, $\sigma$,  of the
fluctuations to get a feeling for the distribution of the survival
probabilities \cite{newnoise} as illustrated in Figure \ref{Fig4}.
In these calculations the correlation length $\tau$ is taken to be
very small, of the order of 10 km., to be consistent with the
helioseismic observations of the sound speed (cf. Section 2.5). In the
opposite limit of very large correlation lengths are very interesting
result is obtained \cite{supernoise}, namely the averaged density
matrix is given as an integral 
\begin{equation}
\lim_{\tau_c\to\infty}\langle\hat \rho(r)\rangle =
{1\over{\sqrt{2\pi \beta^2}}} \int_{-\infty}^{\infty} dx
\exp[{-x^2/(2\beta^2)}]
\hat \rho(r,x),
\end{equation}
reminiscent of the channel-coupling problem in nuclear
physics \cite{takigawa}. Even though this limit is not appropriate to
the solar fluctuations it may be applicable to a number of other
astrophysical situations. 

\subsection{Approximate Solutions and Neutrino Tomography}

\begin{figure}[t]
\centerline{\hbox{\epsfxsize=2.5in \epsfbox[66 56 504 
644]{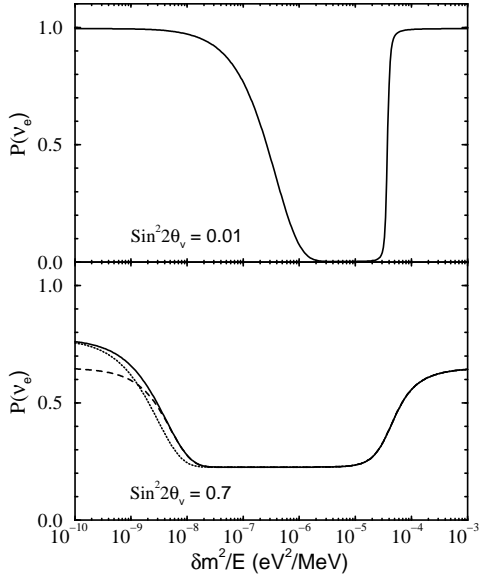}}}
\caption{The electron neutrino survival probability for the Sun. 
 The solid line is calculated using Eq. ~(\ref{a2}).  The
dashed line is the exact (numerical) result.  The dotted line is the
linear Landau-Zener result.  In the top figure, the lines are
indistinguishable. An exponential density with parameters chosen to
approximate the Sun was used.}
\label{Fig1}
\vspace{18pt}
\end{figure}

Exact solutions for the neutrino propagation equations in matter exist
only for a limited class of density profiles. For example, for those
density profiles  that satisfy an integrability condition called shape
invariance neutrino survival probability can be found with algebraic
methods \cite{bahashape}. One can also use various approximation
methods to solve Eq. (53). The standard
semi-classical approximation gives the adiabatic
evolution \cite{baha4}. For a monotonically changing density profile
supersymmetric uniform approximation yields \cite{baha5} the hopping
probability of Eq. (65) to be 
\begin{eqnarray}
P_{hop} &=& \exp (- \pi \Omega ), \nonumber \\
\Omega &=& \frac{i}{\pi} \frac{\delta m^2}{2 E}
\int^{r_0^*}_{r_0} dr
\left[\zeta^2(r) - 2\zeta(r)\cos{2\theta_v} + 1\right]^{1/2}\,,
\label{a2}
\end{eqnarray}
where ${r_0^*}$ and ${r_0}$ are the turning points (zeros) of the
integrand. In this expression we introduced the scaled density 
\begin{equation}
\zeta(r) = \frac{2\sqrt{2} G_F N_e(r)}{\delta m^2/E}\,,
\label{a3}
\end{equation}
where $N_e$ is the number density of electrons in the medium.  By
analytic continuation, this complex integral is primarily sensitive to
densities near the resonance point. The validity of this approximate
expression is illustrated in Figure \ref{Fig1}.
As this figure illustrates the approximation breaks down in the
extreme non-adiabatic limit (i.e., as $\delta m^2 \rightarrow
0$). Hence it is referred to as the quasi-adiabatic approximation. 

The near-exponential form of the density profile in the Sun 
motivates an expansion of the electron number
density scale height, $r_s$, in powers of density:
\begin{equation}
 -r_s \equiv \frac{N_e(r)}{N_e'(r)} = \sum_n b_n N_e^n, 
 \label{a4}
\end{equation}
where prime denotes derivative with respect to $r$. In this expression
a minus sign is introduced because we assumed that density profile
decreases a $r$ increases. (For an exponential
density profile, $ N_e \sim e^{-\alpha x}$, only the $n=0$ term is 
present).
\begin{figure}[t]
\centerline{\hbox{\epsfxsize=2.5 in \epsfbox[17 64 539 
685]{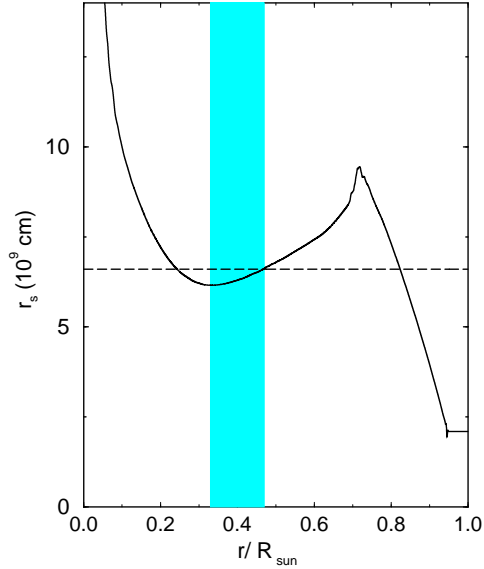}}}
\caption 
{Electron number density scale height (cf. Eq. (\ref{a4})) 
as a function of the radius for the Sun. The dashed line is the
exponential fit over the whole Sun. The shaded are indicates where the
small angle MSW resonance takes place for neutrinos with energies $5 <
E < 15$ MeV.}
\label{Fig1a}
\vspace{18pt}
\end{figure}
To help assess the appropriateness of such an expansion 
the density scale height for the Sun calculated using the Standard
Solar Model density profile is plotted in Figure
\ref{Fig1a}. One observes that there is a significant deviation from a
simple exponential profile over the entire Sun. However the expansion
of Eq.~(\ref{a4}) needs to hold only in the MSW resonance region,
indicated by the shaded area in the figure. Real-time counting
detectors such as SuperKamiokande and Sudbury Neutrino Observatory,
which can get information about energy spectra, are sensitive to
neutrinos with energies greater than about 5 MeV (cf. Section
2.8). For the small angle solution ($\sin 2 \theta \sim 0.01$ and 
$\delta m^2 = 5 \times 10^{-6}$ eV$^2$), the resonance for a 5 MeV
neutrino occurs at about 0.35 
R$_\odot$ and for a 15 MeV neutrino at about 0.45 R$_\odot$ (the shaded
area in the figure). In that region the density profile is 
approximately exponential and one expects that it should be sufficient
to keep only a few terms in the expansion
in Eq.~(\ref{a4}) to represent the density profile of the Standard
Solar Model. 

Inserting the expansion of Eq.~(\ref{a4}) into 
Eq.~(\ref{a2}), and using an integral representation of the Legendre 
functions, one obtains \cite{baha6} 
\begin{eqnarray}
\label{a5}
\Omega &=& -\frac{\delta m^2}{2 E} \left\{ b_0 (1 - \cos{2\theta_v})
\frac{}{} \right. \nonumber \\  &+&   \sum^{\infty}_{n=1}
\left( \frac{ \delta m^2}{2 \sqrt{2} G_F E} \right)^n  \frac{b_n}{2n +
1}  \left. \left[P_{n-1}(\cos{2\theta_v}) -
P_{n+1}(\cos{2\theta_v})\right]\right\}\,, 
\end{eqnarray}
where $P_n$ is the Legendre polynomial of order n.  The $n=0$ term in
Eq.~(\ref{a5}) represents the contribution of the exponential density
profile alone. 
Eq.~(\ref{a5})  directly connects an
expansion of the logarithm of the hopping probability in powers of
$1/E$ to an expansion of the density scale height.  That is, 
 it provides a direct connection between
$N_e(r)$ and $P_\nu(E_\nu)$.  Eq.~(\ref{a5}) provides a
quick and accurate alternative to numerical integration of the MSW
equation for any monotonically-changing 
density profile for a wide range of mixing parameters. 
\begin{figure}[t]
\centerline{\hbox{\epsfxsize=2.5 in \epsfbox[21 36 566 711]
{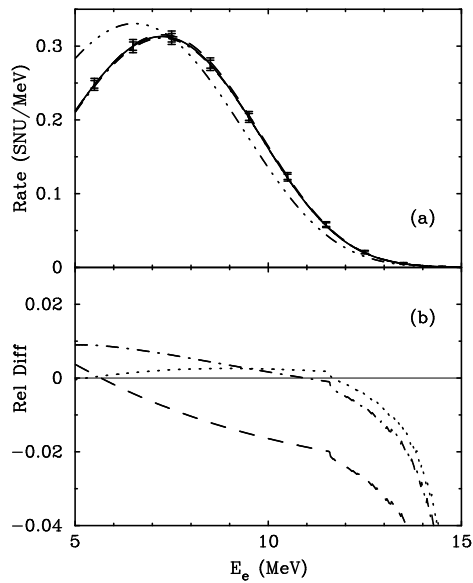}}}
\caption{(a) Spectrum distortion at SNO for the small-angle MSW 
solution ($\delta m^2 \sim 5 \times 10^{-6}$ eV$^2$ and $\sin 2 \theta
\sim 0.01$).  The solid line
is the exact numerical solution.  The dashed, dot-dashed, and dotted
lines result from values of $n$ up to 0, 1, and 2 in Eq.~(\ref{a5}). 
The error bars on the exact numerical result correspond to two and
five years of data collection.  The dot-dot-dot-dashed line is the
spectrum without MSW oscillations, normalized to the same total rate as
with MSW oscillations. Note that on the scale of this figure the $n=1$
and $n=2$ lines are not distinguishable from the exact answer. 
(b) The relative error arising from the use of Eq.~(\ref{a5}).}
\label{Fig2}
\vspace{18pt}
\end{figure}
The accuracy of the expansion of Eq.~(\ref{a5}) is illustrated in
Figure \ref{Fig2} where the spectrum distortion for the small angle
MSW solution is plotted. One observes that for the Sun, where the
density profile is nearly exponential in the MSW resonance region, the
first two terms in the expansion provide an excellent approximation to
the neutrino survival probability. 

\subsection{Oscillations into Sterile Neutrinos}

In describing the MSW mechanism in Section 2.7 we assumed that the
neutrino that mixes with the electron neutrino is an active neutrino
(taken to be the muon neutrino for definiteness). We define an active
neutrino to be one that couples to the Z, i.e. $\nu_e$, $\nu_{\mu}$,
or $\nu_{\tau}$. Data from the solar and atmospheric neutrino
experiments, taken together with the intriguing data from the LSND 
experiment (cf. Section 8) may be viewed as an indication of a fourth
neutrino flavor. Since the Z decay gives the number of active flavors
to be three, if there is fourth flavor, it must be a sterile
neutrino. (We define a sterile neutrino to be one that
does not couple directly to Z and W, but may mix with those neutrinos
that do). Many theories of neutrino mass predict the existence of
sterile neutrinos. These are, however, generally rather heavy. It is
not easy to build a
model with sterile neutrinos light enough to mix with the active
ones. A sterile neutrino may have very interesting implications for
the supernova r-process as well \cite{gail1}.

MSW evolution equations describing the mixing of active and sterile
neutrinos take a different form than Eq. (53), In writing down
Eq. (53) we used the fact that both active neutrinos acquire similar
contributions to their effective masses from the neutral current
interactions on neutrons. Such a term then only contributes to an
overall phase which does not effect neutrino evolution. Since a
sterile neutrino has neither charged- nor neutral-current interactions
with ordinary hadrons we need to keep the neutron-term for the active
flavor. Hence the MSW evolution equation takes the form
\begin{equation}
i\hbar \frac{\partial}{\partial r} \left[\begin{array}{cc} \Psi_e(r)
\\ \\ \Psi_s(r) \end{array}\right] = \left[\begin{array}{cc}
\varphi_e(r) & \frac{\delta m^2}{4 E}\sin{2\theta_v} \\ \\ 
\frac{\delta m^2}{4 E}\sin{2\theta_v} & -\varphi_e(r)
\end{array}\right]
\left[\begin{array}{cc} \Psi_e(r) \\ \\ \Psi_s(r)
  \end{array}\right]\,,
\label{1}
\end{equation}
where
\begin{equation}
  \label{2} \varphi_e(r) = \frac{1}{4 E} \left( \pm
 2 \sqrt{2}\ G_F \left[
  N_e^-(r) - N_e^+(r) - \frac{N_n(r)}{2} \right] E - \delta m^2
  \cos{2\theta_v} \right)
\end{equation}
for the mixing of electron neutrinos (the plus sign on the right-hand
side of the equation) or electron antineutrinos (the minus sign) with
sterile neutrinos. Since matter-enhanced mixing of sterile and active
neutrinos is typically used for supernova environments in
Eq. (\ref{2}) we included the positron ($ N_e^+(r)$) as well as the
electron($ N_e^-(r)$) backgrounds. Note that in what follows, we take
the sterile neutrino to be predominantly the heavier mass eigenstate. 

The mixing of muon and tau neutrinos with sterile neutrinos may be
described similarly.  The evolution Hamiltonian is as for the electron
neutrino species, but with $\varphi_{\mu}$ or $\varphi_{\tau}$
replacing $\varphi_e$ in Eq.~(\ref{1}) as appropriate, where
\begin{equation}
  \label{2a} \varphi_{\mu,\tau}(r) = - \frac{1}{4 E} \left( \pm
  \sqrt{2}\ G_F N_n(r) E + \delta m^2 \cos{2\theta_v}
  \right).
\end{equation}
As before, the $+$ sign corresponds to neutrino mixing, and the $-$
sign to antineutrino mixing.

For astrophysical applications it is useful to express these equations
in terms of the electron fraction. For a neutral medium we have $Y_p =
Y_e$ and $Y_n = 1 - Y_e$, where
$Y_p$ and $Y_n$ give the number of {\it all} protons or neutrons (free
as well as those bound in nuclei), respectively, relative to baryons.
The electron fraction $Y_e$ is given by
\begin{equation}
  \label{4}
Y_e (r)= \frac{N_e^-(r)-N_e^+(r) }{N_e^-(r)-N_e^+(r)+N_n(r)}.
\end{equation}
Inserting Eq.~(\ref{4}) into Eq.~(\ref{2}) one obtains the diagonal
terms in the evolution operator to be
\begin{equation}
  \label{5}
\varphi_e(r) = \pm \frac{3 G_F \rho (r)}{2 \sqrt{2}m_N} \left( Y_e -
  \frac{1}{3} \right) - \frac{\delta m^2}{4E} \cos{2\theta_v},
\end{equation}
and
\begin{equation}
  \label{5a}
\varphi_{\mu,\tau}(r) = \pm \frac{G_F \rho (r)}{2 \sqrt{2}m_N} \left(
Y_e - 1 \right) - \frac{\delta m^2}{4E} \cos{2\theta_v},
\end{equation}
where $\rho(r)$ is the matter density and $m_N$ is the nucleon mass.

One should also point out that for the standard active-active MSW 
evolution only either neutrinos or antineutrinos (depending the sign
of $\delta m^2$) undergo a resonance. The situation for active-sterile
mixing is different. Eq.~(\ref{5}) indicates that, with appropriate
neutrino parameters (i.e. $\delta m^2 >0$) and 
matter density, for $Y_e > 1/3$ electron neutrinos and for $Y_e <
1/3$ electron antineutrinos can undergo an active-sterile MSW
resonance. The possibility of matter-enhanced conversion of both
$\nu_e$'s and $\bar{\nu}_e$'s can have interesting consequences in an
astrophysical environment \cite{gail1}. 

In the presence of neutrino fluxes (\lq\lq background\rq\rq\ 
neutrinos) the neutrino amplitude evolution Hamilton and the effective
mass in Eq.~(\ref{1}) will have an additional term due to
neutrino-neutrino neutral current forward exchange scattering. In the
case of active-active neutrino evolution, the neutrino background,
because of flavor mixing, contributes to both diagonal and
off-diagonal terms \cite{pantale} in the flavor basis amplitude
evolution Hamiltonian, Eq.~(\ref{1}). However, for active-sterile
mixing the off-diagonal terms are identically zero \cite{sigl}.

\section{Neutrino Oscillation Constraints from the r-process}

As we discussed in Section 5 understanding neutrino transport in a
supernova is an essential part of understanding supernova dynamics. 
Neutrino transport in a medium like supernova is a complicated 
process which needs to be treated numerically taking into account 
many different pieces of physics. 
In a core-collapse  driven supernova, the inner core
collapses subsonically, but the outer part of  the core
supersonically. At some point during the collapse, when the nuclear
equation of state stiffens, the inner part of the core bounces, but
the outer  core continues falling in. The shock wave generated at the
boundary loses its  energy as it expands by dissociating material
falling through it into free  nucleons and alpha particles. For a
large initial core mass, the shock wave  gets stalled at $\sim$ 200 to
500 km away from the center of the proto-neutron  star
\cite{mayle}. Meanwhile, the proto-neutron star, shrinking under its
own  gravity, loses energy by emitting neutrinos, which only interact
weakly and  can leak out on a relatively long diffusion time
scale. One question to be investigated then is  the possibility of
regenerating the shock by neutrino heating.
The  density at the neutrinosphere is $\sim 10^{12}$g cm$^{-3}$
and the density at the position of the stalled shock is \cite{mayle}
$\sim 2 \times 10^7$ g cm$^{-3}$. Writing the MSW resonance density in
appropriate units:
\begin{equation}
\rho_{\rm res} = 1.31 \times 10^7 \left( {\delta m^2 \over {\rm eV}^2}
\right)  \left( { {\rm MeV} \over E_{\nu} } \right) \left( {0.5 \over
{\rm Y}_e}  \right) {\rm g} \> {\rm cm}^{-3},
\end{equation}
one sees that, for small mixing angles, $E_{\nu} \sim 10$ MeV, and 
cosmologically interesting $\delta m^2 \sim 1 - 10^4$ eV$^2$, there
could be an MSW resonance point between the neutrinosphere and the
stalled shock. 

Most neutrinos emitted from the core are produced by a neutral current
process, and so the luminosities are approximately the same for all
flavors.  The energy spectra are approximately Fermi-Dirac with a zero
chemical potential characterized by a neutrinosphere temperature. The
$\nu_{\tau}, {\overline \nu}_{\tau}, \nu_{\mu}, {\overline \nu}_{\mu}$
interact with matter only via neutral current interactions. These
decouple at relatively small radius and end up with somewhat high
temperatures, about 8 MeV. The ${\overline \nu}_e$'s decouple at a
larger radius because of the additional charged current interactions
with the protons, and consequently have a somewhat lower temperature,
about 5 MeV. Finally, since they undergo charged current interactions
with more abundant neutrons, $\nu_e$'s decouple at the largest radius
and end up with the lowest temperature, about 3.5 to 4 MeV.  An MSW
resonance between the neutrinosphere and the stalled shock can then
transform $\nu_{\tau} \leftrightarrow \nu_e$, cooling $\nu_{\tau}$'s,
but heating $\nu_e$'s. Since the interaction cross section of electron
neutrinos with the matter in the stalled shock increases with
increasing energy, it may be possible to regenerate the shock. Fuller
{\it et al.} \cite{mayle} found that for small mixing angles between
$\nu_{\tau}$ and $\nu_e$ one can get a 60\% increase in the explosion
energy. 

There is another implication of the $\nu_{\tau}$ and $\nu_e$ mixing in
the supernovae. Supernovae are possible r-process sites (cf. Section
5.4), which requires a neutron-rich environment, i.e., the ratio of
electrons to baryons, $Y_e$, should be less than one half. $Y_e$ in
the nucleosynthesis region is given approximately \cite{fuller} by 
\begin{equation}
Y_e \simeq {1 \over 1+ \lambda_{{\overline \nu}_e p} / \lambda_{ \nu_e
n}}  \simeq {1 \over 1 + T_{{\overline \nu}_e} / T_{ \nu_e}}, 
\end{equation}
where $\lambda_{ \nu_e n}$, etc. are the capture rates. Hence if
$T_{{\overline \nu}_e} > T_{\nu_e}$, then the medium is
neutron-rich. As we discussed above, without matter-enhanced neutrino
oscillations, the neutrino temperatures satisfy the inequality $T_{
\nu_{\tau}} >T_{{\overline \nu}_e} > T_{ \nu_e}$. But the MSW effect,
by heating $\nu_e$ and cooling $\nu_{\tau}$ can reverse the direction
of inequality, making the medium proton-rich instead. Hence the
existence of neutrino mass and mixings puts severe constraints on
heavy-element nucleosynthesis in supernova. In turn requiring
supernova to be an r-process site implies constraints on the neutrino
parameters.  These constraints are investigated in Ref. 54. 

If the supernova is an r-process site it is also desirable to have
neutron to seed nucleus ratio ${\ 
  \lower-1.2pt\vbox{\hbox{\rlap{$>$}\lower5pt\vbox{\hbox{$\sim$}}}}\ }
100$ in order that the heavier $r$-process species ({\it i.e.}, those
in the $A=195$ peak) can be produced. This ratio is basically
determined by three quantities: i) the expansion
rate; ii) the electron fraction $Y_e$; and iii) the entropy per
baryon. Though different calculations \cite{hotbub,r1} disagree on the
value of the entropy in the neutrino-driven wind during the r-process
nucleosynthesis, several models can produce values of these three
parameters that yield a high enough neutron-to-seed nucleus ratio at
freeze-out to effect a reasonable $r$-process. Unfortunately there are
neutrino-induced processes operating during or immediately after
freeze-out which can work to greatly reduce the neutron-to-seed
nucleus ratio to the point where acceptable $r$-process
nucleosynthesis in this site would be impossible.  These
neutrino-induced $r$-process destroyers are: i) neutrino neutral
current spallation of alpha particles; and ii) the $\nu_e+n\rightarrow
p+e^-$ reaction accompanying the formation of alpha particles, also
known as the \lq\lq alpha effect.\rq\rq\ 

Meyer pointed out that previously neglected neutrino spallation
reactions on the alpha particles tend to inhibit the $r$-process by
allowing the assembly of too many seed nuclei \cite{meyeralpha}. This
process is especially effective at wrecking the $r$-process where the
entropy is high. A simple steady-state wind model survey of the
thermodynamic parameters in neutrino-heated outflow was conducted by
Qian and Woosley \cite{qw}. These authors concluded that the entropy
in such models should be $\sim100 k$ per baryon, as opposed to Mayle
and Wilson's model with an entropy of $\sim 400 k$ per baryon. 
In turn, this result might argue against the
effectiveness of neutrino-induced alpha particle spallation in
lowering the neutron to seed nucleus ratio. However, lower entropies
in general imply a lower value of this ratio since there will be more
seed nuclei in these conditions. At best, the neutron to seed nucleus
ratios obtained in lower entropy models are marginal for the
production of the neutron-rich $r$-process species
\cite{putthisin1,meyer97}.

The alpha effect occurs at the epoch of alpha particle formation. As
the temperature drops, essentially all the protons and most of the
neutrons in the ejecta lock themselves into alpha particles which have
a large binding energy. This phenomenon ultimately will tend to push
the electron fraction higher, towards $Y_e=0.5$. The increase in $Y_e$
comes about because protons produced by electron neutrino capture on
neutrons will in turn capture more neutrons to bind into alpha
particles, reducing the number of free neutrons available for the
$r$-process \cite{alpha}. This effect has been shown to be the biggest
impediment to achieving an acceptable $r$-process yield \cite{MMF}. 
Matter-enhanced
neutrino transformation between electron neutrinos and other active
species worsen this problem as it tends to increase electron neutrino
energies. Hence one would like to avoid active-active neutrino
mixing. Indeed this is how the bounds of Ref. 54 on neutrino
parameters were obtained. 

One way to avoid or reduce the efficacy of the alpha effect is to
reduce the flux of electron neutrinos at some point above the surface
of the neutron star.  However, in models of the neutrino-driven wind a
large flux of electron neutrinos is required to lift the material off
the surface of the neutron star. In fact since nucleons are
gravitationally bound by about $\sim100\,{\rm MeV}$ near the surface
of the neutron star, and since each neutrino has an energy $\sim
10\,{\rm MeV}$, each nucleon must suffer some $\sim 10$ neutrino
interactions to be ejected to infinity. So if we are to reduce the
$\nu_e$ flux we must do so only at relatively large radius, so that
effective neutrino heating already can have occurred. This can be
achieved by the matter-enhanced
active-sterile neutrino transformation in the $\nu_e
\rightleftharpoons \nu_s$ and $\bar\nu_e \rightleftharpoons
\bar{\nu}_s$ channels. In such a scheme \cite{gail1} the lightest
  sterile neutrino would be heavier than the $\nu_e$ and split from it
  by a vacuum mass-squared difference of 3 eV$^2 {\ 
    \lower-1.2pt\vbox{\hbox{\rlap{$<$}\lower5pt\vbox{\hbox{$\sim$}}}}\ 
    } \delta m^2_{es} {\ 
    \lower-1.2pt\vbox{\hbox{\rlap{$<$}\lower5pt\vbox{\hbox{$\sim$}}}}\ 
    }$ 70 eV$^2$ with vacuum mixing angle $\sin^2 2\theta_{es} >
  10^{-4}$.
Further details for this scenario is given in Ref. 52. A similar
mechanism exploiting different active-sterile mixing channels is
presented in Ref. 83.

\section{Other Neutrino Oscillation Experiments}

An introduction to neutrino oscillations in vacuum was given in
Section 2.6. In this section we briefly summarize some recent results
from several neutrino oscillation experiments. From Eq. (47) one can
write the appearance probability of the other flavor
\begin{equation}
  \label{last}
  P = \sin^2 2\theta_v \sin^2 (1.27 \delta m^2 L /E). 
\end{equation}
In Eq. (\ref{last}) $\delta m^2$ is measured
in eV$^2$ and $L/E$ in m/MeV. Neutrino oscillation experiments are
somewhat arbitrarily divided into two classes: short-baseline and
long-baseline.\footnote{Atmospheric neutrino experiments can be
  considered very-long baseline experiments.}  
As Eq. (\ref{last}) indicates, the longer is the
baseline, $L$, the more sensitive the experiment is to smaller values
of $\delta m^2$.

The Liquid Scintillator Neutrino Detector (LSND)
collaboration at Los Alamos National Laboratory reported a significant
``oscillation-like'' excess \cite{lsnd}. This detector sits 30 meters
away from the beam dump at the old Los Alamos Meson Physics Facility
(LAMPF). The experimental apparatus is designed to produce a beam of
muon antineutrinos with as little contamination as possible from the 
electron antineutrinos. If these muon antineutrinos oscillate into
electron antineutrinos, such formed electron antineutrinos would interact
with protons in the detector, creating a positron and a neutron. This
neutron, after some time, binds with a proton to form a deuteron,
giving a photon with a characteristic energy of 2.2 MeV. The
experiment observes these photons as well as the positron's Cerenkov
track. When they identify both signatures together, the LSND group
first reported seeing nine events versus an expected background of two
events coming from the electron antineutrinos from sources other than
the muon antineutrino oscillation. This signal received a lot of
attention and a fair amount of criticism. Indeed, a dissenting member
of the LSND team has performed a data analysis of his own which finds
no positive signal above the expected background \cite{hill}. The LSND
result is unlikely to be a statistical fluctuation. The excess
persists after analyzing more data \cite{newlsnd}, including $\mu$
decay in flight, which has different systematics and
backgrounds from the decay at rest analysis.
However, the KARMEN 
collaboration, carrying out a similar (but not identical) experiment
at Rutherford Laboratory in England, reported no evidence for neutrino
oscillations in a parameter space which largely overlaps with that of
LSND \cite{karmen}. 

Recently CHOOZ, a long-baseline reactor-neutrino vacuum-oscillation
experiment, announced \cite{chooz} first results. This experiment,
operating in a reactor in France detects 
electron antineutrinos by a liquid scintillation calorimeter located
at a distance of about 1 km from the source.  From the statistical
agreement between detected and expected neutrino event rates, they
find (at 90
in the electron antineutrino disappearance mode for the parameter
region given approximately by $\delta m^2 > 0.9 \cdot 10^{-3}$ eV$^2$
for maximum mixing and $\sin^2 2 \theta > 0.18$ for large values of 
$\delta m^2$. The measurement
of the atmospheric electron and muon neutrino zenith angle
distributions at SuperKamiokande, taken together with the
CHOOZ data presents a strong evidence for the mixing of muon
neutrinos with either tau neutrinos or sterile neutrinos. 

This work was supported in part by the National Science Foundation,
the US Department of Energy, and the University of Wisconsin Research 
Committee with funds granted by the Wisconsin Alumni Research
Foundation.

\end{document}